\documentclass[12pt,a4paper,twoside]{article}

\usepackage{amsmath,amssymb,amsfonts,amscd}
\usepackage{cancel}
\usepackage{mathtools}
\usepackage{theorem}
\usepackage{color}
\usepackage{xcolor}
\usepackage{graphicx}
\usepackage{mdframed}
\usepackage{helvet}
\usepackage{verbatim}
\usepackage{enumitem}
\usepackage{mathrsfs}
\usepackage{stackrel}
\usepackage{enumitem}
\usepackage{framed}
\usepackage{blindtext}
\usepackage{subdepth}
\usepackage{accents}
\usepackage{amsmath}
\usepackage{amssymb}
\usepackage{theorem}
\usepackage{subdepth}
\usepackage{graphicx}
\usepackage{tikz,pgfplots}
\usetikzlibrary{positioning}

\newcommand{\R}{{\mathord{\mathbb R}}}
\newcommand{\Z}{{\mathord{\mathbb Z}}}
\newcommand{\N}{{\mathord{\mathbb N}}}
\newcommand{\C}{{\mathord{\mathbb C}}}
\newcommand{\T}{{\mathord{\mathbb T}}}
\newcommand{\mA}{\mathcal A}
\newcommand{\mB}{\mathcal B}
\newcommand{\mC}{\mathcal C}
\newcommand{\mD}{\mathcal D}
\newcommand{\mE}{\mathcal E}
\newcommand{\mF}{\mathcal F}

\newcommand{\mG}{\mathcal G}
\newcommand{\mH}{\mathcal H}

\newcommand{\mK}{\mathcal K}
\newcommand{\mL}{\mathcal L}
\newcommand{\mM}{\mathcal M}

\newcommand{\mP}{\mathcal P}

\newcommand{\mS}{\mathcal S}

\newcommand{\mZ}{\mathcal Z}

\newcommand{\e}{{\rm e}}

\newcommand{\pf}{{\rm pf}}

\newcommand{\ran}{{\rm ran\,}}		
\newcommand{\sign}{{\rm sign}}
\newcommand{\sgn}{{\rm sgn}}
\newcommand{\spa}{{\rm span\,}}
\newcommand{\tr}{{\rm tr}}
\newcommand{\res}{{\rm res}}
\def\slim{\mathop{\rm s-lim}}
\def\srlim{\mathop{\rm sr-lim}}

\def\Blim{\mathop{\rm \mB-lim}}
\newcommand{\Aut}{{\rm Aut}}
\newcommand{\card}{{\rm card}}

\newcommand{\rd}{\hspace{-0.5mm}{\rm d}}

\newcommand{\diag}{{\rm diag}}

\newcommand{\fh}{{{\mathfrak h}}}

\newcommand{\fH}{{{\mathfrak H}}}

\newcommand{\ff}{{{\mathfrak f}}}

\newcommand{\fF}{{\mathfrak F}}

\newcommand{\fA}{{\mathfrak A}}

\newcommand{\fP}{{\mathfrak P}}

\newcommand{\kp}{\kappa}
\newcommand{\lm}{\lambda}
\newcommand{\veps}{{\varepsilon}}
\newcommand{\dom}{{\rm dom}}
\newcommand{\eig}{{\rm eig}}
\newcommand{\spec}{{\rm spec}}

\newcommand{\vi}{\varphi}

\colorlet{shadecolor}{gray!25}

\renewcommand{\Re}{{\rm Re}}
\renewcommand{\Im}{{\rm Im}}

\renewcommand{\l}{\langle}
\renewcommand{\r}{\rangle}

\newcommand{\ii}{{\rm i}}

\newtheorem{thm}{Theorem}
\newtheorem{proposition}[thm]{Proposition}
\newtheorem{lemma}[thm]{Lemma}
\newtheorem{definition}[thm]{Definition}
{\theorembodyfont{\upshape} \newtheorem{remark}[thm]{\it Remark}}
\newtheorem{example}[thm]{Example}
\newtheorem{assumption}[thm]{Assumption}
\newtheorem{setting}[thm]{Setting}
\newtheorem{corollary}[thm]{Corollary}
\newcommand{\bd}{\begin{definition}}
\newcommand{\ed}{\end{definition}\vspace{1mm}}
\newcommand{\bt}{\begin{thm}}
\newcommand{\et}{\end{thm}\vspace{2mm}}
\newcommand{\bc}{\begin{corollary}}
\newcommand{\ec}{\end{corollary}\vspace{2mm}}
\newcommand{\bl}{\begin{lemma}}
\newcommand{\el}{\end{lemma}\vspace{2mm}}
\newcommand{\bp}{\begin{proposition}}
\newcommand{\ep}{\end{proposition}\vspace{2mm}}
\newcommand{\bx}{\begin{example}}
\newcommand{\ex}{\end{example}}
\newcommand{\br}{\begin{remark}}
\newcommand{\er}{\end{remark}}
\newcommand{\bass}{\begin{assumption}}
\newcommand{\eass}{\end{assumption}\vspace{2mm}}
\newcommand{\bset}{\begin{setting}}
\newcommand{\eset}{\end{setting}\vspace{1mm}}
\newcommand{\bprf}{\noindent{\it Proof.}\hspace{2mm}}
\newcommand{\eprf}{\hfill $\Box$\vspace{5mm}}

\newcommand{\ie}{i.e.}
\def\bas#1\eas{\begin{align*}#1\end{align*}}
\def\ba#1\ea{\begin{align}#1\end{align}}
\newcommand{\bn}{\begin{enumerate}}
\newcommand{\en}{\end{enumerate}}

\newcommand{\Ev}{{\rm Ev}}
\newcommand{\Od}{{\rm Od}}

\newcommand{\bcm}{}

\newcommand{\Int}{\int_{-\pi}^\pi\frac{\rd k}{2\pi}\hspace{0.5mm}}

\DeclareMathOperator{\sinc}{sinc}

\begin{document}
\pagestyle{myheadings}
\markboth{Walter H. Aschbacher}{Heat flux in quasifree R/L mover systems}
\title{Heat flux in general quasifree fermionic\\ right mover/left mover systems}
\author{Walter H. Aschbacher\footnote{walter.aschbacher@univ-tln.fr}
\\ \\
Universit\'e de Toulon, Aix Marseille Univ, CNRS, CPT, Toulon, France
}
\date{}
\maketitle

\begin{abstract}
With the help of time-dependent scattering theory on the observable algebra of infinitely extended quasifree fermionic chains, we introduce a general class of so-called right mover/left mover states which are inspired by the nonequilibrium steady states for the prototypical nonequilibrium configuration of a finite sample coupled to two thermal reservoirs at different temperatures. Under the assumption of spatial translation invariance, we relate the 2-point operator of such a right mover/left mover state to the asymptotic velocity of the system and prove that the system is thermodynamically nontrivial in the sense that its entropy production rate is strictly positive. Our study of these not necessarily gauge-invariant systems covers and substantially generalizes well-known quasifree fermionic chains  and opens the way for a more systematic analysis of the heat flux in such systems.
\end{abstract}



\section{Introduction}
\label{sec:intro}
The rigorous study, from first  principles, of open quantum systems is of fundamental importance for a deepened understanding of their thermodynamic properties in and out of equilibrium. Since, by definition, open quantum systems have a very large number of degrees of freedom and since the finite accuracy of any feasible experiment does not allow an empirical distinction between an infinite system and a finite system with sufficiently many degrees of freedom, a powerful strategy consists in approximating (in a somewhat reversed sense) the actual finite system by an idealized one with infinitely many degrees of freedom (see \cite{Primas} for an extensive discussion of this idealization and its implications). Furthermore, it is conceptually more appealing and often mathematically more rigorous to treat the idealized system from the outset in a framework designed for infinite systems rather than taking the thermodynamic limit at an intermediate or the final stage.

One of the most important axiomatic frameworks for the study of such idealized  infinite systems is the so-called algebraic approach to quantum mechanics based on operator algebras. Indeed, after having been heavily used from as early as the 1960s on, in particular for the quantum statistical description of quantum systems in thermal equilibrium (see, for example, \cite{Emch2009, Sewell2014, BR, AJP2006}), the benefits of this framework have again started to unfold more recently in the physically much more general situation of open quantum systems out of equilibrium. Although the most interesting phenomena which emerge on the macroscopic level are not restricted to systems in thermal equilibrium but, quite the contrary, often occur out of equilibrium, our general theoretical understanding of nonequilibrium order and phase transitions is substantially less developed since, in particular, the effect of the dynamics becomes much more important out of equilibrium. 

Most of the rather scarce mathematically rigorous results have been obtained for the so-called nonequilibrium steady states (NESSs) introduced by Ruelle in \cite{Ruelle2001} by means of scattering theory on the algebra of observables. An important role in the construction of such NESSs is played by the so-called quasifree fermionic systems, and this is true not only because of their mathematical accessibility but also when it comes to real physical applications. Indeed, from a mathematical point of view, these systems allow for a simple and powerful representation independent description since scattering theory on the fermionic algebra of observables boils down to scattering theory on the underlying 1-particle Hilbert space over which the fermionic algebra is constructed. This restriction of the dynamics to the 1-particle sector opens the way for a rigorous mathematical analysis of many purely quantum mechanical properties which are of fundamental physical interest. But, beyond their importance due to their mathematical accessibility, quasifree fermionic systems effectively describe nature: aside from the various electronic systems in their independent electron approximation, they also play an important role in the rigorous approach to physically realizable spin systems. An important member of the family of Heisenberg spin chains is the so-called XY model, introduced in 1961 in  \cite{LSM1961} (see also \cite{Nambu1950} for the so-called isotropic case),  for which a physical realization has already been identified in the late 1960s (see \cite{CSP1969} for example).  The impact of the XY model on the experimental, numerical, theoretical, and mathematical research activity in the field of low-dimensional magnetic systems is ongoing ever since (see \cite{MK2004} for example). 

In the present paper, we consider a 1-dimensional quantum mechanical system whose configuration space is the 2-sided infinite discrete line $\Z$ and whose  algebra of observables is the CAR (canonical anticommutation relations) algebra over the 1-particle Hilbert space of the square-summable functions over $\Z$. Using scattering theory on the 1-particle Hilbert space, we then introduce a class of states over the CAR algebra which we call the class of right mover/left mover states (R/L movers). For a given Hamiltonian on the 1-particle Hilbert space, an R/L mover is specified by a 2-point operator whose main part consists of a mixture of two independent species stemming from the asymptotic right and left side of $\Z$, carrying the inverse temperatures $\beta_R$ and $\beta_L$, respectively, of the right and left reservoir with configuration spaces 
\ba
\Z_R
&:=\{x\in\Z\,|\, x\ge x_R+1\},\\
\Z_L
&:=\{x\in\Z\,|\, x\le x_L-1\},
\ea
where $x_R, x_L\in\Z$ are fixed and satisfy $x_L\le x_R$. Moreover, the finite piece 
\ba
\label{ZS}
\Z_S
:=\{x\in\Z\,|\, x_L\le x\le x_R\},
\ea
containing $n_S:=x_R-x_L+1\ge 1$ sites, plays the role of the configuration space of the confined sample. The prototypical example of such an R/L-mover is the NESS constructed as the large time limit of the averaged trajectory of a time-evolved initial state which is the decoupled product of three thermal equilibrium states over the corresponding configuration spaces (as somewhat degenerate examples, thermal equilibrium states and ground states are also covered by our setting). The first rigorous construction of such a NESS by means of time-dependent scattering theory has been carried out in \cite{AP2003} for the XY model (see also \cite{AH2000} for the special case of the so-called isotropic XY model [or XX model] using different asymptotic approximation methods) and, to the best of our knowledge, very few models have been rigorously studied within the framework of Ruelle's scattering approach since then (see, for example, \cite{MO2003, AJPP2007, A2011, A2016, A2019, AJR2020}). The setting of the present paper, being thus kept, at various places, at a mathematically rather general level in order to highlight the structural dependence on the different ingredients (and in view of future generalizations), allows for the study of more general and not necessarily gauge-invariant fermionic systems covering and generalizing several well-known models of spin chains (as, for example, the NESS for the XY model from \cite{AP2003}, the Suzuki model, etc.). Furthermore, under the additional assumption of translation invariance of the Hamiltonian (whose breaking will be studied elsewhere, but see also \cite{A2011, A2016, A2019}) and substantially generalizing the approach of \cite{AP2003}, the 2-point operator of the R/L mover is explicitly linked to the asymptotic velocity of the system allowing for a rigorous and detailed study of the heat flux in general quasifree R/L mover systems whose sample is coupled to the reservoirs through short range forces across the boundaries. As a consequence of the structural form of the heat flux, we obtain strict positivity of the entropy production, i.e., thermodynamical nontriviality,  for the whole class of such quasifree R/L mover systems.

\vspace{2mm}

The paper is organized as follows.

\vspace{1mm}

{\it Section \ref{sec:infinite} (Infinite fermionic systems)}\, We introduce the framework for the  systems to be studied, \ie, the CAR algebra of observables, its selfdual generalization, the quasifree dynamics generated by selfdual Hamiltonians, and the states on the observable algebra with their corresponding 2-point operators. 

{\it Section \ref{sec:LR} (Right mover/left mover states)}\, In order to be able to define R/L movers, we introduce the asymptotic projections for the underlying right/left geometry and general Fermi functions. The class of R/L mover 2-point operators is defined under simple assumptions on the selfdual Hamiltonian and for general so-called initial 2-point operators. We also discuss thermal equilibrium and ground states with respect to this framework and the special case of states with gauge-invariant 2-point operators which frequently occurs in practice.

{\it  Section \ref{sec:ness} (Nonequilibrium steady states)}\, 
This section is devoted to the definition and the construction of NESSs, in the nonequilibrium setting at hand,  using Ruelle's time-dependent scattering approach. This class of states serves as the main motivation for the introduction of the R/L movers of Section \ref{sec:LR}. 

{\it Section \ref{sec:ti} (Asymptotic velocity)}\, 
Using the fundamental assumption of translation invariance for the selfdual Hamiltonian, we rigorously determine the asymptotic velocity of the system. The latter is the key ingredient of the so-called R/L mover generator which, together with the selfdual Hamiltonian, determines the main part of the 2-point operator of the R/L mover. 

{\it Section \ref{sec:Ep} (Heat flux)}\, 
We introduce the notion of heat flux and entropy production rate in the R/L mover state. Under the additional assumption that the range of the selfdual Hamiltonian is bounded by the size of the sample, \ie, that there is no direct coupling between the two reservoirs, the R/L mover heat flux is explicitly determined in general and for typical special cases appearing in practice. We also provide examples of several well-known models of spin chains covered by the formalism and explicitly determine instances of new ones. Moreover, under suitable monotonicity conditions on the Fermi function, we prove that the  R/L mover heat flux is nonvanishing and the entropy production strictly positive, \ie, that the system under consideration is thermodynamically nontrivial.

{\it Appendix \ref{app:spectral} (Spectral theory)}\, 
We present a brief summary of the somewhat different approach to spectral theory used in the main body of the paper. Due to the attempt to be, at least conceptually, self-contained, we rather explicitly carry out most of the necessary arguments in this appendix (and also in the main body of the text).

{\it Appendix  \ref{app:mmo} (Matrix multiplication operators)}\, 
Based on Appendix \ref{app:spectral}, we study the functional calculus for the matrix multiplication operators describing selfdual translation invariant observables and derive a criterion for their absolute continuity.

{\it Appendix  \ref{app:TP} (Real trigonometric polynomials)}\, 
We carry out the computations in the ring of  real trigonometric polynomials which, in particular, are needed in the study of the examples in Section \ref{sec:Ep}.

{\it Appendix  \ref{app:cntrJac} (Heat flux contributions)}\, 
This appendix contains some of the lengthy computations from the proof of the main theorem. 

{\it Appendix  \ref{app:Fermi-Pauli} (Hamiltonian densities)}\, 
We display the selfdual second quantization of the local first, second, and third Pauli coefficient of $H$ in the fermionic and the spin picture (the selfdual second quantization of the zeroth Pauli coefficient of $H$ is given in the main body of the paper).

\section{Infinite fermionic systems}
\label{sec:infinite}
In this section, we introduce the operator algebraic setting used to describe the fermionic system under consideration whose extension is infinite. Recall that, in the operator algebraic approach to quantum statistical mechanics, the three fundamental ingredients of a physical system, \ie, the observables, the time evolution, and the states, are given by a $C^\ast$-algebra, by a 1-parameter group of $\ast$-automorphisms, and by normalized positive linear functionals on the observable algebra, respectively  (see \cite{Emch2009, Sewell2014, BR} for example). In order to introduce the notation, let $\mH$ be any separable complex Hilbert space and let $\mL(\mH)$ stand for the bounded linear operators on $\mH$. Moreover, $\mL^\infty(\mH)$, $\mL^1(\mH)$, and $\mL^0(\mH)$ denote the compact operators, the trace class operators, and the operators of finite rank on $\mH$, respectively, and $\|\cdot\|_1$ stands for the trace norm on $\mL^1(\mH)$. Furthermore, if $a_{ij}\in\mL(\mH)$ for all $i,j\in\l1,2\r$, we denote by $A:=[a_{ij}]_{i,j\in\l1,2\r}\in\mL(\mH\oplus\mH)$ the operator on $\mH\oplus\mH$ whose entries are given by $a_{ij}$, where here and in the following, for all $x,y\in\Z$ with $x\le y$, we set 
\ba
\l x,y\r:=
\begin{cases}
\{x,x+1,\ldots,y\}, & x<y,\\
\hfill\{x\}, & x=y.
\end{cases}
\ea
Instead of using the standard basis, it is often useful to expand with respect to the Pauli matrices, \ie, if $a_\alpha\in\mL(\mH)$ for all $\alpha\in\l0,3\r $, we define the operator $A\in\mL(\mH\oplus\mH)$ by
\ba
\label{PauliExp}
A
:=a_0\sigma_0+a\sigma,
\ea
where $\{\sigma_0, \sigma_1, \sigma_2, \sigma_3\}\subseteq \C^{2\times 2}$ stands for the Pauli basis of $\C^{2\times 2}$ which consists of the usual Pauli matrices  $\sigma_1, \sigma_2, \sigma_3\in\C^{2\times 2}$ and the identity $\sigma_0\in\C^{2\times 2}$, 
\ba
\label{pm} 
\sigma_0
:=\left[\begin{array}{cc}
1 & 0
\\ 0&1
\end{array}\right],\quad
\sigma_1
:=\left[\begin{array}{cc}
0 & 1\\
1& 0
\end{array}\right],\quad
\sigma_2
:=\left[\begin{array}{cc}
0 &-\ii\\ 
\ii & 0
\end{array}\right],\quad
\sigma_3
:=\left[\begin{array}{cc}
1 & 0
\\ 0& -1
\end{array}\right].
\ea
Here and in the following, for all $n\in\N:=\{1,2,\ldots\}$, we denote by $\C^{n\times n}$ the complex $n\times n$ matrices and by $\C^{n\times n}_a$ the skew-symmetric complex $n\times n$ matrices.  Moreover, $a\in\mL(\mH)^3$ and $\sigma\in (\C^{2\times 2})^3$ are written as $a:=[a_1,a_2,a_3]$ and $\sigma:=[\sigma_1,\sigma_2,\sigma_2]$, and we set $a\sigma:=a_1\sigma_1+a_2\sigma_2+a_3\sigma_3$, where, for all $b\in\mL(\mH)$  and all $M\in\C^{2\times 2}$ having entries $m_{ij}$ with $i,j\in\l1,2\r$, the operator $bM\in\mL(\mH\oplus\mH)$ is defined by $(bM)_{ij}:=m_{ij}b$ for all $i,j\in\l1,2\r $ (and we have $(bM)(cN)=(bc)(MN)$ for all $b,c\in\mL(\mH)$ and all $M,N\in \C^{2\times 2}$). Conversely, any $A\in\mL(\mH\oplus\mH)$ can be written uniquely in the form \eqref{PauliExp} which we call the Pauli expansion of $A$.  Moreover, if $A, B\in\mL(\mH\oplus\mH)$ have the Pauli expansions $A=a_0\sigma_0+a\sigma$ and $B=b_0\sigma_0+b\sigma$, their product has the Pauli expansion
\ba
\label{prod}
AB
=(a_0b_0+ab)\sigma_0 +(a_0b+ab_0+\ii a\wedge b)\sigma,
\ea
where we set $ab:=a_1b_1+a_2b_2+a_3b_3$, $a_0b:=[a_0b_1, a_0b_2,a_0b_3]$, $ab_0:=[a_1b_0,a_2b_0,a_3b_0]$, and the vector $a\wedge b\in\mL(\mH)^3$ is given by $(a\wedge b)_i:=\sum_{j,k\in\l1,3\r}\veps_{ijk}
\hspace{0.5mm} a_jb_k$ for all $i\in\l1,3\r$, and $\veps_{ijk}$ with $i,j,k\in\l1,3\r$ is the usual Levi-Civita symbol. Of course, all the foregoing considerations can be analogously applied to the case of antilinear operators which we denote by $\bar\mL(\mH)$. Finally, $\ell^2(\Z)$ will stand for the usual separable complex Hilbert space of square-summable complex-valued functions on $\Z$, and for elements $A$ and $B$ in the various sets in question below, the commutator and the anticommutator of $A$ and $B$ are denoted as usual by $[A,B]:=AB-BA$ and $\{A,B\}:=AB+BA$, respectively.

In the following, we will make use of the so-called selfdual setting. For our case, the 1-particle Hilbert space in the selfdual setting is the direct sum of the usual 1-particle Hilbert space with itself. Here and there, we will make brief remarks about this underlying general framework.

\bd[Observables]
\label{def:obs}
\hfill
\bn[label=(\alph*),ref={\it (\alph*)}]
\setlength{\itemsep}{0mm}
\item 
\label{def:obs-a}
The 1-particle position Hilbert space and its doubling are defined by
\ba
\label{1ptcl}
\fh&:=\ell^2(\Z),\\
\label{1ptcl-dbld}
\fH
&:=\fh\oplus\fh.
\ea
Abusing notation, the usual scalar products, the corresponding induced norms (as well as the corresponding operator norms) on both $\fh$ and $\fH$ are all denoted by $(\hspace{0.2mm}\cdot\hspace{0.4mm},\cdot)$ and $\|\cdot\|$, respectively.

\item
\label{def:obs-b}
Let the map $\zeta\in\bar\mL(\fh)$ be given by $\zeta f:=\bar f$ for all $f\in\fh$, where $\bar f$ is the usual complex conjugation of $f$. The antiunitary involution $\Gamma\in\bar\mL(\fH)$, defined by the block anti-diagonal lifting of $\zeta$ to $\fH$ as
\ba
\label{def:Gamma}
\Gamma
:=\zeta\sigma_1,
\ea
is called the conjugation of $\fH$.

\item
\label{def:obs-c}
The algebra of observables, denoted by $\fA$, is defined to be the CAR algebra over $\fh$, 
\ba
\fA
:={\rm CAR}(\fh).
\ea
The generators are denoted, as usual, by $1$, $a(f)$, and $a^\ast(f)$ for all $f\in\fh$ (and the $C^\ast$-norm of $\fA$, as many other norms within their context, by $\|\cdot\|$).

\item
\label{def:obs-e}
The complex linear map $B:\fH\to\fA$, defined, for all $F:=f_1\oplus f_2 \in\fH$, by
\ba
B(F)
:=a^\ast(f_1)+a(\zeta f_2),
\ea
satisfies the selfdual CARs
\ba
\label{B2}
B^\ast(F)
&=B(\Gamma F),\\
\label{B3}
\{B^\ast(F),B(G)\}
&=(F,G)\hspace{0.4mm}1.
\ea
The elements $B(F)$ for all $F\in\fH$ are called the selfdual generators of $\fA$.

\item
\label{def:obs-d}
The complex linear map $b:\mL^1(\fH)\to\fA$, called the selfdual second quantization, is defined, for all $A\in\mL^0(\fH)$,  by
\ba
\label{sdsq}
b(A)
:=\sum_{i=1}^m B(G_i)B^\ast(F_i),
\ea
where $m\in\N$ and $\{F_i, G_i\}_{i\in\l1,m\r}\subseteq\fH$ are such that $A:=\sum_{i=1}^m(F_i,\hspace{0.5mm}\cdot\hspace{0.5mm})G_i$. Moreover, for all $A\in\mL^1(\fH)$, the selfdual second quantization is defined by
\ba
\label{bL1}
b(A)
:=\lim_{n\to\infty} b(A_n),
\ea
where the sequence $(A_n)_{n\in\N}$ in $\mL^0(\fH)$ is such that $\lim_{n\to\infty}\|A-A_n\|_1=0$.
\en
\ed

In the following, for all $r,s\in\R$, we denote by $\delta_{r,s}$ the usual Kronecker symbol and we call Kronecker basis of $\fh$ the complete orthonormal system $\{\delta_y\}_{y\in\Z}\subseteq\fh$, where, for all $y\in\Z$, the function $\delta_y\in\fh$ is defined by $\delta_y(x):=\delta_{x,y}$ for all $x\in\Z$. Moreover, we set $a_x:=a(\delta_x)$ and $a_x^\ast:=a^\ast(\delta_x)$ for all $x\in\Z$. 

\br
\label{rem:SDC-XY}
The algebra of observables $\fA$ is $\ast$-isomorphic to $\overline{\rm SDC}(\fH,\Gamma)$, the $C^\ast$-completed (with respect to the $C^\ast$-norm) selfdual CAR algebra over $\fH$ and $\Gamma$.  The selfdual setting is a very useful general framework which has been  introduced and developed in \cite{Araki1968, Araki1971, Araki1987} for general $\fH$ and $\Gamma$ not necessarily of the form \eqref{1ptcl-dbld} and \eqref{def:Gamma} (see there for a more detailed description of the selfdual objects used in the following).
In particular, the selfdual setting allows for the description of general non gauge-invariant quasifree fermionic systems such as, for example, the prominent XY model from \cite{LSM1961} whose Hamiltonian density has the form
\ba
\label{XYDensity}
(1+\gamma)\,\sigma_1^{(x)}
\sigma_1^{(x+1)}+(1-\gamma)\,\sigma_2^{(x)}\sigma_2^{(x+1)},
\ea
where the superscripts denote the sites in $\Z$ of the local Hilbert space of the spin chain on which the Pauli matrices act. In order to establish a bridge between the spin picture and the fermionic picture, the generalization from \cite{Araki1984} of the Jordan-Wigner transformation for 1-dimensional systems whose configuration space extends infinitely in both directions makes use of the so-called crossed product of the algebra $\fP$ of the Pauli spins over $\Z$ (a Glimm or UHF [\ie, uniformly hyperfinite] algebra as is $\fA$) by the involutive automorphism $\alpha\in\Aut(\fP)$ describing the rotation around the 3-axis by an angle of $\pi$ of the observables on the nonpositive sites (and, thereby, makes it mathematically rigorous for the Jordan-Wigner transformation to be anchored at minus infinity). Up to $\ast$-isomorphism equivalence, it is given by the $C^\ast$-subalgebra 
\ba
\bigg\{\begin{bmatrix} A & B\\ \alpha(B) &\alpha(A)\end{bmatrix}\bigg| A,B \in\fP\bigg\}
\subseteq \fP^{2\times 2},
\ea
where $ \fP^{2\times 2}$ stands for the $C^\ast$-algebra of all $2\times 2$ matrices with entries in $\fP$ (with respect to the naturally generalized matrix operations and the Hilbert $C^\ast$-module norm).
Using this bridge, \eqref{XYDensity} can be expressed in the fermionic picture and becomes (up to a global prefactor)
\ba
\label{FDensity}
a_x^\ast a_{x+1} + a_{x+1}^\ast a_x
+\gamma (a_x^\ast a_{x+1}^\ast +a_{x+1}a_x).
\ea
In order to treat the anisotropic case $\gamma\neq 0$, i.e., the case in which there is an asymmetry between the first and the second term in \eqref{XYDensity}, the selfdual setting is most natural since gauge invariance is broken in \eqref{FDensity}. Hence, due to the presence of the $\gamma$-term, the anisotropy Hamiltonian acquires non-diagonal components with respect to $\fH=\fh\oplus\fh$ (see Example \ref{ex:XY} in Section \ref{sec:Ep}). In many respects, the truly anisotropic XY model is substantially more complicated than the isotropic one (see \cite{A2019} for example).
\er

\br
\label{rem:prop-b(A)}
For all $F\in\fH$, the selfdual generator $B(F)\in\fA$ has the norm (see \cite{Araki1987})
\ba
\label{norm-B(F)}
\|B(F)\|
=\frac{1}{\sqrt{2}}\sqrt{\|F\|^2+\sqrt{\|F\|^4-|(F,\Gamma F)|^2}},
\ea
from which we can infer that
\ba
\label{norm-B(F)-est}
\frac{1}{\sqrt{2}}\hspace{0.2mm}\|F\|
\le\|B(F)\|\le \|F\|.
\ea
Furthermore, the selfdual second quantization \eqref{sdsq} is well-defined since $b(A)$ does not depend on the choice of the functions $F_1,\ldots, G_m\in\fH$ which represent $A\in\mL^0(\fH)$.
As to \eqref{bL1}, the limit exists and is independent of the sequence $(A_n)_{n\in\N}$ in $\mL^0(\fH)$ which approximates $A\in\mL^1(\fH)$ in the trace norm (such a sequence exists since $\mL^0(\fH)$ is dense in $\mL^1(\fH)$ with respect to the trace norm). Moreover, for all $A\in\mL^1(\fH)$, it holds that $b(A)^\ast=b(A^\ast)$ (and $\mL^1(\fH)$ is a 2-sided $\ast$-ideal of $\mL(\fH)$). If $A\in\mL^1(\fH)$ satisfies the condition $\Gamma A\Gamma=-A^\ast$, we have
\ba
\label{norm-b(A)}
\frac14\hspace{0.2mm}\|A\|_1
\le \|b(A)\|
\le \|A\|_1.  
\ea
If, in addition, $A$ is selfadjoint, we even have $\|b(A)\|=\|A\|_1$.
\er

In the following, the set of $\ast$-automorphisms on the algebra of observables $\fA$ is denoted by $\Aut(\fA)$. 

The so-called Bogoliubov $\ast$-automorphisms to be defined next play an important role in the theory of quasifree fermionic systems.

\bd[Bogoliubov $\ast$-automorphism]
\label{def:Bog}
\hfill
\bn[label=(\alph*),ref={\it (\alph*)}]
\setlength{\itemsep}{0mm}
\item 
\label{def:Bog-a}
A unitary operator $U\in\mL(\fH)$ is called a Bogoliubov operator if $[U,\Gamma]=0$. 

\item
\label{def:Bog-b}
 Let $U\in\mL(\fH)$ be a Bogoliubov operator. The $\ast$-automorphism $\tau_U\in\Aut(\fA)$ defined, for all $F\in\fH$, by
\ba
\tau_U(B(F))
:= B(UF),
\ea
and suitably extended to the whole of $\fA$, is called the Bogoliubov $\ast$-automorphism (induced by $U$).
\en
\ed

\br
Note that $\tau_U$ preserves the properties of Definition \ref{def:obs} \ref{def:obs-e}, \ie,  $B':=\tau_U\circ B:\fH\to\fA$ is complex linear and satisfies the selfdual CARs \eqref{B2}-\eqref{B3}. 
\er

The following 1-particle Hilbert space isometries will be frequently used in the sequel.

\bd[Isometries]
\label{def:iso}
\hfill
\bn[label=(\alph*),ref={\it (\alph*)}]
\setlength{\itemsep}{0mm}
\item 
\label{def:iso-a}
The (right) translation $\theta\in\mL(\fh)$ is defined by $(\theta f)(x):=f(x-1)$ for all $f\in\fh$ and all $x\in\Z$. Its lifting to $\fH$ is given by the Bogoliubov operator $\Theta:=\theta\sigma_0\in\mL(\fH)$.

\item 
\label{def:iso-b}
The parity $\xi\in\mL(\fh)$ is defined by $(\xi f)(x):=f(-x)$  for all $f\in\fh$ and all $x\in\Z$. 

\item 
\label{def:iso-c}
For all $\vi\in\R$, the Bogoliubov operator $U_\vi\in\mL(\fH)$ defined by
\ba
\label{rem:gauge-1}
U_\vi
:=\e^{\ii\vi}1\oplus \e^{-\ii\vi}1,
\ea
is called a gauge transformation.
\en
\ed

In the following, the map $\tau:\R\to\Aut(\fA)$, written as $t\to\tau^t$,  is called a dynamics on $\fA$ if it is a group homomorphism between the additive group $\R$ and the group $\Aut(\fA)$ (with respect to composition) and if, for all $A\in\fA$, the map $\R\ni t\mapsto\tau^t(A)\in\fA$ is continuous with respect to the $C^\ast$-norm on $\fA$ (the pair $(\fA, \tau)$ is a sometimes called a $C^\ast$-dynamical system). 

\bd[Quasifree dynamics]
\label{set:obs}
\hfill
\bn[label=(\alph*),ref={\it (\alph*)}]
\setlength{\itemsep}{0mm}

\item
\label{set:obs-a}
An operator $H\in\mL(\fH)$ is called a Hamiltonian if
\ba
\label{H1}
H^\ast
&=H,\\
\label{H2}
\Gamma H \Gamma
&=-H.
\ea

\item
\label{set:obs-b}
Let $H\in\mL(\fH)$ be a Hamiltonian. The dynamics $\tau:\R\to\Aut(\fA)$ defined, for all $t\in\R$ and all $F\in\fH$, by
\ba
\label{qfd-H}
\tau^t(B(F))
:=B(\e^{\ii t H}F),
\ea
and suitably extended to the whole of $\fA$, is called the quasifree dynamics (generated by $H$).
\en
\ed

\br
Due to \eqref{H1}-\eqref{H2}, the map $\R\ni t\mapsto \tau^t\in\Aut(\fA)$ given by \eqref{qfd-H} is a 1-parameter group of Bogoliubov $\ast$-automorphisms induced by the 1-parameter group of Bogoliubov operators $\R\ni t\mapsto \e^{\ii t H}\in\mL(\fH)$ (see \eqref{eithGmm} in the proof of Proposition \ref{prop:sym} \ref{prop:sym-2} below).
\er

The following class of operators characterizes the expectation values of all the quadratic observables in the states we are interested in. The set of states over the observable algebra $\fA$ is denoted by $\mE_\fA$.

\bd[2-point operator]
\label{def:2pt-op}
An operator $T\in\mL(\fH)$ having the properties
\ba
\label{def:2pt-op-1}
T^\ast
&=T,\\
\label{def:2pt-op-2}
\Gamma T \Gamma
&=1-T,\\
\label{def:2pt-op-3}
0
\le 
&\hspace{1.5mm}T
\le 1,
\ea
is called a 2-point operator.
\ed

\br
Since, by definition, a state $\omega\in\mE_\fA$ is a normalized positive linear functional on $\fA$, \eqref{norm-B(F)-est} yields $|\omega(B^\ast(F)B(G))|\le \|F\|\|G\|$ for all $F,G\in\fH$.
Hence, the map $\fH\times\fH\ni (F,G)\mapsto \omega(B^\ast(F)B(G))\in\C$ is a bounded sesquilinear form on $\fH\times\fH$ and Riesz's lemma implies that there exists a unique $T\in\mL(\fH)$ such that, for all $F, G\in\fH$, 
\ba
\label{2pt-op}
\omega(B^\ast(F)B(G))
=(F,TG).
\ea
Moreover, due to the positivity of $\omega$, we get $T\ge 0$ and, hence, $T^\ast=T$. Since $\omega$ is normalized, \eqref{B2}-\eqref{B3} yield $\Gamma T\Gamma=1-T$. Finally, since $T\ge 0$, we have $\Gamma T\Gamma\ge 0$, \ie, $1-T\ge 0$, and it follows that the operator $T$ which characterizes the 2-point function in \eqref{2pt-op} is a 2-point operator.
If $\omega\in\mE_\fA$ satisfies \eqref{2pt-op} for a 2-point operator $T\in\mL(\fH)$, we use the notation $\omega_T$.
\er

We next introduce the class of quasifree states, \ie, the states in $\mE_\fA$ whose many-point correlation functions factorize in Pfaffian form. For this purpose, recall that, for all $m\in\N$, the Pfaffian $\pf:\C^{2m\times 2m}\to\C$ is defined, for all $A\in\C^{2m\times 2m}$, by
\ba
\label{pfaff}
\pf(A)
:=\sum_{\pi\in \mP_{2m}}\sgn(\pi) \prod_{i\in\l1,m\r} A_{\pi(2i-1)\hspace{0.2mm}\pi(2i)},
\ea
where the sum is running over all the $(2m)!/(2^{m}m!)$ pairings of the set $\l1,2m\r$, \ie, $\mP_{2m}:=\{\pi\in\mS_{2m}\,|\, \pi(2i-1)<\pi(2i+1) \mbox{ for all } i\in\l1,m-1\r \mbox{ and } \pi(2i-1)<\pi(2i) \mbox{ for all } i\in\l1,m\r\}$, and  $\mS_{2m}$ stands for the symmetric group on $\l1,2m\r$, \ie, the set of all bijections (permutations) $\l1,2m\r\to\l1,2m\r$, 
see Figure \ref{fig:pairings}. Moreover, as above, for all $n\in\N$, we denote by $A=[a_{ij}]_{i,j\in\l1,n\r}$ the matrix $A\in\C^{n\times n}$ with entries $a_{ij}\in\C$ for all $i,j\in\l1,n\r $.

\begin{center}
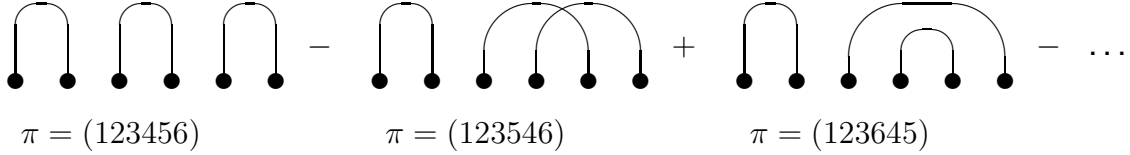
\begin{figure}
\setlength{\unitlength}{13.7mm}
\begin{center}
\begin{picture}(22,2)
\multiput(1,0)(0.5,0){6}{\circle*{0.15}}
\put(1.25,0){\oval(0.5,1.5)[t]}
\put(2.25,0){\oval(0.5,1.5)[t]}
\put(3.25,0){\oval(0.5,1.5)[t]}
\put(1.05,-0.6){$\pi=(123456)$}
\put(3.8,0.25){$-$}
\multiput(4.5,0)(0.5,0){6}{\circle*{0.15}}
\put(4.75,0){\oval(0.5,1.5)[t]}
\put(6,0){\oval(1,1.5)[t]}
\put(6.5,0){\oval(1,1.5)[t]}
\put(4.55,-0.6){$\pi=(123546)$}
\put(7.3,0.25){$+$}
\multiput(8,0)(0.5,0){6}{\circle*{0.15}}
\put(8.25,0){\oval(0.5,1.5)[t]}
\put(9.75,0){\oval(1.5,1.5)[t]}
\put(9.75,0){\oval(0.5,1.0)[t]}
\put(8.05,-0.6){$\pi=(123645)$}
\put(10.8,0.25){$-$}
\put(11.3,0.25){\ldots}
\end{picture}
\end{center}
\vspace{8mm}
\caption{\label{fig:pairings} Some of the 15 pairings for $m=3$. The total 
number of intersections $I$ relates to the signature of the permutation
$\pi$ as $\sgn(\pi)=(-1)^I$.}
\end{figure}
\end{center}

\bd[Quasifree state]
\label{def:qfs}
Let $\omega_T\in\mE_\fA$ be a state with 2-point operator $T\in\mL(\fH)$. If, for all $n\in\N$ and all $\{F_i\}_{i\in\l1,n\r}\subseteq\fH$, it holds that
\ba
\label{qfs}
\omega_T\Big(\prod\nolimits_{i\in\l1,n\r} B(F_i)\Big)
=\begin{cases}
\pf\big([\omega_T(B(F_i)B(F_j))]_{i,j\in\l1,n\r}\big), & \mbox{$n$ even},\\
\hfill 0, & \mbox{$n$ odd},
\end{cases}
\ea
the state $\omega_T$ is called the quasifree state (induced by $T$). 
\ed

\section{Right mover/left mover states}
\label{sec:LR}

In this section, we introduce the R/L mover states whose definition is based on the geometric decomposition of the configuration space into a right part, a central part, and a left part.  To begin with, this decomposition gives rise to the so-called R/L mover generator whose role is to specify the temperatures carried by the R/L movers stemming from the corresponding reservoirs.

In the following, let $\ell^\infty(\Z)$ stand for the usual complex Banach space of bounded complex-valued functions on $\Z$.  For all  $u\in\ell^\infty(\Z)$, the multiplication operator $m[u]\in\mL(\fh)$ is defined, for all $f\in\fh$, by
\ba
\label{mu}
m[u]f
:=uf.
\ea
In view of the Pauli expansion \eqref{PauliExp}, we also write $m[v]:=[m[v_1], m[v_2], m[v_3]]\in\mL(\fh)^3$ for all  $v:=[v_1,v_2,v_3]\in\ell^\infty(\Z)^3$. Moreover, for all $M\subseteq\R$, we denote by $1_M$ the usual characteristic function of $M$, and we use the abbreviations $1_L:=1_{\Z_L}$, $1_S:=1_{\Z_S}$, and $1_R:=1_{\Z_R}$, and $1_\lambda:=1_{\{\lambda\}}$ for all $\lambda\in\R$. Finally, for all $n\in\N$, the family $\{P_1,\ldots, P_n\}\subseteq\mL(\fH)$ is called an orthogonal family of projections if $P_iP_j=\delta_{i,j}P_i$ for all $i,j\in\l1,n\r $. If, in addition, $\sum_{i=1}^n P_i=1$, the family is said to be complete.

\bd[1-sided projections]
\label{def:osp}
The operators $q_L, q_R\in\mL(\fh)$ defined by 
\ba
\label{def:QL}
q_L
:=m[1_L],\\
\label{def:QR}
q_R
:=m[1_R],
\ea
and their liftings to $\fH$ by $Q_L:=q_L\hspace{0.2mm}\sigma_0$ and $Q_R:=q_R\hspace{0.2mm}\sigma_0$ are called the 1-sided projections.
\ed

\br
\label{rem:cf}
Setting $q_S:=m[1_S]\in\mL(\fh)$ and $Q_S:=q_S\hspace{0.2mm}\sigma_0$, it follows that $\{Q_L, Q_S, Q_R\}\subseteq\mL(\fH)$ is a complete orthogonal family of orthogonal projections.
\er

In the following, if $H\in\mL(\fH)$ is a Hamiltonian, we denote by $1_{sc}(H), 1_{ac}(H), 1_{pp}(H)\in\mL(\fH)$ the orthogonal projections onto the singularly continuous, the absolutely continuous, and the pure point subspace of $H$, respectively. Moreover, we denote by $\spec(H)$ and $\eig(H)$ the spectrum and the set of all eigenvalues of $H$. Finally, for all $\kp,\lambda\in\{L,S,R\}$, we write
\ba
\label{Hk}
H_\kp
&:=Q_\kp H Q_\kp,\\
\label{Hkl}
H_{\kp\lambda}
&:=Q_\kp HQ_{\lambda}.
\ea

In the course of our study, one or several of the following conditions on the Hamiltonian of the system will be used. For the case of a translation invariant system, \ie, if Assumption \ref{ass:H} \ref{HTheta} below holds, we will also rely on further conditions which we will discuss in Section \ref{sec:ti}.

\newpage

\bass[Hamiltonian]
\label{ass:H}
Let $H\in\mL(\fH)$ be a Hamiltonian. 
\bn[label=(\alph*),ref={\it (\alph*)}]
\setlength{\itemsep}{0mm}
\item
\label{1sc=0}
$1_{sc}(H)=0$

\item
\label{HTheta}
$[H,\Theta]=0$ 

\item
\label{L1}
$H_{LR}\in\mL^1(\fH)$ 
\item
\label{HLR=0}
$H_{LR}=0$

\item
\label{nontrv}
$H\neq z1$ for all $z\in\C$

\en
\eass

\br
\label{rem:nontrv}
If Assumption \ref{ass:H} \ref{nontrv} does not hold, \ie, if there exists $z\in\C$ such that $H=z1$, \eqref{H1}-\eqref{H2} imply that $z=0$. 
\er

In the following, $\slim$ stands for the limit with respect to the strong operator topology on $\mL(\fH)$.

In order to define the R/L mover states, we make use of the large time asymptotic behavior of the 1-sided projections. 

\bd[Asymptotic projections]
\label{def:Asmpt}
Let $H\in\mL(\fH)$ be a Hamiltonian satisfying Assumption \ref{ass:H} \ref{L1} and let $\beta_L, \beta_R\in\R$ be the inverse reservoir temperatures.
\bn[label=(\alph*),ref={\it (\alph*)}]
\setlength{\itemsep}{0mm}
\item
\label{def:Asmpt-1}
The operators $P_L, P_R\in\mL(\fH)$, defined by
\ba
\label{PL}
P_L
&:=\slim_{t\to\infty} \e^{-\ii t H}Q_L \e^{\ii t H}1_{ac}(H),\\
\label{PR}
P_R
&:=\slim_{t\to\infty} \e^{-\ii t H}Q_R \e^{\ii t H}1_{ac}(H),
\ea
are called the asymptotic projections (for $H$).

\item
\label{def:Asmpt-2}
The operator $\Delta\in\mL(\fH)$, defined by
\ba
\label{R/LGen}
\Delta
:=\beta_L P_L+\beta_R P_R,
\ea
is called the R/L mover generator (for $H$ and $\beta_L, \beta_R$).
\en
\ed

For the following, we define $p_{x,y}\in\mL^0(\fH)$ by $p_{x,y}:=(\delta_x,\,\cdot\,)\,\delta_y\in\mL^0(\fh)$ for all $x,y\in\Z$. Moreover, the domain, the range, and the kernel of a given map will be denoted by $\dom, \ran, \ker$, respectively. For any separable complex Hilbert space $\mH$, the commutant of any $\mA\subseteq\mL(\mH)$ is defined by $\mA':=\{A\in\mL(\mH)\,|\, [A,B]=0\mbox{ for all }B\in\mA\}$. Furthermore, $\mB(\R)$ stands for the Borel functions and $\mM(\R)$ for the Borel sets on $\R$ as given in Definition \ref{def:Bf} \ref{Bf:Bsets} (we frequently refer to the appendices in the following). For any $M\in\mM(\R)$, we denote by $|M|$ the (non-complete) Borel-Lebesgue measure of $M$. Finally, for all $M\subseteq\R$, we set $-M:=\{-x\,|\,x\in M\}$.\\

\bp[Asymptotic projections]
\label{prop:sym}
Let $H\in\mL(\fH)$ be a Hamiltonian satisfying Assumption \ref{ass:H} \ref{L1}. Then:
\bn[label=(\alph*),ref={\it (\alph*)}]
\setlength{\itemsep}{0mm}

\item 
\label{prop:sym-1}
The asymptotic projections $P_L, P_R\in\mL(\fH)$ exist.  Moreover, $\{P_L, P_R\}$ is a (not necessarily complete) orthogonal family of orthogonal projections satisfying
\ba
\label{sym:SumP}
P_L+P_R
&=1_{ac}(H).
\ea

\item
\label{prop:sym-2}
For all $\kp\in\{L,R\}$, we have
\ba
\label{sym:PH}
[P_\kp,H]
&=0,\\
\label{sym:PGamma}
[P_\kp,\Gamma]
&=0.
\ea

\item
\label{prop:sym-3}
If, in addition, Assumption \ref{ass:H} \ref{HTheta} is satisfied, we also have
\ba
\label{sym:PTheta}
[P_\kp,\Theta]
=0.
\ea
\en
\ep

\bprf
\ref{prop:sym-1}\, 
Let $\dom(\mu_H):=\{A\in\mL(\fH)\,|\,\slim_{t\to\infty}\e^{-\ii t H}A\,\e^{\ii t H}1_{ac}(H)\mbox{ exists}\}$ be the so-called wave algebra (see \cite{BW} for example) and let the so-called wave morphism $\mu_H:\dom(\mu_H) \subseteq\mL(\fH)\to\mL(\fH)$ be defined, for all $A\in\dom(\mu_H)$, by 
\ba
\label{muH}
\mu_H(A)
:=\slim_{t\to\infty}\e^{-\ii t H}A\,\e^{\ii t H}1_{ac}(H).
\ea
Due to Remark \ref{rem:cf}, we have the decomposition
\ba 
\label{partsH}
H
=H_L+H_S+H_R+H_{LS}+H_{SL}+H_{RS}+H_{SR}+H_{LR}+H_{RL},
\ea
 from which it follows that $Q_LH=H_L+H_{LS}+H_{LR}$ and $HQ_L=H_L+H_{SL}+H_{RL}$. Hence, due to Assumption \ref{ass:H} \ref{L1} (and since $\mL^0(\fH)$ is a 2-sided $\ast$-ideal of $\mL(\fH)$, too), 
we get
\ba
[Q_L,H]
&=H_{LS}+H_{LR}-(H_{SL}+H_{RL})\nonumber\\
&\in\mL^1(\fH),
\ea
where we used that $H_{LS}\in\mL^0(\fH)$ since $Q_S\in\mL^0(\fH)$ (and analogously for $[Q_R,H]\in\mL^1(\fH)$).  Hence, the Kato-Rosenblum theorem from the trace class approach to scattering theory implies that $Q_L, Q_R\in \dom(\mu_H)$,
\ie, we get the first conclusion of part \ref{prop:sym-1}.

We next show that $\{P_L, P_R\}\subseteq\mL(\fH)$ is an orthogonal family of orthogonal projections (which, in general, is incomplete). For this purpose, we note that, since $Q_\kp^2=Q_\kp^\ast=Q_\kp$ for all $\kp\in\{L,R\}$ due to Remark  \ref{rem:cf}, since $\mu_H$ is an algebra homomorphism,
and since $\mu_H(A^\ast)=\mu_H(A)^\ast$ for all $A\in\dom(\mu_H)$ for 
which $A^\ast\in\dom(\mu_H)$ 
(note that, in general,  $\mu_H$ is not a $\ast$-algebra homomorphism because
$\dom(\mu_H)$ is not a $\ast$-algebra),
we find that $P_\kp=\mu_H(Q_\kp)$ is an orthogonal projection for all $\kp\in\{L,R\}$. Moreover, since $Q_LQ_R=0$, we also get $P_LP_R=0$. Finally, since $Q_L+Q_R=1-Q_S$ and since $\mu_H(Q_S)=0$ because we know that $Q_S\in\mL^0(\fH)\subseteq\mL^\infty(\fH)\subseteq\ker(\mu_H)$,
we get 
\ba
P_L+P_R
&=\mu_H(1)-\mu_H(Q_S)\nonumber\\
&=1_{ac}(H).
\ea

\vspace{1mm}
\ref{prop:sym-2}\, 
If $B\in\ran(\mu_H)$, there exists $A\in\dom(\mu_H)$ such that $B=\mu_H(A)$ and, since the strong limit is translation invariant, we get $\e^{-\ii sH}A \e^{\ii sH}\in\dom(\mu_H)$ and $B=\mu_H(\e^{-\ii sH}A \e^{\ii sH})$ for all $s\in\R$.  Hence, $\e^{-\ii sH}B\e^{\ii sH}=\e^{-\ii sH}\mu_H(A)\e^{\ii sH}=\mu_H(\e^{-\ii sH}A \e^{\ii sH})=B$ for all $s\in\R$ and it follows that $\ran(\mu_H)\subseteq\{H\}'$ (we actually know that $\ran(\mu_H)=\{B\in \{H\}'\,|\, 1_{ac}(H)B=B1_{ac}(H)=B\}$), 
\ie, we get \eqref{sym:PH}. As for \eqref{sym:PGamma}, we first note that Lemma \ref{lem:SpecId} \ref{lem:SpecId-a} and Remark \ref{rem:rA} yield $\Gamma E_H(\e_t)\Gamma=E_{\Gamma H\Gamma}(\zeta \e_t)=E_{-H}(\e_{-t})=E_H(\e_t)$ for all $t\in\R$, where, for all $t\in\R$, the function $\e_t\in\mB(\R)$ is defined by $\e_t(x):=\e^{\ii t x}$ for all $x\in\R$ (and $E_A$ stands for resolution of the identity of the selfadjoint operator $A$ as discussed in Appendix \ref{app:spectral}), \ie, we get, for all $t\in\R$, 
\ba 
\label{eithGmm}
[\e^{\ii t H},\Gamma]
=0.
\ea 
Moreover, we also note that, for all $\kp\in\{L,R\}$, 
\ba
\label{QGmm}
[Q_\kp, \Gamma]
=0. 
\ea
Finally, since the absolutely continuous subspace of $H$ is given by $\ran(1_{ac}(H))=\{F\in\fH\,|\, (F,1_M(H)F)=0 \mbox{ for all } M\in\mM(\R) \mbox{ with } |M|=0\}$, since,  again due to Lemma \ref{lem:SpecId} \ref{lem:SpecId-a} and Remark \ref{rem:rA}, we can write that $(\Gamma F,1_M(H)\Gamma F)=(\Gamma E_H(1_M)\Gamma F,F)=(F,1_{-M}(H)F)$ for all $M\in\mM(\R)$ and all $F\in\fH$, and since the Borel-Lebesgue measure is reflection invariant, \ie, since $-M\in\mM(\R)$ and $|\hspace{-1mm}-\hspace{-1mm}M|=|M|$ for all $M\in\mM(\R)$, 
we get $\Gamma F\in\ran(1_{ac}(H))$ for all $F\in\ran(1_{ac}(H))$. This implies that $1_{ac}(H) \Gamma 1_{ac}(H)= \Gamma 1_{ac}(H)$, and since the (anti-linear) adjoints of $1_{ac}(H) \Gamma 1_{ac}(H), \Gamma 1_{ac}(H)\in\bar\mL(\fH)$ are given by $1_{ac}(H) \Gamma 1_{ac}(H)$ and $1_{ac}(H)\Gamma$, respectively, we get
\ba
\label{1acGmm}
[1_{ac}(H), \Gamma]=0.
\ea
Hence, using \eqref{eithGmm}-\eqref{1acGmm}, we arrive at \eqref{sym:PGamma}.

\vspace{1mm}
\ref{prop:sym-3}\, 
We first note that Assumption \ref{ass:H} \ref{HTheta} and the proof of Lemma \ref{lem:SpecId} \ref{lem:SpecId-d} imply that $[\chi(H),\Theta]=0$ for all $\chi\in\mB(\R)$. Hence, since $\e_t\in\mB(\R)$ for all $t\in\R$ and since we know that $1_{ac}(H)=E_H(1_{M_{ac}})$ for some $M_{ac}\in\mM(\R)$, 
we get $[\e^{\ii t H}, \Theta]=0$ for all $t\in\R$ and $[1_{ac}(H), \Theta]=0$. This implies that, for all $\kp\in\{L,R\}$, 
\ba
[P_\kp,\Theta]
=\mu_H([Q_\kp,\Theta]).
\ea
Now, since $[Q_\kp,\Theta]=[q_\kp,\theta]\sigma_0$ for all $\kp\in\{L,R\}$ and since $[q_L,\theta]=-p_{x_L-1,x_L}\in\mL^0(\fh)$ and $[q_R,\theta]=p_{x_R,x_R+1}\in\mL^0(\fh)$, we get  $[Q_\kp,\Theta]\in\mL^0(\fH)\subseteq\ker(\mu_H)$ as in part \ref{prop:sym-1}.
\eprf

In the following, we also denote by $\xi$ the parity operation from Definition \ref{def:iso} \ref{def:iso-b} when applied to a function $\chi:M\to\C$, where $M\subseteq\R$ satisfies $M=-M$. Moreover, the even and odd parts of such a $\chi$ are written as
\ba
\label{Ev}
\Ev(\chi)
&:=\frac12(\chi+\xi\chi),\\
\label{Od}
\Od(\chi)
&:=\frac12(\chi-\xi\chi).
\ea

In order to define our L/R mover states, we introduce the following class of functions.

\bd[Fermi function]
\label{def:Ff}
If $\rho\in\mB(\R)$ has the properties
\ba
\label{def:Ff-1}
\rho
&\ge 0,\\
\label{def:Ff-2}
\Ev(\rho)
&=\frac12,
\ea
it is called a Fermi function.
\ed

\br
Since $\rho=\Ev(\rho)+\Od(\rho)$ for all  $\rho\in\mB(\R)$ and since $\mB(\R)$ is a $\ast$-algebra due to Definition \ref{def:Bf} \ref{Bf:min}, $\rho$ is a Fermi function if and only if there exists an odd function $\mu\in\mB(\R)$ with $-1\le \mu \le 1$ such that $\rho=(1+\mu)/2$.
\er

The following assumption will be used at the end of Section \ref{sec:Ep}.

\bass[Strict positivity]
\label{ass:mono}
Let $\rho\in\mB(\R)$ be a Fermi function and let  $\beta_L, \beta_R\in\R$ be the inverse reservoir temperatures.
\hfill
\bn[label=(\alph*),ref={\it (\alph*)}]
\setlength{\itemsep}{0mm}
\item 
\label{ass:mono-a}
$\rho(x)>\rho(y)$ for all $x,y\in\R$ with $x>y$

\item 
\label{ass:mono-aa}
$\rho'(x)\ge c$ for some $c>0$ and almost all $x\in\R$ 

\item
\label{ass:mono-b}
$\beta_L<\beta_R$
\en
\eass

\br
\label{rem:Lbsg}
Due to Lebesgue's theorem on the differentiability of monotone functions, the derivative
$\rho'$ exists almost everywhere on $\R$ if Assumption \ref{ass:mono} \ref{ass:mono-a} holds.
\er

For the following, recall that, since $\fH$ is a separable Hilbert space, $\eig(H)$ is a countable subset of $\R$, 
and we write 
\ba
\eig(H)
=\{\lambda_i\}_{i\in I},
\ea
where the index set $I$ is empty, finite, or countably infinite (in the following notations, we  stick to the case of a countably infinite number of eigenvalues, \ie, we set $I=\N$). Also recall Definition \ref{def:Asmpt} \ref{def:Asmpt-2} for the R/L generator $\Delta$. 

With the help of the asymptotic projections and the class of Fermi functions, we next define what we call the R/L mover states. 

\bd[R/L mover state]
\label{def:R/L}
Let $H\in\mL(\fH)$ be a Hamiltonian satisfying Assumption \ref{ass:H} \ref{1sc=0} and \ref{L1}. Moreover,  let $T_0\in\mL(\fH)$ be a 2-point operator, called the initial  2-point operator, let $\rho\in\mB(\R)$ be a Fermi function, and let $\beta_L, \beta_R\in\R$ be the inverse reservoir temperatures.
\bn[label=(\alph*),ref={\it (\alph*)}]
\setlength{\itemsep}{0mm}
\item
\label{R/L-1}
An operator $T\in\mL(\fH)$ of the form $T:=T_{ac}+T_{pp}$, where $T_{ac}, T_{pp}\in\mL(\fH)$ are given by
\ba
\label{Rac}
T_{ac}
&:=\rho(\Delta H)1_{ac}(H),\\
\label{Rpp}
T_{pp}
&:=\sum_{\lambda\in\eig(H)}
1_\lambda(H) \hspace{0.2mm} T_0\hspace{0.5mm} 1_\lambda(H),
\ea
is called an R/L mover 2-point operator (for $H$, $T_0$, $\rho$, and $\beta_L, \beta_R$). Moreover, the right hand side of \eqref{Rpp} is defined by $\slim_{N\to\infty}\sum_{i\in\l1,N\r}1_{\lambda_i}(H) \hspace{0.2mm} T_0\hspace{0.5mm} 1_{\lambda_i}(H)$.

\item
\label{R/L-2}
A state whose 2-point operator is an R/L mover 2-point operator is called  an R/L mover state.
\en
\ed

\br
\label{rem:uncond}
First, note that \eqref{Rac} is well-defined since, due to  Proposition \ref{prop:sym} \ref{prop:sym-1} and \eqref{sym:PH}, we have $\Delta^\ast=\Delta$ and $[\Delta,H]=0$, \ie, $\Delta H$ is selfadjoint.
As for \eqref{Rpp}, since, for all $N\in\N$, we have  $\chi_N:=1_{\{\lambda_1,\ldots, \lambda_N\}}\in\mB(\R)$  and $|\chi_N|_\infty\le 1$, we get $\Blim_{N\to\infty}\chi_N=1_{\eig(H)}\in\mB(\R)$ and, hence, Proposition \ref{prop:ext} \ref{prop:ext-2} yields
\ba 
\label{rem:uncond-1}
\slim_{N\to\infty}\chi_N(H)=1_{\eig(H)}(H).
\ea
Moreover, since $\chi_N(H)=\sum_{i\in\l1,N\r }1_{\lambda_i}(H)$ and since  $\{1_{\lambda_i}(H)\}_{i\in\N}$ is an orthogonal family of orthogonal projections, we can write $\|\chi_N(H)F-\chi_M(H)F\|^2=\sum_{i\in\l M+1,N\r }\|1_{\lambda_i}(H)F\|^2$ for all $N,M\in\N$ with $N>M$ and all $F\in\fH$. Hence, due to \eqref{rem:uncond-1}, $\sum_{i\in\l M+1,N\r }\|1_{\lambda_i}(H)F\|^2$ vanishes for sufficiently large $N$ and $M$, \ie, the numerical series  $\sum_{i\in\N}\|1_{\lambda_i}(H)F\|^2:=\lim_{N\to \infty}\sum_{i\in\l1,N\r }\|1_{\lambda_i}(H)F\|^2$ converges absolutely and, hence, unconditionally. Setting $S_N^\pi:=\sum_{i\in\l1,N\r } 1_{\lambda_{\pi(i)}}(H) \hspace{0.2mm} T_0\hspace{0.5mm} 1_{\lambda_{\pi(i)}}(H)\in\mL(\fH)$ for all $N\in\N$ and all $\pi\in \mS_\N$,  where $\mS_\N$ denotes the symmetric group of $\N$, we can write, for all $N,M\in\N$ with $N>M$, all  $\pi\in \mS_\N$,  and all $F\in\fH$,
\ba
\label{rem:uncond-2}
\|S_N^\pi F-S_M^\pi F\|^2
&=\sum_{i\in\l M+1,N\r }\|1_{\lambda_{\pi(i)}}(H)T_01_{\lambda_{\pi(i)}}(H)F\|^2\nonumber\\
&\le \|T_0\|^2\sum_{i\in\l M+1,N\r }\|1_{\lambda_{\pi(i)}}(H)F\|^2,
\ea
where we used that, for all $\pi\in\mS_\N$, the family $\{1_{\lambda_{\pi(i)}}(H)\}_{i\in\N}$ is again an orthogonal family of orthogonal projections. Since the series $\sum_{i\in\N}\|1_{\lambda_i}(H)F\|^2$ is unconditionally convergent, the right hand side of \eqref{rem:uncond-2} vanishes for sufficiently large $N$ and $M$. Hence, the strong limit of $(S_N^\pi)_{N\in\N}$ exists for all $\pi\in\mS_\N$.
Since, in addition, we know that it is independent of $\pi$,
the notation on the right hand side of \eqref{Rpp} is well-motivated.
\er

\br
\label{rem:diffTac}
Due to Proposition \ref{prop:sym} \ref{prop:sym-1}, \eqref{sym:PH}, and \eqref{SpecId-1} in the proof of Lemma \ref{lem:SpecId} \ref{lem:SpecId-c}, we have $\rho(\Delta H)=\rho(\beta_L HP_L+\beta_R H P_R)=\rho(\beta_L H)P_L+\rho(\beta_R H)P_R+\rho(0)(1-1_{ac}(H))$. Hence, we get
\ba
\label{diffTac}
T_{ac}-\rho(\Delta H)
=-\frac12\hspace{0.3mm}1_{pp}(H),
\ea
where we used Assumption \ref{ass:H} \ref{1sc=0}, the fact that the family $\{1_{pp}(H), 1_{ac}(H), 1_{sc}(H)\}$ is a complete orthogonal family of orthogonal projections, and $\rho(0)=1/2$ from \eqref{def:Ff-2}.
\er

\bp[R/L mover 2-point operator]
\label{prop:R/L2pt}
Let $H\in\mL(\fH)$ be a Hamiltonian satisfying Assumption \ref{ass:H} \ref{1sc=0} and \ref{L1}, $T_0\in\mL(\fH)$ an initial  2-point operator, $\rho\in\mB(\R)$ a Fermi function, and $\beta_L, \beta_R\in\R$ the inverse reservoir temperatures. Moreover, let $T\in\mL(\fH)$ be the R/L mover 2-point operator for $H$, $T_0$, $\rho$, and $\beta_L, \beta_R$. Then:
\bn[label=(\alph*),ref={\it (\alph*)}]
\setlength{\itemsep}{0mm}

\item 
\label{prop:R/L2pt-a}
$T$ is a 2-point operator.

\item
\label{prop:R/L2pt-b}
$[T,H]=0$
\en
\ep

\bprf
\ref{prop:R/L2pt-a}\,
We have to verify \eqref{def:2pt-op-1}-\eqref{def:2pt-op-3} for $T=T_{ac}+T_{pp}$, where $T_{ac}$ and $T_{pp}$ are defined in  \eqref{Rac} and \eqref{Rpp}, respectively. As for \eqref{def:2pt-op-1}, we first note that, due to \eqref{chiApsiB} from Lemma \ref{lem:SpecId} \ref{lem:SpecId-d},
\ba
\label{com1ac}
[1_{ac}(H), \rho(\Delta H)]
=0, 
\ea
where we used that $1_{ac}(H)=E_H(1_{M_{ac}})$ for some $M_{ac}\in\mM(\R)$ as in the proof of Proposition \ref{prop:sym} \ref{prop:sym-2} (and that $[\Delta, H]=0$ as discussed at the beginning of Remark \ref{rem:uncond}).  Hence, we get $T_{ac}^\ast=1_{ac}(H)\rho(\Delta H)=T_{ac}$. 
As for the selfadjointness of $T_{pp}$, since $T_{pp}=\slim_{N\to\infty} S_N$, where $S_N:=\sum_{i\in\l1,N\r}1_{\lambda_i}(H) \hspace{0.2mm} T_0\hspace{0.5mm} 1_{\lambda_i}(H)\in\mL(\fH)$ for all $N\in\N$, and since $S_N^\ast=S_N$ for all $N\in\N$, we have $(T_{pp}^\ast F,G)=(F, T_{pp}G)=\lim_{N\to\infty}(S_N F,G)=(T_{pp}F,G)$ for all $F,G\in\fH$, \ie, $T_{pp}^\ast=T_{pp}$. Hence, $T$ satisfies \eqref{def:2pt-op-1}. As for \eqref{def:2pt-op-2}, using Lemma \ref{lem:SpecId} \ref{lem:SpecId-a}, Remark \ref{rem:rA}, and applying \eqref{sym:PGamma} and \eqref{def:Ff-1}-\eqref{def:Ff-2}, we get 
$\Gamma \rho(\Delta H)\Gamma=E_{\Gamma\Delta H\Gamma}(\zeta\rho)=E_{-\Delta H}(\rho)=E_{\Delta H}(1-\rho)=1-\rho(\Delta H)$. Hence, with the help of \eqref{1acGmm}, we get
\ba
\label{GTacG}
\Gamma T_{ac}\Gamma
=1_{ac}(H)-T_{ac}.
\ea
As for the contribution $T_{pp}$, since we know that $1_{pp}(H)=E_H(1_{\eig(H)})$,
\eqref{rem:uncond-1}  yields 
\ba
\label{1ppsum}
1_{pp}(H)
=\sum_{\lambda\in\eig(H)} 1_\lambda(H),
\ea
where we used the notation $\sum_{\lambda\in\eig(H)} 1_\lambda(H):=\slim_{N\to\infty}\sum_{i\in\l1,N\r }1_{\lambda_i}(H)$ (and, as in Remark \ref{rem:uncond}, we note that the strong convergence of the corresponding Banach space series is unconditional). Next, using again Lemma \ref{lem:SpecId} \ref{lem:SpecId-a} and \ref{lem:SpecId-b}, we get, for all $\lambda\in\eig(H)$, 
\ba
\label{G1lamG}
\Gamma 1_\lambda(H)\Gamma
=1_{-\lambda}(H).
\ea 
Moreover,  since  $H(\Gamma F)=-\Gamma HF=-\lambda (\Gamma F)$ for all $F\in\ran(1_\lambda(H))$ with $F\neq 0$ (and, hence, $\Gamma F\neq 0$ because $\|\Gamma F\|=\|F\|$ for all $F\in\fH$), we find
\ba
\label{eig-sym}
\eig(H)
=-\hspace{0.5mm}\eig(H).
\ea
Therefore, using \eqref{G1lamG}, the property \eqref{def:2pt-op-2} for $T_0$, \eqref{eig-sym}, and \eqref{1ppsum}, we get
 \ba
 \label{GTppG}
 \Gamma T_{pp}\Gamma 
 &=\slim_{N\to\infty} \sum_{i\in\l1,N\r } 1_{-\lambda_i}(H)(1-T_0) 1_{-\lambda_i}(H)\nonumber\\
 &=1_{pp}(H)-T_{pp}.
 \ea
Due to Assumption \ref{ass:H} \ref{1sc=0}, we have $1=1_{ac}(H)+1_{pp}(H)$ 
and, hence, $T$ also satisfies \eqref{def:2pt-op-2}. Finally, we turn to \eqref{def:2pt-op-3}. Using \eqref{com1ac}, we get, for all $F\in\fH$, 
\ba
\label{FTacF}
(F, T_{ac} F)
&=(1_{ac}(H)F, \rho(\Delta H) 1_{ac}(H)F),\\
\label{FTppF}
(F, T_{pp} F)
&=\lim_{N\to\infty}\sum_{i\in\l1,N\r } (1_{\lambda_i}(H)F, T_0 1_{\lambda_i}(H)F).
\ea
Hence, with the help of \eqref{def:Ff-1} and Proposition \ref{prop:ext} \ref{prop:ext-1} for $\rho(\Delta H)$ in \eqref{FTacF}, and \eqref{def:2pt-op-3} for $T_0$ in \eqref{FTppF}, we arrive at $T\ge 0$. Furthermore, since, as shown above, \eqref{def:2pt-op-1} and \eqref{def:2pt-op-2} hold for $T$, we can write $(F,(1-T)F)=(F,\Gamma T\Gamma F)=(T\Gamma F,\Gamma F)=(\Gamma F,T\Gamma F)\ge 0$ for all $F\in\fH$, \ie, we also find $T\le 1$. 

\vspace{1mm}
\ref{prop:R/L2pt-b}\,
With the help of the first part in the proof of Lemma \ref{lem:SpecId} \ref{lem:SpecId-d}, we can write $[T_{ac},H]=\linebreak\rho(\Delta H) [1_{ac}(H),H] +[\rho(\Delta H),H]1_{ac}(H)=0$. Moreover, for all $\lm\in\eig(H)$, Remark \ref{rem:rA} yields $H1_\lambda(H)=E_H(\kp_11_{\spec(H)}1_\lm)=\lm 1_\lambda(H)$ and $[H,1_\lambda(H)]=0$ and, hence, for all $F\in\fH$, we get
\ba
\label{TppH}
[T_{pp},H]F
&=\lim_{N\to\infty}\sum_{i\in\l1,N\r }[1_{\lambda_i}(H)T_01_{\lambda_i}(H),H]F\nonumber\\
&=\lim_{N\to\infty}\sum_{i\in\l1,N\r }[1_{\lambda_i}(H)T_01_{\lambda_i}(H),\lambda_i 1]F\nonumber\\
&=0.
\ea
\eprf

We next give some examples of important states which fit into the foregoing framework of R/L mover states. However, the prototypical example of a non degenerate R/L mover state will be treated separately in Section \ref{sec:ness}.

\bx[Thermal equilibrium state]
\label{ex:KMS}
Let $H$ be a Hamiltonian satisfying Assumption \ref{ass:H} \ref{1sc=0} and \ref{L1}. Moreover, set $\beta_L:=\beta$ and  $\beta_R:=\beta$ for some $\beta\in\R$, let $\rho$ be a Fermi function, and let the initial  2-point operator be defined by $T_0:=\rho(\beta H)$ (it can be verified as in the proof of Proposition \ref{prop:R/L2pt} \ref{prop:R/L2pt-a} that $T_0$ is indeed a 2-point operator). Then, the R/L mover 2-point operator $T$ (for $H$, $T_0$, $\rho$, and $\beta_L, \beta_R$) has the form
\ba
\label{T-KMS}
T
=\rho(\beta H), 
\ea
where we made use of Lemma \ref{lem:SpecId} \ref{lem:SpecId-c}  yielding $T_{ac}=E_{\beta H 1_{ac}(H)}(\rho)1_{ac}(H)=\rho(\beta H)1_{ac}(H)$. Moreover, with the help of Lemma \ref{lem:SpecId} \ref{lem:SpecId-d} and \eqref{1ppsum}, we get $T_{pp}=\rho(\beta H)1_{pp}(H)$. The so-called $(\tau,\beta)$-KMS state, or thermal equilibrium state, where $\tau$ is the quasifree dynamics generated by $H$ from Definition \ref{set:obs} \ref{set:obs-b} and where $\beta$ plays the role of an inverse physical temperature (if $\beta>0$ and the Boltzmann constant $k_B$ set to unity), is the quasifree state whose 2-point operator has the form \eqref{T-KMS}  and the Fermi function is given, for all $x\in\R$, by the classical Fermi-Dirac (Pauli) distribution
\ba
\label{ex:KMS-2}
\rho(x)
:=\frac{1}{1+\e^{-x}}.
\ea
Moreover, the $(\tau,\beta)$-KMS state is unique if $1_{0}(H)=0$ (see \cite{Araki1971}
for the foregoing and other sufficient conditions).
\ex

Note that, in contrast to the gauge-invariant case discussed next (which frequently occurs in practice), there is a minus sign in \eqref{ex:KMS-2} (see Lemma \ref{lem:gauge} \ref{lem:gauge-d} below). 

\bd[Gauge invariance]
A state $\omega\in \mE_\fA$ is called gauge-invariant if it is invariant under the 1-parameter group of Bogoliubov $\ast$-automorphisms $\R\ni\vi\mapsto \tau_{U_\vi}\in\Aut(\fA)$ induced by the 1-parameter group of Bogoliubov operators $\R\ni\vi\mapsto U_\vi\in\mL(\fH)$ given in Definition \ref{def:iso} \ref{def:iso-c}, \ie, if $\omega\circ\tau_{U_\vi}=\omega$ for all $\vi\in\R$. 
\ed

Gauge invariance leads to the following properties.

\bl[Gauge-invariant 2-point operator]
\label{lem:gauge}
Let $T\in\mL(\fH)$. Then:
\bn[label=(\alph*),ref={\it (\alph*)}]
\setlength{\itemsep}{0mm}

\item 
\label{lem:gauge-a}
If $\omega\in\mE_\fA$ is a gauge-invariant state with 2-point operator $T$, we have, for all $\vi\in\R$, 
\ba
\label{rem:gauge-2}
[T,U_\vi]
=0.
\ea
Any $T\in\mL(\fH)$ satisfying \eqref{rem:gauge-2} is called gauge-invariant.

\item
\label{lem:gauge-b}
$T$ is a gauge-invariant 2-point operator if and only if there exists an operator $s\in\mL(\fh)$ with $0\le s\le 1$ such that
\ba
\label{rem:gauge-7}
T
=(1-s)\oplus \zeta s\zeta.
\ea

\item
\label{lem:gauge-c}
If $\omega\in\mE_\fA$ is a state with 2-point operator $T$ and if $\eta\in\mE_\fA$ satisfies, for some $s\in\mL(\fh)$ with $0\le s\le 1$ and all $f,g\in\fh$, 
\ba
\label{lem:gauge-c1}
\eta(a^\ast(f)a(g))
=(g,s f),
\ea
we have, for all $F, G\in\fH$, 
\ba
\label{lem:gauge-c2}
\omega(B^\ast(F)B(G))
=\eta(B^\ast(F)B(G)),
\ea
if and only if $T$ has the form \eqref{rem:gauge-7}.

\item
\label{lem:gauge-d}
Let $T=\rho(\beta H)$, where $\rho\in\mB(\R)$ has the form \eqref{ex:KMS-2}, $H\in\mL(\fH)$ is a Hamiltonian, and $\beta\in\R\setminus\{0\}$. If $T$ is gauge invariant, there exists $h\in\mL(\fh)$ with $h^\ast=h$ such that 
\ba
\label{Hgauge}
H
=h\oplus (-\zeta h\zeta).
\ea
Moreover, if \eqref{Hgauge} holds, $T$ has the form \eqref{rem:gauge-7} and 
\ba
s
=\rho(-\beta h).
\ea
\en
\el

\br
\label{rem:state-gauge}
Note that, if $s\in\mL(\fh)$ with $0\le s\le 1$ and if $\omega_s\in\mE_\fA$ is a state having the property $\omega_s(a^\ast(f)a(g))=(g,sf)$ for all $f,g\in\fh$, we have, for all $F,G\in\fH$, that
\ba
\omega_s(B^\ast(F)B(G))
=\omega_T(B^\ast(F)B(G)),
\ea
where $\omega_T\in\mE_\fA$ is a state whose 2-point operator $T\in\mL(\fH)$ has the form \eqref{rem:gauge-7}.
\er

\vspace{5mm}

\bprf
\ref{lem:gauge-a}\,
Since $\omega(\tau_{U_\vi}(B^\ast(F)B(G)))=\omega(B^\ast(F)B(G))$ for all $\vi\in\R$ and all $F,G\in\fH$,  \eqref{2pt-op} yields $[T,U_\vi]=0$ for all $\vi\in\R$.

\vspace{1mm}
\ref{lem:gauge-b}\,
If we write $T=[t_{ij}]_{i,j\in\l1,2\r }$, where $t_{ij}\in\mL(\fh)$ for all $i,j\in\l1,2\r $, \eqref{rem:gauge-2}  is equivalent to
\ba
\label{rem:gauge-3}
t_{12}
&=0,\\
\label{rem:gauge-4}
t_{21}
&=0,
\ea
since $[T,U_\vi]
=2\ii \sin(\vi)
\begin{bmatrix}
0 & -t_{12}\\
t_{21} & 0
\end{bmatrix}$
for all $\vi\in\R$. Moreover, since $T$ is a 2-point operator, it satisfies \eqref{def:2pt-op-1}, \eqref{def:2pt-op-2}, and \eqref{def:2pt-op-3}, respectively equivalent to $t_{11}^\ast=t_{11}$, $t_{22}^\ast=t_{22}$, and $t_{12}^\ast=t_{21}$, to 
\ba
\label{rem:gauge-5}
t_{22}
&=1-\zeta t_{11}\zeta,\\
\label{rem:gauge-6}
t_{21}
&=-\zeta t_{21}\zeta,
\ea
and to the two conditions that, for all $f_1, f_2\in\fh$, 
\ba
\label{0<T}
0
&\le (f_1,t_{11}f_1+t_{12}f_2)+(f_2,t_{21}f_1+t_{22}f_2),\\
\label{T<1}
0
&\le (f_1,(1-t_{11})f_1-t_{12}f_2)-(f_2,t_{21}f_1-(1-t_{22})f_2).
\ea
Hence, setting $s:=1-t_{11}\in\mL(\fh)$, \eqref{rem:gauge-3}-\eqref{rem:gauge-6} are equivalent to $T=(1-s)\oplus \zeta s\zeta$. Moreover, \eqref{0<T} and \eqref{T<1} are equivalent to $0\le s\le 1$. 

\ref{lem:gauge-c}\,
Note that \eqref{lem:gauge-c2} is equivalent to the condition that, for all $f_1, f_2, g_1, g_2\in\fh$, 
\ba
\label{omega-eta}
 (f_1,t_{11}g_1+t_{12}g_2)+(f_2,t_{21}g_1+t_{22}g_2)
 =(f_1,(1-s)g_1)+(\zeta g_2,s\zeta f_2),
\ea
where we used the same notation as in part \ref{lem:gauge-b}. Hence, plugging $f_2=g_2=0$ into \eqref{omega-eta}, we get $t_{11}=1-s$, and \eqref{rem:gauge-5} implies that $t_{22}=\zeta s\zeta$. Moreover, plugging $f_2=g_1=0$ into \eqref{omega-eta}, we get $t_{12}=0$, and \eqref{rem:gauge-6} yields $t_{21}=0$.
Conversely, if $T$ has the form \eqref{rem:gauge-7}, since $s^\ast=s$ and $(f,\zeta g)=(g,\zeta f)$ for all $f,g\in\fh$, \eqref{omega-eta} is satisfied.

\vspace{1mm}
\ref{lem:gauge-d}\, 
If $T$ satisfies \eqref{rem:gauge-2}, part \ref{lem:gauge-b} implies that there exists $s\in\mL(\fh)$ with $0\le s\le 1$ such that $T$ has the form \eqref{rem:gauge-7}. Next, note that $\rho:\R\to (0,1)$ is strictly monotonically increasing and let $\rho^{-1}:(0,1)\to\R$ be its inverse function (\ie, $\rho^{-1}(x)=\log(x/(1-x))$ for all $x\in (0,1)$). Since $1=E_{\beta H}(1_{\spec(\beta H)})$ and since $\spec(\beta H)\subseteq  [-|\beta| r,|\beta| r]$, where $r:=\|H\|$, we can write $T=E_{\beta H}(\rho 1_{[-|\beta| r,|\beta| r]})$. Defining $\psi\in\mB(\R)$ by $\psi(x):=\rho^{-1}(x)$ for all $x\in [\rho(-|\beta| r),\rho(|\beta| r)]$ and $\psi(x):=0$ for all $x\in \R\setminus[\rho(-|\beta| r),\rho(|\beta| r)]$, Lemma \ref{lem:SpecId} \ref{lem:SpecId-b} yields, on one hand, 
\ba
\psi(T)
&=E_{\beta H}(\psi\circ(\rho 1_{[-|\beta| r,|\beta| r]}))\nonumber\\
&=\beta H,
\ea
where we used that $\psi\circ(\rho 1_{[-|\beta| r, |\beta| r]})=\kappa_1 1_{[-|\beta| r,|\beta| r]}$. On the other hand, since, for all $\chi\in\mB(\R)$ and all selfadjoint $a,b\in\mL(\fh)$, we have $\chi(a\oplus b)=\chi(a)\oplus \chi(b)$ (which can be proven as, for example, in the proof of Lemma \ref{lem:SpecId} \ref{lem:SpecId-c}),
we get $\psi(T)=\psi((1-s)\oplus \zeta s\zeta)=\psi(1-s)\oplus \psi(\zeta s\zeta)$, \ie, $H$ is block diagonal. Hence, writing $H=[h_{ij}]_{i,j\in\l1,2\r }$ and setting $h:=h_{11}\in\mL(\fh)$, the fact that $H^\ast=H$ implies that $h^\ast=h$. Moreover, since $\Gamma H\Gamma=-H$ (which is equivalent to $h_{22}=-\zeta h_{11}\zeta$ and $h_{21}=-\zeta h_{12}\zeta$), 
we get $h_{22}=-\zeta h\zeta$, \ie, $H$ has the form \eqref{Hgauge}. 

Moreover, if $H$ is given by \eqref{Hgauge} for some $h\in\mL(\fh)$ with $h^\ast=h$, \eqref{def:Ff-1}-\eqref{def:Ff-2} imply 
\ba
T
&=E_{\beta h}(\rho)\oplus E_{-\beta\zeta h\zeta}(\rho)\nonumber\\
&=E_{\beta h}(1-\xi\rho)\oplus E_{-\beta\zeta h\zeta}(\rho)\nonumber\\
&=(1-E_{\beta h}(\xi\rho))\oplus \zeta E_{-\beta h}(\rho)\zeta\nonumber\\
&=(1-E_{-\beta h}(\rho))\oplus \zeta E_{-\beta h}(\rho)\zeta,
\ea
where we used that $\zeta\chi(a)\zeta=(\zeta\chi)(\zeta a\zeta)$ for all selfadjoint $a\in\mL(\fh)$ and all $\chi\in\mB(\R)$ (which follows from Lemma \ref{lem:SpecId} \ref{lem:SpecId-a} for the case $A=a\oplus 0$) and Remark \ref{rem:rA} for $\xi\rho=\rho_{-1}$. 
\eprf

\bx[Ground state]
\label{ex:ground}
Let $H$ be as in Example \ref{ex:KMS}.
The so-called $\tau$-ground state is the quasifree state whose 2-point operator has the form \eqref{T-KMS} (with $\beta=1$), where the Fermi function, denoted by $\rho_\infty$, is given by
\ba
\rho_\infty
:=1_{(0,\infty)}+\frac12\hspace{0.5mm}1_0.
\ea
Moreover, the $\tau$-ground state is unique if $1_{0}(H)=0$ (if $1_{0}(H)\neq 0$, a 2-point operator of the form $T=1_{(0,\infty)}(H)+ 1_0(H)S1_0(H)$, where $S\in\mL(\fH)$ is any 2-point operator, specifies  a $\tau$-ground state, see \cite{AM1985}).
Furthermore, note that $\Blim_{\beta\to\infty}\rho(\beta\hspace{0.5mm}\cdot\hspace{0.5mm}) =\rho_\infty$, where $\rho$ is given by \eqref{ex:KMS-2}. 
Hence, Proposition \ref{prop:ext} \ref{prop:ext-2} yields
\ba
\slim_{\beta\to\infty} \rho(\beta H)
=\rho_\infty(H),
\ea
\ie, if $1_0(H)=0$,  the 2-point operator of the unique $(\tau,\beta)$-KMS state converges strongly to the 2-point operator of the unique $\tau$-ground state.
\ex

\section{Nonequilibrium steady states}
\label{sec:ness}

In this section, we construct a special class of R/L movers, the so-called nonequilibrium steady states (NESSs) discussed in the introduction. They serve as the main motivation for the introduction of the R/L mover states in the foregoing section.\\

For the following, recall the definitions \eqref{Hk}-\eqref{Hkl} from Section \ref{sec:LR}.

\bd[Initial system]
\label{def:init}
Let $H\in\mL(\fH)$ be a Hamiltonian, let $\rho\in\mB(\R)$ be a Fermi function, and let $\beta_L, \beta_R\in\R$ the inverse reservoir temperatures. Moreover, let $\beta_S\in\R$ be the inverse sample temperature. 
\bn[label=(\alph*),ref={\it (\alph*)}]
\setlength{\itemsep}{0mm}
\item 
\label{def:init-a}
The operator $H_0\in\mL(\fH)$, defined by
\ba
H_0
:=H_L+H_S+H_R,
\ea
is called the initial Hamiltonian (for $H$).

\item
\label{def:init-b}
The quasifree dynamics generated by $H_0\in\mL(\fH)$ is called the initial dynamics.

\item
\label{def:init-c}
The quasifree state $\omega_0\in\mE_\fA$ whose 2-point operator 
$T_0\in\mL(\fH)$ has the form
\ba
T_0
:=\rho(\Delta_0 H_0),
\ea
where $\Delta_0\in\mL(\fH)$ is defined by
\ba
\Delta_0
:=\beta_L Q_L+\beta_SQ_S+\beta_R Q_R,
\ea
is called the initial state (for $\rho$ and $\beta_L, \beta_S, \beta_R$).
\en
\ed

\br
Since, due to Remark \ref{rem:cf}, the family $\{Q_L,Q_S,Q_R\}\subseteq\mL(\fH)$ is a complete orthogonal family of orthogonal projections and since $[Q_\kp, \Gamma]=0$ for all $\kp\in\{L,S,R\}$, the operators $H_L$, $H_S$, $H_R$, and $H_0$ satisfy \eqref{H1}-\eqref{H2}. Moreover, $\Delta_0 H_0\in\mL(\fH)$ is selfadjoint and, as in the proof of Proposition \ref{prop:R/L2pt} \ref{prop:R/L2pt-a}, property \eqref{def:Ff-2}, Definition \ref{def:resolution}, Lemma \ref{lem:SpecId} \ref{lem:SpecId-a}, and Remark \ref{rem:rA} yield that $T_0$ is a 2-point operator.
\er

For the setting at hand, the NESS discussed in the introduction is defined as follows.

\bd[NESS]
Let $H\in\mL(\fH)$ be a Hamiltonian, let $\rho\in\mB(\R)$ be a Fermi function, and let $\beta_L, \beta_R\in\R$ be the inverse reservoir temperatures. Moreover, let $\beta_S\in\R$ be the inverse sample temperature, let $\omega_0\in\mE_\fA$ be the initial state for $\rho$ and $\beta_L, \beta_S, \beta_R$, and let $\R\ni t\mapsto \tau^t\in\Aut(\fA)$ be the quasifree dynamics generated by $H$.  The state $\omega\in\mE_\fA$ defined, for all $A\in\fA$, by 
\ba
\label{ness}
\omega(A)
:=\lim_{t\to\infty}\frac1t\int_0^t\rd s\hspace{1.5mm}\omega_0(\tau^s(A)),
\ea
is called the NESS (for $H$, $\rho$, and $\beta_L, \beta_S, \beta_R$).
\ed

\br
The general definition stems from \cite{Ruelle2001} and defines the NESSs as the limit points in the weak-$\ast$ topology of the net defined by the ergodic mean between $0$ and $t>0$ of the given initial state time-evolved by the dynamics of interest (note that, due to the Banach-Alaoglu theorem, the set of such NESSs is not empty). In general, the averaging procedure enables us to treat a nonvanishing contribution to the point spectrum of the Hamiltonian which generates the full time evolution (see Theorem \ref{thm:ness} below).
\er

The following ingredients from the time-dependent approach to Hilbert 
space scattering theory will be used for the construction of our NESS.

\bd[Wave operators]
\label{def:wave}
Let $H\in\mL(\fH)$ be a Hamiltonian satisfying Assumption \ref{ass:H} \ref{L1} and let $H_0\in\mL(\fH)$ be the initial Hamiltonian for $H$.
\bn[label=(\alph*),ref={\it (\alph*)}]
\setlength{\itemsep}{0mm}
\item 
\label{def:wave-a}
The operator $W\in\mL(\fH)$, defined by
\ba
\label{def:W}
W
:=\slim_{t\to\infty}\hspace{0.5mm}\e^{-\ii t H_0} \e^{\ii t H} 1_{\rm ac}(H),
\ea
is called the wave operator (for $H$ and $H_0$).

\item 
\label{def:wave-b}
The operators $W_L, W_R\in\mL(\fH)$, defined by
\ba
\label{def:Wpm}
W_L
:=\slim_{t\to\infty}\hspace{0.5mm} \e^{-\ii t H_L}Q_L\e^{\ii t H} 1_{\rm ac}(H),\\
W_R
:=\slim_{t\to\infty}\hspace{0.5mm} \e^{-\ii t H_R}Q_R\e^{\ii t H} 1_{\rm ac}(H),
\ea
are called the partial wave operators (for $H$ and $H_L$, and $H$ and $H_R$, respectively).
\en
\ed

For the following, let us denote by $AP(\R)$ the complex-valued functions on $\R$ which are almost-periodic (in the sense of H. Bohr [the brother of N. Bohr], \ie, the uniformly almost-periodic functions, see \cite{Katznelson2004} for example). Also recall the definitions of the asymptotic projections $P_L, P_R$ and of the R/L mover generator $\Delta$ from  Definition \ref{def:Asmpt}. Moreover, for all $A=[a_{ij}]_{i,j\in\l1,n\r}\in\C^{n\times n}$, the Euclidean matrix norm is denoted by $|A|_2:=(\sum_{i,j\in\l1,n\r}|a_{ij}|^2)^{1/2}$.\\

\bt[NESS]
\label{thm:ness}
Let $H\in\mL(\fH)$ be a Hamiltonian satisfying Assumption \ref{ass:H} \ref{1sc=0} and \ref{L1}, let $\rho\in\mB(\R)$ be a Fermi function, and let $\beta_L, \beta_R\in\R$ be the inverse reservoir temperatures. Moreover, let $\beta_S\in\R$ be the inverse sample temperature, let $\omega_0\in\mE_\fA$ be the initial state for $\rho$ and $\beta_L, \beta_S, \beta_R$, and let $\R\ni t\mapsto \tau^t\in\Aut(\fA)$ be the quasifree dynamics generated by $H$. Then:
\bn[label=(\alph*),ref={\it (\alph*)}]
\setlength{\itemsep}{0mm}
\item 
\label{thm:ness-a}
The NESS $\omega\in\mE_\fA$ for $H$, $\rho$, and $\beta_L, \beta_S, \beta_R$ exists.

\item 
\label{thm:ness-b}
The 2-point operator $T\in\mL(\fH)$ of $\omega$ is given by
\ba
T
=T_{ac}+T_{pp},
\ea
where $T_{ac}, T_{pp}\in \mL(\fH)$ are defined by
\ba
T_{ac}
&:=\rho(\Delta H)1_{ac}(H),\\
T_{pp}
&:=\sum_{\lambda\in\eig(H)}
1_\lambda(H) \hspace{0.2mm} T_0\hspace{0.5mm} 
1_\lambda(H).
\ea
\en
\et

\bprf 
\ref{thm:ness-a}\,
We start off by studying \eqref{ness} for elements of $\fA$ of the form $\prod_{i\in\l1,2n\r}B(F_i)$ for all $n\in\N$ and all $\{F_i\}_{i\in\l1,2n\r}\subseteq\fH$. Since $\omega_0$ is quasifree, the expectation value with respect to $\omega_0$ of such elements propagated in time by means of the quasifree dynamics generated by $H$ has the Pfaffian factorization property from Definition \ref{def:qfs}, \ie, for all $t\in\R$, 
\ba
\label{pfO}
\omega_0\Big(\prod\nolimits_{i\in\l1,2n\r}\tau^t\big(B(F_i)\big)\Big)
=\pf(\Omega(t)),
\ea
where, for all $n\in\N$ and all $\{F_i\}_{i\in\l1,2n\r}\subseteq\fH$, the entries of the matrix-valued map $\Omega:\R\to\C^{2n\times 2n}$ are defined, for all $i,j\in\l1,2n\r$ and all $t\in\R$, by
\ba
\Omega_{ij}(t)
:=(\Gamma\e^{\ii t H}F_i,T_0 \e^{\ii t H}F_j).
\ea
In the following, let $n\in\N$ and $\{F_i\}_{i\in\l1, 2n\r}\subseteq\fH$ be fixed. Since, due to Assumption \ref{ass:H} \ref{1sc=0}, we can write $1=1_{ac}(H)+1_{pp}(H)\in\mL(\fH)$, we have, for all $i,j\in\l1,2n\r$ and all $t\in\R$,
\ba
\Omega_{ij}(t)
=\Omega_{ij}^{aa}(t)+\Omega_{ij}^{ap}(t)+\Omega_{ij}^{pa}(t)
+\Omega_{ij}^{pp}(t),
\ea
where the entries of the matrix-valued maps $\Omega^{aa}, \Omega^{ap}, \Omega^{pa}, \Omega^{pp}\hspace{-0.6mm}:\R\to\C^{2n\times 2n}$ are defined, for all $i,j\in\l1,2n\r$ and all $t\in\R$, by
\ba
\label{Omega-aa}
\Omega_{ij}^{aa}(t)
&:=(\e^{\ii t H}1_{ac}(H) \Gamma F_i,T_0\hspace{0.2mm}\e^{\ii t H}1_{ac}(H) F_j),\\
\label{Omega-ap}
\Omega_{ij}^{ap}(t)
&:=(\e^{\ii t H}1_{ac}(H) \Gamma F_i,T_0 \hspace{0.2mm}\e^{\ii t H}1_{pp}(H) F_j),\\
\label{Omega-pa}
\Omega_{ij}^{pa}(t)
&:=(\e^{\ii t H}1_{pp}(H) \Gamma F_i,T_0 \hspace{0.2mm}\e^{\ii t H}1_{ac}(H) F_j),\\
\label{Omega-pp}
\Omega_{ij}^{pp}(t)
&:=(\e^{\ii t H}1_{pp}(H) \Gamma F_i,T_0 \hspace{0.2mm}\e^{\ii t H}1_{pp}(H)F_j),
\ea
and we used \eqref{eithGmm} and \eqref{1acGmm}. We next study the large time averages of \eqref{Omega-aa}-\eqref{Omega-pp}. 

As for \eqref{Omega-aa}, since the family $\{Q_L,Q_S,Q_R\}\subseteq\mL(\fH)$ is complete, we can insert $1=Q_L+Q_S+Q_R$ in front of the propagators on both sides of \eqref{Omega-aa}. Hence, for all $i,j\in\l1,2n\r$ and all $t\in\R$, we get
\ba
\label{Oaadec}
\Omega_{ij}^{aa}(t)
=\sum_{\kp\in\{L,S,R\}}\Omega_{ij}^{aa, \kp}(t),
\ea
where, for all $\kp\in\{L,S,R\}$,  the entries of the matrix-valued maps $\Omega^{aa, \kp}\hspace{-0.6mm}:\R\to\C^{2n\times 2n}$ are defined, for all $i,j\in\l1,2n\r$ and all $t\in\R$, by
\ba
\label{Okplm}
\Omega_{ij}^{aa, \kp}(t)
&:=(Q_\kp\e^{\ii t H}1_{ac}(H) \Gamma F_i, T_0 Q_\kp\e^{\ii t H}1_{ac}(H) F_j),
\ea
and we used that, since $[Q_\kp, \Delta_0 H_0]=[Q_\kp, \beta_L H_L+\beta_S H_S+\beta_R H_R]=0$ for all $\kp\in\{L,S,R\}$, Lemma \ref{lem:SpecId} \ref{lem:SpecId-d}  yields $[Q_\kp, T_0]=0$ for all $\kp\in\{L,S,R\}$. In order to determine the large time limit of $\Omega^{aa, L}$, we note that, again due to Lemma \ref{lem:SpecId} \ref{lem:SpecId-d},  we have $[\e^{-\ii t H_0}, T_0]=0$ for all $t\in\R$ since $[H_0, \Delta_0 H_0]=0$. Hence, for all  $i,j\in\l1,2n\r$ and all $t\in\R$, we can write
\ba
\Omega_{ij}^{aa, L}(t)
&=(\e^{-\ii t H_0}Q_L\e^{\ii t H}1_{ac}(H) \Gamma F_i, 
T_0\e^{-\ii t H_0}Q_L\e^{\ii t H}1_{ac}(H) F_j)\nonumber\\
&=(\e^{-\ii t H_L}Q_L\e^{\ii t H}1_{ac}(H) \Gamma F_i,
\rho(\beta_L H_L)\e^{-\ii t H_L}Q_L\e^{\ii t H}1_{ac}(H) F_j),
\ea
where we used that $\rho(\Delta_0 H_0)Q_L=\rho(\beta_L H_L)Q_L$ which follows from Lemma \ref{lem:SpecId} \ref{lem:SpecId-c}. 
Since, due to Assumption \ref{ass:H} \ref{L1}, we have 
\ba
H_L Q_L-Q_LH
&=-(H_{LS}+H_{LR})\nonumber\\
&\in\mL^1(\fH), 
\ea
the Kato-Rosenblum theorem guarantees the existence of the partial wave operator $W_L$  as in the proof of  Proposition \ref{prop:sym} \ref{prop:sym-1}.
Moreover, the Kato-Rosenblum theorem also implies the existence of the wave operator $W_L'\in\mL(\fH)$ given by
\ba
W_L'
:=\slim_{t\to\infty} \e^{-\ii t H}Q_L\e^{\ii t H_L} 1_{\rm ac}(H_L).
\ea
Hence, since the adjoint property for wave operators 
yields $W_L^\ast=W_L'$, since Remark \ref{rem:rA} and the intertwining property for wave operators
imply that $W_L'\rho(\beta_L H_L)=\rho(\beta_L H)W_L'$, and since the chain rule for wave operators 
results in $W_L' W_L=P_L$, we get, for all $i,j\in\l1,2n\r$, 
\ba
\label{limOaapp}
\lim_{t\to\infty}\Omega_{ij}^{aa,L}(t)
&=(W_L\Gamma F_i, \rho(\beta_L H_L )W_L F_j)\nonumber\\[-2.0mm]
&=(\Gamma F_i, W_L' \rho(\beta_L H_L)W_LF_j)\nonumber\\
&=(\Gamma F_i, \rho(\beta_L H) W_L' W_L F_j)\nonumber\\
&=(\Gamma F_i, \rho(\beta_L H)P_L F_j).
\ea
Interchanging $L$ and $R$, we also get $\lim_{t\to\infty}\Omega_{ij}^{aa, R}(t)=(\Gamma F_i, \rho(\beta_R H)P_R F_j)$ for all $i,j\in\l1,2n\r$. Moreover, since, for all $i,j\in\l1,2n\r$ and all $t\in\R$, we have
\ba 
|\Omega_{ij}^{aa, S}(t)|
\le \|T_0 Q_S\e^{\ii t H}1_{ac}(H) F_j\|\|F_i\|,
\ea
and since $T_0Q_S\in\mL^0(\fH)$, we know that $\lim_{t\to\infty}\Omega_{ij}^{aa, S}(t)=0$ for all $i,j\in\l1,2n\r$.
Therefore, \eqref{Oaadec}, \eqref{limOaapp}, and the foregoing arguments yield, for all $i,j\in\l1,2n\r$, 
\ba
\label{Oaa}
\lim_{t\to\infty}\Omega_{ij}^{aa}(t)
&=(\Gamma F_i, (\rho(\beta_L H)P_L+\rho(\beta_R H)P_R)F_j)\nonumber\\
&=(\Gamma F_i, \rho(\Delta H)1_{ac}(H) F_j),
\ea
where, in the last equality, we used Lemma \ref{lem:SpecId} \ref{lem:SpecId-c} and \eqref{sym:SumP}.

We next turn to \eqref{Omega-ap}. First, we note that, since $\fH$ is separable, there exists a sequence $(\eta_N)_{N\in\N}\subseteq\mL^0(\fH)$ of orthogonal projections satisfying $[\eta_N,H]=0$ for all $N\in\N$ and $\slim_{N\to\infty}\eta_N=1_{pp}(H)$ (pick an orthonormal basis $\{E_n\}_{n\in\N}$ of eigenvectors of $H$ for the closed subspace $\ran(1_{pp}(H))$ of $\fH$ and set $\eta_N:=\sum_{n\in\l1,N\r}(E_n,\hspace{0.2mm}\cdot\hspace{0.2mm})E_n\in\mL^0(\fH)$ for all $N\in\N$).
Inserting $1=\eta_N+(1-\eta_N)$ for any $N\in\N$ after the second propagator on the right hand side of \eqref{Omega-ap}, we get, for all $i,j\in\l1,2n\r$, all $t\in\R$, and all $N\in\N$, 
\ba
\Omega_{ij}^{ap}(t)
=\Omega_{ij}^{ap,N}(t)+\Omega_{ij}^{ap, N_{\!\perp}}(t),
\ea
where the entries of the matrix-valued maps $\Omega^{ap, N}, \Omega^{ap, N_{\!\perp}}\hspace{-1mm}:\R\to\C^{2n\times 2n}$ are defined, for all $i,j\in\l1,2n\r$, all $N\in\N$, and all $t\in\R$, by
\ba
\label{Omega-ap0}
\Omega_{ij}^{ap, N}(t)
&:=(\e^{\ii t H}1_{ac}(H) \Gamma F_i, T_0 \e^{\ii t H}\eta_N 1_{pp}(H) F_j),\\
\label{Omega-apperp}
\Omega_{ij}^{ap, N_{\!\perp}}(t)
&:=(\e^{\ii t H}1_{ac}(H) \Gamma F_i, T_0 \e^{\ii t H} (1-\eta_N) 1_{pp}(H) F_j).
\ea
Using that $[\eta_N,H]=0$ for all $N\in\N$ and Lemma \ref{lem:SpecId} \ref{lem:SpecId-d} for \eqref{Omega-ap0}, we get, for all $i,j\in\l1,2n\r$, all $N\in\N$, and all $t\in\R$, 
\ba
\label{bound-Oap0}
|\Omega_{ij}^{ap, N}(t)|
&\le \|\eta_N T_0\e^{\ii t H}1_{ac}(H) \Gamma F_i\| \|F_j\|,\\
\label{bound-Oapperp}
|\Omega_{ij}^{ap, N_{\!\perp}}(t)|
&\le \|(1-\eta_N) 1_{pp}(H) F_j\|\|T_0\|\|F_i\|.
\ea
As above, since $\eta_NT_0\in\mL^0(\fH)$ for all $N\in\N$, \eqref{bound-Oap0} implies that $\lim_{t\to\infty} \Omega_{ij}^{ap, N}(t)=0$ for all $i,j\in\l1,2n\r$ and all $N\in\N$.
Moreover, $\lim_{N\to\infty} \Omega_{ij}^{ap, N_{\!\perp}}(t)=0$ for all $i,j\in\l1,2n\r$ and all $t\in\R$ due to \eqref{bound-Oapperp}. Hence, for all $i,j\in\l1,2n\r$, we get
\ba
\label{limOap}
\lim_{t\to\infty} \Omega_{ij}^{ap}(t)
=0.
\ea
The term \eqref{Omega-pa} is treated analogously leading to $\lim_{t\to\infty} \Omega_{ij}^{pa}(t)=0$ for all $i,j\in\l1,2n\r$.

We finally turn to \eqref{Omega-pp}. Setting $\chi_N:=\sum_{i\in\l1,N\r}1_{\lm_i}(H)$ for all $N\in\N$, we know from \eqref{rem:uncond-1} in Remark \ref{rem:uncond} and from $1_{pp}(H)=1_{\eig(H)}(H)$ that $\slim_{N\to\infty}\chi_N=1_{pp}(H)$. Next, we define the entries of the matrix-valued map $\Omega^{pp, N}\hspace{-1mm}:\R\to\C^{2n\times 2n}$, for all $i,j\in\l1,2n\r$, all $N\in\N$, and all $t\in\R$, by
\ba
\label{def:OppN}
\Omega_{ij}^{pp, N}(t)
:=(\e^{\ii t H}\chi_N \Gamma F_i, T_0 \e^{\ii t H}\chi_NF_j),
\ea
and we note $\Omega_{ij}^{pp, N}\in AP(\R)$ for all $i,j\in\l1,2n\r$ and all $N\in\N$ because \eqref{def:OppN} defines a trigonometric polynomial on $\R$ due to the fact that $\e^{\ii t H}1_\lm(H)=\e^{\ii t \lm}1_{\lm}(H)$ for all $t\in\R$ and all $\lm\in\eig(H)$.
Moreover, since, for all $i,j\in\l1,2n\r$, all $N\in\N$, and all $t\in\R$, we have
\ba
|\Omega_{ij}^{pp, N}(t)-\Omega_{ij}^{pp}(t)|
\le  \|T_0\|\|\chi_N F_j\| \|(\chi_N-1_{pp}(H))F_i\|+ \|T_0\|\|F_i\| \|(\chi_N-1_{pp}(H))F_j\|,
\ea  
and since the sequence $(\|\chi_N F\|)_{N\in\N}$ is bounded for all $F\in\fH$, we get, for all $i,j\in\l1,2n\r$,
\ba
\label{OppGlm}
\lim_{N\to\infty} |\Omega_{ij}^{pp, N}-\Omega_{ij}^{pp}|_\infty
=0,
\ea
which implies that $\Omega_{ij}^{pp}\in AP(\R)$ for all $i,j\in\l1,2n\r$ since $ AP(\R)$ is closed with respect to the norm $|\cdot|_\infty$ (given in \eqref{supnrm} of Appendix \ref{app:spectral}). Defining the entries of the matrix $\Lambda^{aa}\in\C^{2n\times 2n}$ by the right hand side of \eqref{Oaa}, \ie, for all $i,j\in\l1,2n\r$, by
\ba
\Lambda^{aa}_{ij}
:=(\Gamma F_i, \rho(\Delta H)1_{ac}(H) F_j), 
\ea
noting that the Pfaffian is a polynomial function of the entries of the matrix on which it acts, and recalling that $AP(\R)$ is an algebra (with respect to the usual pointwise addition, scalar multiplication, and multiplication), we get that the function $\vartheta:\R\to\C$, defined by $\vartheta(t):=\pf(\Lambda^{aa}+\Omega^{pp}(t))$ for all $t\in\R$, satisfies $\vartheta\in AP(\R)$. Therefore, we know that the large time average $\lim_{t\to\infty}\int_0^t\rd s\hspace{0.5mm} \vartheta(s)/t$ exists,
and we want to show that it is equal to the large time average of \eqref{pfO}.
To this end, let the entries of the range of the linear map $\C^{2n\times 2n}\ni A=[a_{ij}]_{i,j\in\l1,2n\r}\mapsto A^a\in\C_a^{2n\times 2n}$ be defined, for all $i,j\in\l1,2n\r$,  by $[A^a]_{ij}:=a_{ij}$ if $i<j$, $[A^a]_{ii}:=0$, and $[A^a]_{ij}:=-a_{ji}$ if $i>j$. Using Hadamard's inequality $|\det(A)|\le \prod_{i\in\l1,n\r}(\sum_{j\in\l1,n\r}|a_{ij}|^2)^{1/2}$ for all $A\in\C^{n\times n}$
and the Cayley-Muir lemma $(\pf(A))^2=\det(A)$ for all $A\in\C_a^{2n\times 2n}$, 
we get $|\pf(A)|\le |A|_2^n$ for all $A\in\C^{2n\times 2n}_a$.
Hence, since the function $[0,\infty)\ni r\mapsto r^n\in\R$ is monotonically increasing, we know that $|\pf(A)-\pf(B)|\le |A-B|_2 (|A|_2+|B|_2+1)^n$ for all $A, B\in\C^{2n\times 2n}_a$ (see \cite{Simon1977} for example). Moreover, we note that $\pf(A)=\pf(A^a)$ due to \eqref{pfaff} and that $|A^a|_2^2=2 \sum_{i,j\in\l1,2n\r, \hspace{0.3mm}i<j} |a_{ij}|^2\le 2|A|_2^2$ for all $A\in\C^{2n\times 2n}$. 
Hence, since, for all $t\in\R$, we have $|\Omega(t)|_2\le C_1$ and $|\Lambda^{aa}+\Omega^{pp}(t)|_2\le C_2$, where $C_1:=\|T_0\| (\sum_{i,j\in\l1,2n\r}\|F_i\|^2\|F_j\|^2)^{1/2}$ and $C_2:=\sqrt{2}(\sum_{i,j\in\l1,2n\r}|\Lambda^{aa}_{ij}|^2+C_1^2)^{1/2}$, we get, for all $t\in\R$, 
\ba
\label{EstOOaaOpp}
|\pf(\Omega(t))-\pf(\Lambda^{aa}+\Omega^{pp}(t))|
&\le C(|\Omega^{aa}(t)-\Lambda^{aa}|_2+|\Omega^{ap}(t)|_2+|\Omega^{pa}(t)|_2),
\ea
where $C:=\sqrt{2}(1+\sqrt{2}(C_1+C_2))^n$. 
Therefore, since the right hand side of \eqref{EstOOaaOpp} vanishes for $t\to\infty$, its large time average also vanishes 
and we get
\ba
\label{limpoly}
\lim_{t\to\infty}\frac1t\int_0^t\rd s\hspace{1mm}
\omega_0\Big(\prod\nolimits_{i\in\l1,2n\r}\tau^s\big(B(F_i)\big)\Big)
&=\lim_{t\to\infty}\frac1t\int_0^t\rd s\hspace{1mm}
\pf(\Omega(s))\nonumber\\
&=\lim_{t\to\infty}\frac1t\int_0^t\rd s\hspace{1mm}
\pf(\Lambda^{aa}+\Omega^{pp}(s)).
\ea

Finally, we have to show that the limit on the right hand side of \eqref{ness} exists for all $A\in\fA$. To this end, let $A\in\fA$ be fixed. Since, by definition, $\fA$ is the $C^\ast$\hspace{-0.5mm}-completion (with respect to the $C^\ast$-norm $\|\cdot\|$) of the $\ast$-algebra generated by the selfdual generators from Definition \ref{def:obs} \ref{def:obs-e}, 
there exists a sequence $(P_n)_{n\in\N}$ of polynomials in these generators such that $\lim_{n\to\infty}\|A-P_n\|=0$.
For all $B\in\fA$, defining the function $F_B:(0,\infty)\to\C$ by $F_B(t):=\int_0^t\rd s\hspace{0.5mm}\omega_0(\tau^s(B))/t$ for all $t\in(0,\infty)$ and noting that $|\omega_0(\tau^t(B))|\le \|B\|$ for all $t\in\R$ and all $B\in\fA$, we get, for all $t, t'\in\R$ and all $n\in\N$, 
\ba
\label{CauchyFA}
|F_A(t)-F_A(t')|
\le 2\|A-P_n\|+|F_{P_n}(t)-F_{P_n}(t')|.
\ea
Hence, since the limit for $t\to\infty$ of $F_{P_n}(t)$ exists for all $n\in\N$ due to \eqref{limpoly},  \eqref{CauchyFA} implies the existence of the desired limit on the right hand side of  \eqref{ness}.

\vspace{1mm}
\ref{thm:ness-b}\,
Let $F_1, F_2\in\fH$ be fixed. Due to \eqref{ness}, \eqref{limpoly}, and \eqref{2pt-op}, we have
\ba
\label{2ptness}
(\Gamma F_1,TF_2)
&=\omega(B(F_1) B(F_2))\nonumber\\
&=\lim_{t\to\infty}\frac1t\int_0^t\rd s\hspace{1mm}
\omega_0(\tau^s(B(F_1)B(F_2)))\nonumber\\
&=(\Gamma F_1, \rho(\Delta H) 1_{ac}(H)F_2)+\lim_{t\to\infty}\frac1t\int_0^t\rd s\hspace{1mm}\Omega^{pp}_{12}(s),
\ea
where we recall that $\Omega^{pp}_{12}$ from \eqref{Omega-pp} satisfies $\Omega^{pp}_{12}\in AP(\R)$ and that the limit on the right hand side of \eqref{2ptness} thus exists. Moreover, we have, for all $t\in \R^+:=(0,\infty)$ and all  $N\in\N$, 
\ba
\label{intON}
\frac1t\int_0^t\rd s\hspace{1mm}\Omega^{pp,N}_{12}(t)
&=\sum_{i\in\l1,N\r}(\Gamma F_1, 1_{\lm_i}(H)T_0 1_{\lm_i}(H)F_2)\nonumber\\ &\quad+\sum_{\substack{i,j\in\l1,N\r\\ i\neq j}}\frac{\e^{\ii t(\lm_j-\lm_i)}-1}{\ii t(\lm_j-\lm_i)}(\Gamma F_1, 1_{\lm_i}(H)T_0 1_{\lm_j}(H)F_2),
\ea
from which it follows that, for all $N\in\N$, 
\ba
\label{limOppN}
\lim_{t\to\infty} \frac1t\int_0^t\rd s\hspace{1mm}\Omega^{pp,N}_{12}(s)
=\Big(\Gamma F_1, \sum\nolimits_{i\in\l1,N\r} 1_{\lm_i}(H)T_0 1_{\lm_i}(H)F_2\Big).
\ea
Hence, due to \eqref{limOppN}, and since \eqref{OppGlm} implies that the sequence of functions $\R^+\ni t\mapsto \int_0^t\rd s\hspace{1mm}\Omega^{pp,N}_{12}(s)/t$ converges, for $N\to\infty$, uniformly in $t\in\R^+$ to the function $\R^+\ni t\mapsto \int_0^t\rd s\hspace{1mm}\Omega^{pp}_{12}(s)/t$, the limit operations for $t\to\infty$ and $N\to\infty$ can be interchanged and the second term on the right hand side of \eqref{2ptness} becomes
\ba
\label{limTpp}
\lim_{t\to\infty}\frac1t\int_0^t\rd s\hspace{1mm}\Omega^{pp}_{12}(s)
&=\lim_{t\to\infty}\frac1t\int_0^t\rd s\hspace{1mm}\lim_{N\to\infty}\Omega^{pp,N}_{12}(s)\nonumber\\
&=\lim_{t\to\infty}\lim_{N\to\infty}\frac1t\int_0^t\rd s\hspace{1mm}\Omega^{pp,N}_{12}(s)\nonumber\\
&=\lim_{N\to\infty}\lim_{t\to\infty}\frac1t\int_0^t\rd s\hspace{1mm}\Omega^{pp,N}_{12}(s)\nonumber\\
&=\lim_{N\to\infty}\Big(\Gamma F_1, \sum\nolimits_{i\in\l1,N\r} 1_{\lm_i}(H)T_0 1_{\lm_i}(H)F_2\Big)\nonumber\\
&=\Big(\Gamma F_1, \sum\nolimits_{\lm\in\eig(H)} 1_{\lm}(H)T_0 1_{\lm}(H)F_2\Big),
\ea
where, in the last equality, we used Definition \ref{def:R/L} \ref{R/L-1}. 
\eprf

\br
Due to Assumption \ref{ass:H} \ref{L1}, we have $H-H_0=H_{LS}+H_{SL}+H_{RS}+H_{SR}+H_{LR}+H_{RL}\in\mL^1(\fH)$. Hence, the Kato-Rosenblum theorem again implies the existence of the wave operator $W\in\mL(\fH)$ from Definition \ref{def:wave} \ref{def:wave-a}. Moreover, as in the proof of Theorem \ref{thm:ness} \ref{thm:ness-b} (or by noting that $W=W_L+W_R$ 
and by using Lemma \ref{lem:SpecId} \ref{lem:SpecId-c}), inserting $1=\e^{\ii t H_0}\e^{-\ii t H_0}$ for all $t\in\R$ in front of $T_0$ in \eqref{Omega-aa} and using that $[\e^{-\ii t H_0}, T_0]=0$ for all $t\in\R$ directly leads to 
\ba
T_{ac}=W^\ast T_0 W.
\ea 
\er

\section{Asymptotic velocity}
\label{sec:ti}
In this section, we implement translation invariance and study its consequences. In particular, we construct the so-called asymptotic velocity and derive the action of the R/L generator as a matrix multiplication operator.

In the following, we will resort to the usual Fourier Hilbert space isomorphism 
$\ff$ between the 1-particle position Hilbert space $\fh$ over $\Z$ and 
the 1-particle momentum Hilbert space $\hat\fh$ over $\T:=[-\pi,\pi]$ defined by 
\ba
\label{L2}
\hat\fh
:=L^2(\T).
\ea
Here and in the following, for all $p\in\R$ with $p\ge 1$, we denote by $L^p(\T)$ the space of equivalence classes of functions $\vi:\T\to\C$ which are measurable with respect to $\mM(\T)$ and for which $|\vi|^p$ is integrable with respect to the Borel-Lebesgue measure  (and analogously if $\T$ is replaced by another subinterval of $\R$). As usual, the equivalence relation identifies functions which coincide almost everywhere with respect to the Borel-Lebesgue measure (\ie, on the complement of a subset of a set of Borel-Lebesgue measure zero).
Moreover, $\mM(\T)$ is defined to be the restriction of $\mM(\R)$  to $\T$, \ie, we have $\mM(\T)
:=\{M\cap\T\,|\,M\in\mM(\R)\}=\{M\subseteq\T\,|\, M\in\mM(\R)\}$,
and we recall that $\mM(\R)$ is given in Definition \ref{def:Bf} \ref{Bf:Bsets}.  Abusing notation, for all $M\in\mM(\T)$, we write $|M|$ for the (restriction to $\mM(\T)$ of the) Borel-Lebesgue measure of $M$. Moreover, we denote by $L^\infty(\T)$ the space of equivalence classes of functions $\vi:\T\to\C$ which are measurable with respect to $\mM(\T)$ and almost everywhere bounded on $\T$, and the norm on $L^\infty(\T)$ is denoted by $\|\cdot\|_\infty$.
 
For all $f\in\fh$, the Fourier transform is given by the limit (in $\hat\fh$) $\ff f:=\lim_{N\to\infty}\sum_{|x|\le N}f(x)\e_x$, where, for all $x\in\Z$, the plane wave functions $\e_x\in\hat\fh$ are given by $\e_x(k):=\e^{\ii kx}$ for all $k\in\T$. Furthermore, for all $f\in\fh$ and all $a\in\mL(\fh)$, we set $\hat f:=\ff f\in\hat\fh$ and $\hat a:=\ff a\ff^\ast\in\mL(\hat\fh)$ and, sometimes, we will also write $\check\vi:=\ff^\ast\vi$ for all $\vi\in\hat\fh$. On $\fH=\fh\oplus\fh$, we define $\fF:=\ff\sigma_0:\fH\to\widehat\fH$, where the doubled 1-particle momentum Hilbert space is given by
\ba
\label{def:fH}
\widehat\fH
:=\hat\fh\oplus\hat\fh,
 \ea
and we set $\widehat F:=\fF F\in\widehat\fH$ for all $F\in\fH$ and $\widehat A:=\fF A\fF^\ast\in\mL(\widehat\fH)$ for all $A\in\mL(\fH)$ (the usual scalar products and the corresponding induced norms and operator norms, on both $\hat\fh$ and $\widehat\fH$, are again all denoted by $(\hspace{0.3mm}\cdot,\cdot)$ and $\|\cdot\|$, respectively). 

Furthermore, similarly to \eqref{mu} on position space, for all  $u\in L^\infty(\T)$, the multiplication operator $m[u]\in\mL(\hat\fh)$ on momentum space is defined by $m[u]\vi:=u\vi$ for all $\vi\in\hat\fh$. Moreover, for all $u:=[u_1,u_2,u_3]\in L^\infty(\T)^3$, we define $m[u]\in\mL( \hat\fh)^3$ by
\ba
m[u]
:=[m[u_1], m[u_2], m[u_3]],
\ea
and we note that, for all  $\Phi\in\widehat\fH$ on which it is defined, the operator $U:=m[u_0]\sigma_0+m[u]\sigma$ (using the same matrix operator notation as the one introduced in \eqref{PauliExp}) satisfies the bound
\ba
\label{Ubnd}
\|U\Phi\|
\le C_U\|\Phi\|,
\ea
where we set $C_U:=\sum_{\alpha\in\l0,3\r }\|u_\alpha\|_\infty$ (in particular, if $U$ is defined on the whole of $\widehat\fH$, we have $U\in\mL(\widehat\fH)$).
Finally, for all $u\in L^\infty(\T)$, we denote the real, imaginary, even, and odd part of $u$ (defined almost everywhere) by $\Re(u)$, $\Im(u)$, $\Ev(u)$, and $\Od(u)$, respectively.

We next determine the properties of the Pauli coefficients specifying a translation invariant Hamiltonian in momentum space.

\bp[Translation invariance]
\label{prop:H}
Let $H\in\mL(\fH)$ be a Hamiltonian satisfying Assumption \ref{ass:H} \ref{HTheta}.Then:

\bn[label=(\alph*),ref={\it (\alph*)}]
\setlength{\itemsep}{0mm}
\item
\label{prop:H-a}
There exist $u_0\in L^\infty(\T)$ and $u=[u_1,u_2,u_3]\in L^\infty(\T)^3$ such that 
\ba
\label{Hu}
\widehat H
=m[u_0]\sigma_0+m[u]\sigma.
\ea  

\item
\label{prop:H-b}
For all $\alpha\in\l0,3\r$, we have
\ba
\label{Imu=0}
\Im(u_\alpha)
&=0.
\ea
Moreover, the even and odd parts have the properties that, for all $\alpha\in\l0,2\r$,
\ba
\label{Evu}
\Ev(u_\alpha)
&=0,\\
\label{Oddu}
\Od(u_3)
&=0.
\ea
\en
\ep 

\bprf 
\ref{prop:H-a}\,
Let us first note that, since the Hamiltonian can be written in the form $H=h_0\sigma_0+h\sigma$ with the Pauli coefficients $h_0\in\mL(\fh)$ and $h:=[h_1,h_2,h_3]\in\mL(\fh)^3$, \eqref{H1} and \eqref{H2} respectively yield, for all $\alpha\in\l0,3\r$, 
\ba
\label{hsa}
h_\alpha^\ast
&=h_\alpha,\\
\label{hac}
\zeta h_\alpha\zeta
&=\begin{cases}
-h_\alpha, & \alpha\in\l0,2\r,\\
\hfill h_\alpha, & \alpha=3.
\end{cases}
\ea
On the other hand, Assumption \ref{ass:H} \ref{HTheta} implies that, for all $\alpha\in\l0,3\r$,
\ba
\label{prfH-1}
[h_\alpha,\theta]
=0.
\ea
Hence, we know (see \cite{BS2006} for example) that, for all $\alpha\in\l0,3\r $, there exist $u_\alpha\in L^\infty(\T)$  with
\ba
\label{prfH-2}
\hat h_\alpha
=m[u_\alpha],
\ea
 \ie, we can write $\widehat H=m[u_0]\sigma_0+m[u]\sigma$, where we set  $u:=[u_1,u_2,u_3]\in L^\infty(\T)^3$.

\vspace{1mm}
\ref{prop:H-b}\,
Using \eqref{hsa} and \eqref{prfH-2}, we have $\bar u_\alpha=u_\alpha$ for all $\alpha\in\l0,3\r $, where $\bar\vi$ is the complex conjugation of $\vi\in\hat\fh$. Moreover, since $\ff\zeta f=\hat\xi\hspace{0.5mm} \bar{\hat f}$ for all $f\in\fh$, where we recall that $\xi\in\mL(\fh)$ is the parity from Definition \ref{def:iso}  \ref{def:iso-b}, we get, for all $\alpha\in\l0,3\r $, 
 \ba
 \hat\xi u_\alpha
 &=\begin{cases}
-u_\alpha, & \alpha\in\l0,2\r ,\\
\hfill u_\alpha, & \alpha=3,
\end{cases}
 \ea
 and we note that $(\hat\xi\vi)(k)=\vi(-k)$ for all $\vi\in\hat\fh$ and almost all $k\in\T$.
\eprf

\br
\label{rem:analytic}
Let $\alpha\in\l0,3\r$. Due to a theorem by Bernstein, if, and only if, (the $2\pi$-periodic extension of) $u_\alpha$ is real-analytic, there exist constants $C, a>0$ such that, for all $x\in\Z$,
\ba
|\check u_\alpha (x)|
\le C \e^{-a |x|}. 
\ea
Moreover, under these conditions, the number of zeros of $u_\alpha$ on $\T$ is finite. In Section \ref{sec:Ep}, we will study the special case for which $\check u_\alpha$ has finite support.
\er

In the following, for all functions $\T\ni k\mapsto u(k)\in\C$ and all $k_0\in\T$, we denote by $u'(k_0)$ not only the derivative of $u$ with respect to $k$ at the point $k_0$ if $k_0\in\T\setminus\{\pm\pi\}$ but also the one-sided derivatives if $k_0\in\{\pm\pi\}$ (if all the derivatives in question exist). In this sense, for all $m\in\N$, we denote by $C^m(\T)$ the $m$ times continuously differentiable complex-valued functions on $\T$. Moreover, $C(\T)$ stands for the continuous and $C^\infty(\T)$ for the infinitely differentiable complex-valued functions on $\T$. The analogous notations are used if $\T$ is replaced by $\R$ and/or the target space $\C$ by another Banach space (which is then explicitly indicated).

The following conditions will be used at various places in the sequel.

\bass[Pauli coefficient functions]
\label{ass:reg}
Let $u_\alpha\in L^\infty(\T)$ for all $\alpha\in\l0,3\r$.
\bn[label=(\alph*),ref={\it (\alph*)}]
\setlength{\itemsep}{0mm}
\item
\label{ass:reg-1}

$\Im(u_\alpha)=0$ for all $\alpha\in\l0,3\r$

\item
\label{ass:reg-3}
$u_\alpha\in C^1(\T)$ with $u_\alpha(\pi)=u_\alpha(-\pi)$ and $u_\alpha'(\pi)=u_\alpha'(-\pi)$  for all $\alpha\in\l0,3\r$
\en
\eass

In the following, for all $m\in\N$ and all $u:=[u_1,\ldots,u_m]\in L^\infty(\T)^m$ with real-valued entries, we define the Euclidean norm function $|u|\in L^\infty(\T)$ 
by $|u|:=\sqrt{\sum_{i\in\l1,m\r }u_i^2}$ and the generalized zero set $\mZ_u\in\mM(\T)$ (up to subsets of sets of Borel-Lebesgue measure zero)
by 
\ba
\label{Zu}
\mZ_u
&:=\{k\in\T\,|\,|u|(k)=0\}\\
\label{Zu-2}
&=\bigcap_{i\in\l1,m\r } \mZ_{u_i},
\ea
and we use the notation $\mZ_u^c:=\T\setminus\mZ_u\in\mM(\T)$.
Moreover, for all $i\in\l1,m\r $, the functions $\tilde u_i\in  L^\infty(\T)$  are defined   by
\ba
\label{tildeui}
\tilde u_i
:=\begin{cases}
\displaystyle\frac{u_i}{|u|}, & \mbox{on $\mZ_u^c$},\\
\hfill 0, & \mbox{on $\mZ_u$},
\end{cases}
\ea
and we set $\tilde u:=[\tilde u_1,\ldots, \tilde u_m]\in L^\infty(\T)^m$. Moreover, for all $u:=[u_1,\ldots,u_m]\in L^\infty(\T)^m$ and $v:=[v_1,\ldots,v_m]\in L^\infty(\T)^m$ with real-valued entries, the Euclidean scalar product function $uv\in L^\infty(\T)$ is defined by $uv:=\sum_{i\in\l1,m\r }u_iv_i$ and we set $u^2:=uu$. Finally, for all $u_0\in L^\infty(\T)$ and all $u:=[u_1,\ldots,u_m]\in L^\infty(\T)^m$, we set $u_0u:=[u_0u_1,\ldots,u_0u_m]\in L^\infty(\T)^m$.

The following functions are the basic ingredients in the diagonalization of the Hamiltonian (see Proposition \ref{prop:diag} and  Remark \ref{rem:diag}).

\bd[Eigenvalue functions]
\label{def:eigf}
Let $u_0\in L^\infty(\T)$ and $u:=[u_1,u_2,u_3]\in L^\infty(\T)^3$ satisfy Assumption \ref{ass:reg} \ref{ass:reg-1} and \ref{ass:reg-3}. The eigenvalue functions $e_\pm\in C(\T)\cap C^1(\mZ_u^c)$ are defined by
\ba
\label{def:epm}
e_\pm
:=u_0\pm|u|.
\ea
Moreover, we define the set  $\mZ_\pm\in\mM(\T)$ 
by 
\ba
\label{def:Zpm}
\mZ_\pm
&:=\{k\in\mZ_u^c\,|\, e_\pm'(k)=0\}, 
\ea
and, on $\mZ^c_u$, we have $e_\pm'=u_0'\pm\tilde uu'$.
\ed

\br
Since, due to Assumption \ref{ass:reg} \ref{ass:reg-3}, we have $u_\alpha\in C(\T)$ for all $\alpha\in\l1,3\r$, the set $\mZ_u$ is closed relative to $\T$ 
(and $\R$)
and, hence, $\mZ_u^c$ is open relative to $\T$.
\er

In the following, whenever the symbol $\pm$ appears several times in the same equation, the latter stands for two equations, one of which corresponds to all the upper signs and the other one to all the lower signs (no cross terms).

The following conditions will be used in Section \ref{sec:Ep} and Appendix \ref{app:mmo}.

\bass[Null sets]
\label{ass:nullsets}
Let $u_0\in L^\infty(\T)$ and $u:=[u_1,u_2,u_3]\in  L^\infty(\T)^3$ satisfy Assumption \ref{ass:reg} \ref{ass:reg-1} and \ref{ass:reg-3} and let $M\in\mM(\R)$. 
\bn[label=(\alph*),ref={\it (\alph*)}]
\setlength{\itemsep}{0mm}
\item
\label{ass:reg-4}
$|\mZ_u\cap e_\pm^{-1}(M)|=0$ 

\item
\label{ass:reg-5}
$|\mZ_\pm\cap e_\pm^{-1}(M)|=0$ 

\en
\eass

In the following, $\ell^0(\Z)$ stands for the subspace of $\fh$ of all the complex-valued functions on $\Z$ with finite support. Moreover, let $\dom(q)$ be the subspace of $\fh$ defined by
\ba
\label{def:domq}
\dom(q)
:=\Big\{f\in\fh\,\Big|\,\sum\nolimits_{x\in\Z} x^2 \hspace{0.5mm}|f(x)|^2<\infty\Big\},
\ea
and note that $\dom(q)$ is dense in $\fh$ since the Kronecker basis satisfies $\{\delta_x\}_{x\in\Z}\subseteq\ell^0(\Z)\subseteq\dom(q)$.
Moreover, let $q:\dom(q)\subseteq\fh\to\fh$ stand for the usual position operator on the position space $\fh$ whose action is given,  for all $f\in \dom(q)$ and all $x\in\Z$, by
\ba
\label{def:q}
(qf)(x)
:=xf(x).
\ea
Recall that $q$ is unbounded since, if, for any $f\in\dom(q)$ with $f\notin\ell^0(\Z)$, we set $f_n:=(f-1_{\l-n,n\r}f)/\|f-1_{\l-n,n\r}f\|\in\dom(q)$ for all $n\in\N$, we have $\|qf_n\|\ge n+1$ for all $n\in\N$. 
Moreover,  since $f/(\kp_1\pm\i)\in\dom(q)$ for all $f\in\fh$,
where $\kp_1(x)=x$ for all $x\in\R$ stems from \eqref{kp1}, we have $\ran(q\pm\i1)=\fh$. Hence, since, due to \eqref{def:q}, $q$ is symmetric, the standard criterion for selfadjointness implies that $q^\ast=q$.
Furthermore, its lifting to the doubled 1-particle Hilbert space $\fH$ is defined by 
\ba
\dom(Q)
&:=\dom(q)\oplus\dom(q),\\
Q
&:=q\sigma_0, 
\ea
where, for the case of unbounded operators, we use the same matrix operator notation as the one introduced in \eqref{PauliExp} for the bounded operators  (but acting on the domain of definition of the unbounded operator in question). From the foregoing considerations for $q$, we obtain that $\dom(Q)$ is a dense subspace of $\fH$ and that $Q^\ast=Q$.

Next,  let $\dom(p)$ be the subspace of $\hat\fh$ defined by
\ba
\label{def:domp}
\dom(p)
:=\{\vi\in AC(\T)\,|\, \vi'\in\hat\fh\mbox{ and } \vi(\pi)=\vi(-\pi)\},
\ea
where $ AC(\T)$ stands for the complex-valued absolutely continuous functions on $\T$. Recall that if $\vi\in  AC(\T)$, then $\vi'(k)$ exists for almost all $k\in\T$, $\vi'\in L^1(\T)$, and $\vi(k)=\vi(-\pi)+\int_{[-\pi,k]}\rd s\, \vi'(s)$ for all $k\in\T$. Conversely, if $\psi\in L^1(\T)$, then the function  $\T\ni k\to\vi(k):=\int_{[-\pi,k]}\rd s\, \psi(s)$ satisfies $\vi\in AC(\T)$ and  $\vi'(k)=\psi(k)$ for almost all $k\in\T$.
Now, note that $\ff^\ast\vi\in\dom(q)$ for all $\vi\in\dom(p)$ 
since,  for all $\vi\in\dom(p)$ and all $x\in\Z$, we have
\ba
\label{xffvi}
x(\ff^\ast\vi)(x)
=-\i(\ff^\ast\vi')(x), 
\ea
where we used partial integration in $ AC(\T)$ (and, for example, that $C^1(\T)\subseteq AC(\T)$).
Moreover, we also have $\ff f\in\dom(p)$ for all $f\in\dom(q)$ because, on one hand, $\ff f\in W^{1,2}(\T)$ due to the fact that $\{\vi\in \hat\fh\,|\, \ff^\ast\vi\in\dom(q)\}= W^{1,2}(\T)$ 
(\eqref{xffvi} also holds for $\vi\in  W^{1,2}(\T)$), where $W^{1,2}(\T)$ stands for the usual (periodic) Sobolev space,
and, on the other hand, since we know that $W^{1,2}(\T)=\dom(p)$.  
Hence, since $\ff^\ast\vi\in\dom(q)$ for all $\vi\in\dom(p)$ and since $\ff f\in\dom(p)$ for all $f\in\dom(q)$, the restriction of the unitary operator $\ff:\fh\to\hat\fh$ to $\dom(q)$ is a bijection between $\dom(q)$ and $\dom(p)$. Therefore, $\dom(p)$ being a dense subspace of $\hat\fh$, we define the position operator on momentum space $p:\dom(p)\subseteq\hat\fh\to\hat\fh$,  for all $\vi\in \dom(p)$, by
\ba
\label{def:p}
p\vi
:=-\ii \vi'.
\ea
Moreover, due to \eqref{xffvi}, we can write $p\vi=\ff q \ff^\ast\vi$ for all $\vi\in\dom(p)$ which implies that $p$ is an unbounded selfadjoint operator on momentum space. Finally, as for $Q$ above, the lifting to the doubled 1-particle momentum space $\widehat\fH$ is defined by 
\ba
\label{def:domP}
\dom(P)
&:=\dom(p)\oplus\dom(p),\\
P
&:=p\sigma_0,
\ea
and we again get that $\dom(P)$ is a dense subspace of $\widehat\fH$ and that $P^\ast=P$. Moreover, $\fF$ is a bijection between $\dom(Q)$ and $\dom(P)$ and $P\Phi=\fF Q\fF^\ast\Phi$ for all $\Phi\in\dom(P)$.

We now arrive at the definition of the asymptotic velocity of the system.

\bd[Asymptotic velocity]
\label{def:assvel}
Let $u_0\in L^\infty(\T)$ and $u:=[u_1,u_2,u_3]\in  L^\infty(\T)^3$ satisfy Assumption \ref{ass:reg} \ref{ass:reg-1} and define $U\in\mL(\widehat\fH)$ by 
\ba
\label{Umu}
U:=m[u_0]\sigma_0+m[u]\sigma.
\ea
If $\e^{\ii t U}\dom(P)\subseteq\dom(P)$ for all $t\in\R$ and if, for all $\Phi\in\dom(P)$, the limit for $t\to\infty$ of $\e^{-\ii t U}P\e^{\ii t U}\Phi/t$ exists in $\widehat\fH$, the  operator $V:\dom(P)\to\widehat\fH$ defined, for all $\Phi\in\dom(P)$, by
\ba
\label{def:assvel-0}
V\Phi
:=\lim_{t\to\infty}\frac1t\hspace{1mm}\e^{-\ii t U}P\e^{\ii t U}\Phi, 
\ea
is called asymptotic velocity (with respect to $U$). 
\ed

\br
\label{rem:sa}
Under Assumption \ref{ass:reg} \ref{ass:reg-1}, the operator $U$ from \eqref{Umu} is bounded on $\widehat\fH$ (due to \eqref{Ubnd}) and symmetric. Hence, 
$U^\ast=U$ and the propagator $\e^{\ii t U}$ is well-defined for all $t\in\R$.
\er

Under a simple regularity assumption specific to Section \ref{sec:Ep}, we get the natural explicit form of the asymptotic velocity.

\bp[Asymptotic velocity]
\label{prop:V}
Let $u_0\in L^\infty(\T)$ and $u:=[u_1,u_2,u_3]\in  L^\infty(\T)^3$ satisfy Assumption \ref{ass:reg} \ref{ass:reg-1} and \ref{ass:reg-3} and define $U\in\mL(\widehat\fH)$ by $U:=m[u_0]\sigma_0+m[u]\sigma$. Then:
\bn[label=(\alph*),ref={\it (\alph*)}]
\setlength{\itemsep}{0mm}
\item
\label{prop:V-a}
The asymptotic velocity $V$ with respect to $U$ exists and is a bounded symmetric operator on $\dom(P)$. 

\item
\label{prop:V-b}
The bounded extension $\bar V\in\mL(\widehat\fH)$ of $V$ to $\widehat\fH$ is selfadjoint and has the form 
\ba
\bar V
=m[v_0]\sigma_0+m[v]\sigma, 
\ea
where $v_0\in L^\infty(\T)$ and $v\in L^\infty(\T)^3$ are defined by
\ba
\label{Vas2}
v_0
&:= u_0',\\
\label{Vas3}
v
&:=\begin{cases}
(\tilde uu')\hspace{0.2mm}\tilde u, & \mbox{\textup{on} $\mZ_u^c$},\\
\hfill u', &  \mbox{\textup{on} $\mZ_u$}.
\end{cases}
\ea
\en
\ep

\bprf
\ref{prop:V-a}\,
We first want to show that $\e^{\ii t U}\dom(P)\subseteq\dom(P)$ for all $t\in\R$. To this end, we make use of Proposition \ref{prop:diag} \ref{prop:diag-a} which asserts that, for all $t\in\R$, 
\ba
\label{expa0a}
\e^{\ii t U}
=m[\exp\circ(\ii t u_0) C_t]\sigma_0+\ii t\hspace{0.5mm}m[\exp\circ(\ii t u_0) S_t u]\sigma,
\ea
where the maps $\R\ni t\to C_t\in L^\infty(\T)$ and $\R\ni t\to S_t\in 
L^\infty(\T)$ are given, for all $t\in\R$, by
\ba
\label{def:kt}
C_t
&:=\cos\circ (t|u|),\\
\label{def:st}
S_t
&:=\sinc\circ(t|u|),
\ea
and $\sinc\in C^\infty(\R)$ stands for the usual cardinal sine function (see the beginning of Appendix \ref{app:mmo}). In order to verify the first property in \eqref{def:domp} (and \eqref{def:domP}), we note that, for all $t\in\R$, the functions $C_t$ and $S_t$ are differentiable with respect to $k$ for all $k\in\T$.
Moreover, for all $t\in\R$, the derivatives have the form 
\ba
\label{ct'}
C_t'
&=-t^2\sinc\circ(t|u|) (uu'),\\
\label{st'}
S_t'
&=-\frac{t^2}{2} \big(\sinc\circ (t|u|)+\sinc''\circ (t|u|)\big) (uu'),
\ea
where, in \eqref{st'}, we used the fact that $x\sinc''(x)+2\sinc'(x)+x\sinc(x)=0$ for all $x\in\R$. 
Due to the first part of Assumption \ref{ass:reg} \ref{ass:reg-3}, \eqref{ct'} and \eqref{st'} yield $C_t, S_t\in C^1(\T)$ for all $t\in\R$.
Hence, since  $C^1(\T)\subseteq AC(\T)$ and since $ AC(\T)$ is a $\ast$-algebra,
we get from \eqref{expa0a} that $\e^{\ii t U}\Phi\in AC(\T)^2$ for all $t\in\R$ and all $\Phi\in\dom(P)$. As for the second property in \eqref{def:domp}, we note that, for all $t\in\R$ and all $\Phi\in\dom(P)$, we have (almost everywhere in $\T$)
\ba
\label{Dprop}
\big(\e^{\ii t U}\Phi\big)'
&=\big(m[\exp\circ(\ii t u_0) (\ii t u_0' C_t+C_t')]\sigma_0
+\ii t\hspace{0.5mm}m[\exp\circ(\ii t u_0)((\ii t u_0' S_t +S_t')u+S_t u')]\sigma\big)\Phi\nonumber\\
&\quad+\e^{\ii t U}\Phi',
\ea
where we set $\Phi':=\vi_1'\oplus\vi_2'$ if $\Phi=\vi_1\oplus\vi_2 $. Hence, since $C(\T)\subseteq L^\infty(\T)\subseteq\hat\fh$, the first part of Assumption \ref{ass:reg} \ref{ass:reg-3} implies that $(\e^{\ii t U}\Phi)'\in\widehat\fH$ for all $t\in\R$ and all $\Phi\in\dom(P)$. Finally, due to \eqref{expa0a} and the second part of Assumption \ref{ass:reg} \ref{ass:reg-3}, we also get the third property in \eqref{def:domp}.

Next, let $\Phi\in\dom(P)$ be fixed, let the map $X:\R\to\widehat\fH$  be defined, for all $t\in\R$, by 
\ba
\label{Xt}
X^t
:=\e^{-\ii t U} P \e^{\ii t U}\Phi,
\ea
and let us show that $X\in C^1(\R,\widehat\fH)$. To this end, let $t\in\R$, let $s\in I:=[-1,1]\setminus\{0\}$, and consider the difference quotient 
\ba
\label{DiffQuot}
\frac{X^{t+s}-X^t}{s}
=D_{1,t}^s+D_{2,t}^s,
\ea
where, for all $i\in\l1,2\r $ and all $t\in\R$, the maps $D_{i,t}:I\to \widehat\fH$ are given, for all $s\in I$, by
\ba
\label{D1}
D_{1,t}^s
&:=\e^{-\ii t U}\hspace{0.7mm}\frac{\e^{-\ii s U}-1}{s}\hspace{0.7mm} P \e^{\ii t U}\Phi,\\
\label{D2}
D_{2,t}^s
&:=\e^{-\ii t U}\e^{-\ii s U}P\hspace{0.7mm}\frac{\e^{\ii s U}-1}{s}\hspace{0.7mm}\e^{\ii tU}\Phi.
\ea
Now, since $U\in\mL(\widehat\fH)$, the limit (in $\widehat\fH$) for $s\to 0$ of \eqref{D1} yields, for all $t\in\R$, 
\ba
\label{limD1}
\lim_{s\to 0}D_{1,t}^s
=-\ii \hspace{0.2mm}\e^{-\ii t U} U P\e^{\ii t U}\Phi.
\ea
In order to determine the limit for $s\to 0$ of \eqref{D2}, we make use of \eqref{Dprop} and get, for all $s\in I$ and all $\Psi\in\dom(P)$,
\ba
\label{diffquot}
P\hspace{0.7mm}\frac{\e^{\ii s U}-1}{s}\hspace{0.7mm}\Psi
&=\bigg(\hspace{-0.5mm}m\bigg[\hspace{-0.5mm}\exp\circ(\ii s u_0) \bigg(u_0' C_s-\ii \frac{C_s'}{s}\bigg)\bigg]\sigma_0
+\hspace{0.5mm}m[\exp\circ(\ii s u_0)((\ii s u_0' S_s +S_s')u+S_s u')]\sigma\bigg)\Psi\nonumber\\
&\quad-\ii \hspace{0.7mm}\frac{\e^{\ii s U}-1}{s}\hspace{0.7mm}\Psi'.
\ea
Moreover, since $(U\Psi)'=(m[u_0']\sigma_0+m[u']\sigma)\Psi+U\Psi'$ for all $\Psi\in\dom(P)$, we get, as above, that $U\Psi\in\dom(P)$ for all $\Psi\in\dom(P)$. Hence, for all $s\in I$ and all $\Psi\in\dom(P)$, the decomposition \eqref{diffquot} leads to
\ba
\label{estQ}
\bigg\|P\hspace{0.7mm}\frac{\e^{\ii s U}-1}{s}\hspace{0.7mm}\Psi
-\ii PU\Psi \bigg\|
&\le \sum_{i\in\l1,6\r } A_i(s),
\ea
where, for all $i\in\l1,6\r $, the functions $A_i: I\to\R$ are defined, for all $s\in I$, by
\ba
\label{U1}
A_1(s)
&:=\|m[(\exp\circ(\ii s u_0)C_s-1)u_0']\sigma_0\Psi\|,\\
\label{U2}
A_2(s)
&:=\frac{1}{|s|}\hspace{0.5mm}\|m[C_s']\sigma_0 \Psi\|,\\
\label{U3}
A_3(s)
&:=|s| \|m[u_0'S_s u]\sigma\Psi\|,\\
\label{U4}
A_4(s)
&:=\|m[S_s'u]\sigma\Psi\|,\\
\label{U5}
A_5(s)
&:=\|(m[(\exp\circ(\ii s u_0)S_s-1)u']\sigma\Psi\|,\\
\label{U6}
A_6(s)
&:=\bigg\|\frac{\e^{\ii s U}-1}{s}\hspace{0.7mm}\Psi'-\ii U\Psi'\bigg\|,
\ea
and, in \eqref{U2}-\eqref{U4},  we used that $|\exp\circ(\ii s u_0)|=1$ for all $s\in\R$. In order to estimate \eqref{U1}-\eqref{U6}, we next write $\Psi=\psi_1\oplus\psi_2$ for all $\Psi\in\dom(P)$. As for \eqref{U1}-\eqref{U2}, we have, for all $n\in\l1,2\r$ and all $s\in  I$, that $A_n(s)^2=\sum_{i\in\l1,2\r}\|f_{n,i}^s\|^2$, where, for all $n, i\in\l1,2\r$ and all $s\in  I$, we define $f_{n,i}^s\in\hat\fh$ by 
\ba
\label{f1is}
f_{1,i}^s
&:=(\exp\circ(\ii s u_0)C_s-1)u_0'\psi_i,\\
\label{f2is}
f_{2,i}^s
&:=\frac{C_s'}{s}\hspace{0.5mm}\psi_i.
\ea
Using \eqref{ct'} and the bound $|\sinc(x)|\le 1$ for all $x\in\R$ for \eqref{f2is} (which follows from the representation $\sinc(x)=\int_0^1\rd \lambda\hspace{0.5mm}\cos(\lambda x)$ for all $x\in\R$), we get,  for all $n, i\in\l1,2\r$, that $\lim_{s\to 0}f_{n,i}^s(k)=0$ for all $k\in\T$. Since, in addition, for all $n, i\in\l1,2\r$, we have $|f_{n,i}^s|^2\le C_n |\psi_i|^2\in L^1(\T)$ for all $s\in  I$, where $C_1:=4 \|u_0'\|_\infty^2$ and  $C_2:=3\sum_{\alpha\in\l1,3\r}\|u_\alpha\|_\infty^2\|u_\alpha'\|_\infty^2$, Lebesgue's dominated convergence theorem implies that $\lim_{s\to 0} A_n(s)=0$ for all $n\in\l1,2\r$.  As for \eqref{U3}-\eqref{U5}, for all $n\in\l3,5\r$ and all $s\in  I$,  we can write $A_n(s)^2\le 3\sum_{i\in\l1,2\r}\sum_{\alpha\in\l1,3\r}\|f_{n,i,\alpha}^s\|^2$, where, for all $n\in\l3,5\r$, all $i\in\l1,2\r$, all  $\alpha\in\l1,3\r$, and all $s\in  I$, we define $f_{n,i,\alpha}^s\in\hat\fh$ by 
\ba
\label{f3ias}
f_{3,i,\alpha}^s
&:=s u_0'S_su_\alpha\psi_i,\\
\label{f4ias}
f_{4,i,\alpha}^s
&:=S_s'u_\alpha\psi_i,\\
\label{f5ias}
f_{5,i,\alpha}^s
&:=(\exp\circ(\ii s u_0)S_s-1)u_\alpha'\psi_i.
\ea
Using \eqref{st'} and $|\sinc''(x)|\le 1/3$ for all $x\in\R$ for \eqref{f4ias}, we get,  for all $n\in\l3,5\r$, all $i\in\l1,2\r$, and all  $\alpha\in\l1,3\r$, that $\lim_{s\to 0}f_{n,i,\alpha}^s(k)=0$ for all $k\in\T$. Since, in addition, for all $n\in\l3,5\r$, all $i\in\l1,2\r$, and all $\alpha\in\l1,3\r$, we have $|f_{n,i,\alpha}^s|^2\le C_{n,\alpha} |\psi_i|^2\in L^1(\T)$ for all $s\in  I$, where, for all $\alpha\in\l1,3\r$, we set $C_{3,\alpha}:=\|u_0'\|_ \infty^2\|u_\alpha\|_ \infty^2$, $C_{4,\alpha}:= 3(\sum_{\beta\in\l1,3\r}\|u_\beta\|_\infty^2\|u_\beta'\|_\infty^2)\|u_\alpha\|_\infty^2$, and $C_{5,\alpha}:=4 \|u_\alpha'\|_\infty^2$, Lebesgue's dominated convergence theorem again implies that $\lim_{s\to 0} A_n(s)=0$ for all $n\in\l3,5\r$. 
Moreover, we have $\lim_{s\to 0} A_6(s)=0$ as in \eqref{limD1}.
Finally, since, for all $t\in\R$ and all $s\in I$, we can write
$D_{2,t}^s-\ii\hspace{0.2mm}\e^{-\ii t U}PU\e^{\ii t U}\Phi=\ii\hspace{0.2mm}\e^{-\ii t U}(\e^{-\ii s U}-1)PU\e^{\ii t U}\Phi+\e^{-\ii t U}\e^{-\ii s U}(P(\e^{\ii s U}-1)\e^{\ii t U}\Phi/s-\ii PU\e^{\ii t U}\Phi)$, we get, for all $t\in\R$ and all $s\in I$, 
\ba
\label{D2tconv}
\big\|D_{2,t}^s-\ii\hspace{0.2mm}\e^{-\ii t U}PU\e^{\ii t U}\Phi\big\|
\le\big \|\big(\e^{-\ii s U}-1\big)PU\e^{\ii t U}\Phi\big\|
+\bigg\|P\hspace{0.7mm}\frac{\e^{\ii s U}-1}{s}\hspace{0.7mm}
\e^{\ii t U}\Phi-\ii P U\e^{\ii t U}\Phi\bigg\|,
\ea
which, using \eqref{estQ} and the strong continuity
of the propagator, implies that, for all $t\in\R$, 
\ba
\label{limD2}
\lim_{s\to 0}D_{2,t}^s
=\ii\hspace{0.2mm}\e^{-\ii t U}PU\e^{\ii t U}\Phi.
\ea
Therefore, it follows from \eqref{DiffQuot}, \eqref{limD1}, and \eqref{limD2}, that the map $X$ is differentiable in $\widehat\fH$ at any point in $\R$ and that its derivative $\dot X:\R\to\widehat\fH$, defined, for all $t\in\R$, by $\dot X^t:=\lim_{s\to 0}(X^{t+s}-X^t)/s$, reads, for all $t\in\R$, as
\ba
\label{Xdot}
\dot X^t
=\e^{-\ii t U} U'\e^{\ii t U}\Phi,
\ea
where the commutator $U':\dom(P)\to\widehat\fH$ is defined by $U'\Phi:=-\i(UP\Phi-PU\Phi)$ for all $\Phi\in\dom(P)$. Since, for all $\Phi\in\dom(P)$, we have
\ba
\label{U'}
U'\Phi
=(m[u_ 0']\sigma_0+m[u']\sigma)\Phi, 
\ea
\eqref{Ubnd} implies that $\|U'\Phi\|\le C_{U'}\|\Phi\|$ for all  $\Phi\in\dom(P)$ with $C_{U'}:=\sum_{\alpha\in\l0,3\r}\|u_\alpha'\|_\infty$. Moreover, since, as above, $\dot X^{t+s}-\dot X^t=\e^{-\ii tU}(\e^{-\ii sU}-1)U'\e^{\ii tU}\Phi+\e^{-\ii tU}\e^{-\ii sU}U'\e^{\ii tU}(\e^{\ii sU}-1)\Phi$ for all $s,t\in\R$, we get, for all $s,t\in\R$, 
\ba
\label{estdotX}
\big\|\dot X^{t+s}-\dot X^t\big\|
\le \big\| \big(\e^{-\ii s U}-1 \big) U' \e^{\ii tU}\Phi\big\|
+C_{U'}\big\| \big(\e^{\ii sU}-1 \big)\Phi \big\|.
\ea
The strong continuity of the propagator and \eqref{estdotX} now imply that $\dot X\in C(\R,\widehat\fH)$, \ie, we find that $X\in C^1(\R, \widehat\fH)$ as desired.

We next want to compute the limit (in $\widehat\fH$) for $t\to\infty$ of $X^t/t$. In order to do so, we note that, due to $\dot X\in C(\R,\widehat\fH)$ 
and \eqref{Xdot}, the second fundamental theorem of Banach space-valued Riemann integral calculus yields, 
for all $t\in\R^+$,
\ba
\label{intform}
X^t
&=X^0
+\int_0^t\rd s\hspace{1mm}\dot X^s\nonumber\\
&=P\Phi
+\int_0^t\rd s\hspace{1mm} \e^{-\ii s U} U' \e^{\ii s U}\Phi.
\ea
Using \eqref{expa0a}, \eqref{U'}, and \eqref{prod}, we compute that $\e^{-\ii s U} U' \e^{\ii s U}\Phi=(m[u_0']\sigma_0+m[a^s]\sigma)\Phi$ for all $s\in\R$, where the map $\R\ni s\mapsto a ^s\in L^\infty(\T)^3$ has the form $a^s=\sum_{i\in\l1,3\r}a_i^s$, and, for all $i\in\l1,3\r$, the maps $\R\ni s\mapsto a_i^s\in L^\infty(\T)^3$ are defined, for all $s\in\R$, by
\ba
\label{b1}
a_1^s
&:=C_{2s}\hspace{0.3mm} u',\\
\label{b2}
a_2^s
&:=2sC_sS_s\hspace{0.3mm} (u\wedge u'),\\
\label{b3}
a_3^s
&:=2s^2S^2_s\hspace{0.3mm}  (uu')u.
\ea
As for \eqref{b1}, since for all $i\in\l1,2\r$, all $\alpha\in\l1,3\r$, and all $t\in\R^+$, the Riemann integral $\int_0^t\rd s\hspace{1mm} C_{2s}\hspace{0.3mm} u_\alpha'\vi_i$ exists in $\hat\fh$ due to the fact that $\dot X\in C(\R,\widehat\fH)$,
since, for all $i\in\l1,2\r$, all $\alpha\in\l1,3\r$, all $k\in\T$, and all $t\in\R^+$, the Riemann integral $\int_0^t\rd s\hspace{1mm} C_{2s}(k)\hspace{0.3mm} u_\alpha'(k)\vi_i(k)$ exists in $\C$, and since every sequence which converges in $\hat\fh$ to a limit has a subsequence which converges pointwise always everywhere to the same limit, 
we get, for all $t\in\R^+$,
\ba
\label{intb1}
\int_0^t\rd s\hspace{1mm} (m[a_1^s]\sigma)\Phi
&= t\hspace{0.5mm}(m[S_{2t}u']\sigma)\Phi.
\ea
The terms \eqref{b2} and \eqref{b3} are treated analogously. We find, for all $t\in\R^+$,
\ba
\label{intb2}
\int_0^t\rd s\hspace{1mm} (m[a_2^s]\sigma)\Phi
&=t^2(m[S_t^2(u\wedge u')]\sigma)\Phi,\\
\label{intb3}
\int_0^t\rd s\hspace{1mm} (m[a_3^s]\sigma)\Phi
&=t\hspace{0.5mm}(m[(1-S_{2t})(\tilde uu')\tilde u]\sigma)\Phi,
\ea
where $\tilde u$ is given after \eqref{tildeui}. 
Hence, using \eqref{intform} and \eqref{intb1}-\eqref{intb3}, we get, for all $t\in\R^+$, 
\ba
\label{qas}
\frac{X^t}{t}
=\frac{P\Phi}{t}
+(m[u_0']\sigma_0)\Phi
+(m[S_{2t}u']\sigma)\Phi
+t(m[S_t^2(u\wedge u')]\sigma)\Phi
+(m[(1-S_{2t})(\tilde uu')\tilde u]\sigma)\Phi.
\ea
Since,  on $\mZ_u$, the fourth and fifth term on the right hand side of \eqref{qas} satisfy $tS_t^2(u\wedge u')=(1-S_{2t})(\tilde uu')\tilde u=0$ whereas, for the third term, we have $S_{2t}u'=u'$ on $\mZ_u$, we decompose the latter as $S_{2t}u'=1_{\mZ_u}u'+S_{2t}1_{\mZ_u^c}u'$. Hence, we get, for all $t\in\R^+$,
\ba
\label{Xdivtest}
\bigg\|\frac{X^t}{t}
-\big((m[u_0']\sigma_0)\Phi
+(m[1_{\mZ_u} u']\sigma)\Phi
+(m[(\tilde uu')\tilde u]\sigma)\Phi\big)\bigg\|
\le \sum_{i\in\l1,4\r}B_i(t),
\ea
where, for all $i\in\l1,4\r$, we define $B_i:\R^+\to\R$, for all $t\in\R^+$, by 
\ba
\label{V1}
B_1(t)
&:=\frac{1}{t}\hspace{0.1mm}\|P\Phi\|,\\
\label{V2}
B_2(t)
&:=\|(m[S_{2t}1_{\mZ_u^c} u']\sigma)\Phi\|,\\
\label{V3}
B_3(t)
&:=t\hspace{0.1mm}\|(m[S_t^2(u\wedge u')]\sigma)\Phi\|,\\
\label{V4}
B_4(t)
&:=\|(m[S_{2t}(\tilde uu')\tilde u]\sigma)\Phi\|.
\ea
Setting $\Phi=\vi_1\oplus\vi_2 $ and proceeding as above, we have, for all $n\in\l2,4\r$ and all $t\in\R^+$, that $B_n(t)^2\le 3\sum_{i\in\l1,2\r}\sum_{\alpha\in\l1,3\r}\|g_{n,i,\alpha}^t\|^2$, where, for all $n\in\l2,4\r$, all $i\in \l1,2\r$, all $\alpha\in\l1,3\r$, and all $t\in\R^+$, we define $g_{n,i,\alpha}^t\in\hat\fh$ by 
\ba
g_{2,i,\alpha}^t
&:=S_{2t}1_{\mZ_u^c} u_\alpha'\vi_i,\\
g_{3,i,\alpha}^t
&:=t S_t^2\sum_{\beta,\gamma\in\l1,3\r}\veps_{\alpha\beta\gamma}u_\beta u_\gamma'\vi_i,\\
g_{4,i,\alpha}^t
&:=S_{2t}(\tilde uu')\tilde u_\alpha\vi_i.
\ea
Since, for all $x\in\R\setminus\{0\}$, we have $\lim_{t\to\infty} \sinc(t x)=0$ and $\lim_{t\to\infty} t \sinc^2(t x)=0$, we get, for all $n\in\l2,4\r$, all $i\in\l1,2\r$, and all  $\alpha\in\l1,3\r$, that $\lim_{t\to \infty}g_{n,i,\alpha}^t(k)=0$ for all $k\in\T$. Moreover, for all $n\in\l2,4\r$, all $i\in\l1,2\r$, and all $\alpha\in\l1,3\r$, we have $|g_{n,i,\alpha}^t|^2\le D_{n,\alpha}|\vi_i|^2\in L^1(\T)$ for all $t\in\R^+$,  where, for all $\alpha\in\l1,3\r$, we set $D_{2,\alpha}:=\|u_\alpha'\|_\infty^2$ and, using $|\tilde u_\alpha|\le 1$ for all $\alpha\in\l1,3\r$, we also have $D_{3,\alpha}:=27\sum_{\gamma\in\l1,3\r}\|u_\gamma'\|_\infty^2$ and $D_{4,\alpha}:=3\sum_{\beta\in\l1,3\r}\|u_\beta'\|_\infty^2$ (independent of $\alpha$). 
Hence, Lebesgue's dominated convergence theorem again implies that $\lim_{t\to \infty} B_n(t)=0$ for all $n\in\l2,4\r$ and since $\lim_{t\to \infty} B_1(t)=0$ holds, too, \eqref{Xdivtest} yields the limit in $\widehat\fH$ for $t\to\infty$ of $X^t/t$, \ie, we get, for all $\Phi\in\dom(P)$,
\ba
\label{VPhi}
V\Phi
&=\lim_{t\to\infty}\frac{X^t}{t}\nonumber\\
&=(m[v_0]\sigma_0+m[v]\sigma)\Phi,
\ea
where $v_0\in L^\infty(\T)$ and $v\in L^\infty(\T)^3$ are defined by $v_0:= u_0'$ and $v:=1_{\mZ_u}u'+1_{\mZ_u^c}(\tilde uu')\tilde u$.

Finally, due to \eqref{VPhi} and \eqref{Ubnd}, $V$ is a bounded operator on $\dom(P)$ and, due to Assumption \ref{ass:reg} \ref{ass:reg-1}, $V$ is also symmetric.

\vspace{1mm}
\ref{prop:V-b}\,
Since $V$ is bounded on the dense domain $\dom(P)$, 
we know that the unique bounded extension of $V$ to $\widehat\fH$ is given by $\bar V:=V^{\ast\ast}\in\mL(\widehat\fH)$.
Moreover, since $V$ is symmetric due to part \ref{prop:V-a}, \ie, since $V\subseteq V^\ast$, we get $\bar V=V^{\ast\ast}\subseteq V^{\ast\ast\ast}=\bar V^\ast$ 
and, hence, 
$\bar V^\ast=\bar V$. Finally, since the right hand side of \eqref{VPhi} defines a 
bounded  operator on $\widehat\fH$, the uniqueness of the bounded extension 
implies that the action of $\bar V$ on $\widehat\fH$ is also given by \eqref{VPhi}.
\eprf

\br
\label{rem:rot}
Using the usual group homomorphism between ${\rm SU}(2)$ and ${\rm SO}(3)$ which, for all $\theta\in\R$, all $a=[a_1,a_2,a_3]\in\R^3$ with $\sum_{i\in\l1,3\r}a_ i^2=1$, and all $x\in\R^3$, is given by
\ba
\e^{-\ii \frac\theta2(a\sigma)}(x\sigma)\e^{\ii \frac\theta2(a\sigma)}
=(R(a,\theta)x)\sigma,
\ea
where $R(a,\theta)x:=(ax)a+\cos(\theta)(x-(ax)a)+\sin(\theta)a\wedge x$ stands for the positive rotation of $x$ by the angle $\theta$ around the axis $a$,
 we obtain the geometric interpretation of \eqref{b1}-\eqref{b3}.
\er

The following condition will be used in the sequel.

\bass[Asymptotic velocity]
\label{ass:eig}
Let $H\in\mL(\fH)$ be a Hamiltonian satisfying Assumption \ref{ass:H} \ref{HTheta}, let the Pauli coefficient functions $u_0\in L^\infty(\T)$ and $u\in L^\infty(\T)^3$ of $\widehat H$ satisfy Assumption \ref{ass:reg} \ref{ass:reg-3}, and let $\bar V\in\mL(\widehat\fH)$ be the bounded extension of the asymptotic velocity with respect to $\widehat H$. 
\hfill
\bn[label=(\alph*),ref={\it (\alph*)}]
\setlength{\itemsep}{0mm}
\item 
\label{ass:eig-1}
$0\notin \eig(\bar V)$
\en
\eass

In the following, we use the sign function $\sign:\R\to\{-1,0,1\}$ which is defined according to the convention that $\sign(x):=-1$ if $x<0$, $\sign(0):=0$, and $\sign(x):=1$ if $x>0$. Moreover, recall Definition \ref{def:Asmpt} \ref{def:Asmpt-2} for the R/L generator $\Delta$, Definition \ref{def:eigf} for $e_\pm'$, and define the mean inverse temperature $\beta\in\R$ and (half) the affinity 
$\delta\in\R$, driving the heat flux between the reservoirs, by
\ba
\label{beta}
\beta
&:=\frac{\beta_R+\beta_L}{2},\\
\label{delta}
\delta
&:=\frac{\beta_R-\beta_L}{2}.
\ea

The R/L generator has the following form. 

\bp[R/L generator]
\label{prop:RLgen}
Let $H\in\mL(\fH)$ be a Hamiltonian satisfying Assumption \ref{ass:H} \ref{HTheta} and \ref{L1} and let the Pauli coefficient functions $u_0\in L^\infty(\T)$ and $u\in L^\infty(\T)^3$ of $\widehat H$ satisfy Assumption \ref{ass:reg} \ref{ass:reg-3}. Moreover, let the bounded extension $\bar V\in\mL(\widehat\fH)$ of the asymptotic velocity with respect to $\widehat H$ satisfy Assumption \ref{ass:eig} \ref{ass:eig-1} and let $\beta_L, \beta_R\in\R$ be the inverse reservoir temperatures. Then:

\bn[label=(\alph*),ref={\it (\alph*)}]
\setlength{\itemsep}{0mm}
\item
\label{prop:RLgen-a}
In  momentum space, the R/L generator for $H$ and $\beta_L, \beta_R$ has the form 
\ba
\label{RLgen}
\widehat\Delta
=(\beta 1+\delta\hspace{0.2mm}\sign(\bar V)) 1_{ac}(\widehat H).
\ea

\item
\label{prop:RLgen-b}
The sign function of the asymptotic velocity  can be written as
\ba
\label{signV}
\sign(\bar V)
=m[w_0]\sigma_0+m[w]\sigma,
\ea
where $w_0\in L^\infty(\T)$ and $w\in  L^\infty(\T)^3$ are defined by 
\ba
\label{RL1}
w_0
&:=\frac12
\begin{cases}
\sign\circ e_+'+\sign\circ e_-', & \mbox{\textup{on} $\mZ_u^c$},\\
\hfill \sign\circ f_++\sign\circ f_-, & \mbox{\textup{on} $\mZ_u$}, 
\end{cases}\\
\label{RL2}
w
&:=\frac12
\begin{cases}
(\sign\circ e_+'-\sign\circ e_-')\hspace{0.3mm}\tilde u, & \mbox{\textup{on} $\mZ_u^c$},\\
\hfill(\sign\circ f_+-\sign\circ f_-)\hspace{0.3mm}\widetilde{u'
}, & \mbox{\textup{on} $\mZ_u$},
\end{cases}
\ea
and  $f_\pm\in L^\infty(\T)$ is defined by $f_\pm:=u_0'\pm|u'|$.
\en
\ep

In the following proof, $\srlim$ stands for the convergence in the strong resolvent sense. Recall that a sequence $(A_n)_{n\in\N}$ of (not necessarily bounded) selfadjoint operators on the separable complex Hilbert space $\mH$ converges to a  (not necessarily bounded) selfadjoint operator $A$ on $\mH$ if there exists $z\in (\bigcap_{n\in\N}\res(A_n))\cap\hspace{0.5mm}\res(A)$ (where $\res$ stands for the resolvent set of the operator in question) such that the sequence of the resolvents $((A_n-z1)^{-1})_{n\in\N}\subseteq\mL(\mH)$ converges strongly to $(A-z1)^{-1}\in\mL(\mH)$.

\vspace{5mm}

\bprf 
\ref{prop:RLgen-a}\,
We first note that, due to \eqref{sym:SumP}, the R/L generator can be written as
\ba
\label{RL:Delta}
\Delta
=\beta 1_{ac}(H)+\delta\hspace{0.2mm}P_{RL},
\ea
where $P_{RL}:=P_R-P_L\in\mL(\fH)$. Since $f_0:=(1_R-1_L)-\sign\hspace{-1.5mm}\restriction_\Z\in\ell^0(\Z)$, we have $m[f_0]\in\mL^0(\fh)$ and, hence, using \eqref{def:QL}-\eqref{def:QR} and  \eqref{PL}-\eqref{PR}, we get
\ba
P_{RL}
&=\slim_{t\to\infty} \e^{-\ii tH}(m[1_R-1_L] \sigma_0)\hspace{0.2mm} \e^{\ii tH}1_{ac}(H)\nonumber\\
\label{PRL-1}
&=\slim_{t\to\infty} \e^{-\ii tH}(m[\sign\hspace{-1.5mm}\restriction_\Z] \sigma_0)\hspace{0.2mm} \e^{\ii tH} 1_{ac}(H),
\ea
where, with \eqref{muH}, we used that $\mL^0(\fH)\subseteq\ker(\mu_H)$, \ie, we have $\mu_H(m[f_0]\sigma_0)=0$. In order to express \eqref{PRL-1} be means of the position operator in momentum space, we use the uniqueness property of the resolution of the identity stated in Theorem \ref{thm:spect} \ref{spect-uniq} and define the map $E^m:\mB(\R)\to\mL(\fH)$ by $E^m(\chi):=m[\chi\hspace{-1.5mm}\restriction_\Z]\sigma_0$ for all $\chi\in\mB(\R)$. In order to obtain $E^m=E_Q$, Theorem \ref{thm:spect} \ref{spect-uniq} asserts that it is enough to verify that $\bar E^m(\kappa_1)=Q$ since $E^m$ is a resolution of the identity. Using \eqref{domEx}-\eqref{actionEx}, we have $\dom(\bar E^m(\kappa_1))=\ran(E^m(\kappa_{-1}))=\{ m[\kappa_{-1}\hspace{-1.5mm}\restriction_\Z] f_1\oplus m[\kappa_{-1}\hspace{-1.5mm}\restriction_\Z] f_2  \,|\, f_1, f_2\in\fh\}=\dom(Q)$ because $\ran(m[\kappa_{-1}\hspace{-1.5mm}\restriction_\Z])=\dom(q)$, \ie, for all $F=f_1\oplus f_2 \in\dom(Q)$, there exists $G=g_1\oplus g_2\in\fH$ with $F=E^m(\kappa_{-1})G$, and $\bar E^m(\kappa_1) F=\bar E^m(\kappa_1) E^m(\kappa_{-1})G=E^m(\kappa_1\kappa_{-1})G= (m[(\kappa_1\kappa_{-1})\hspace{-1mm}\restriction_\Z]\sigma_0)G= m[\kappa_1\hspace{-1.5mm}\restriction_\Z] m[\kappa_{-1}\hspace{-1.5mm}\restriction_\Z]g_1\oplus\linebreak m[\kappa_1\hspace{-1.5mm}\restriction_\Z] m[\kappa_{-1}\hspace{-1.5mm}\restriction_\Z]g_2= qf_1\oplus qf_2=QF$. Therefore,  since $\sign\in\mB(\R)$, \eqref{PRL-1} and Theorem \ref{thm:spect} \ref{spect-uniq} imply that 
 \ba
 \label{PRL-2}
P_{RL}
=\slim_{t\to\infty} \e^{-\ii tH} \sign(Q)\hspace{0.5mm} \e^{\ii tH} 1_{ac}(H).
\ea
Note that, here and at various other analogous places, we could have used Theorem \ref{thm:spect} \ref{spect-fourier} and Remark \ref{rem:det} instead of Theorem \ref{thm:spect} \ref{spect-uniq} (see the proof of Lemma \ref{lem:SpecId} \ref{lem:SpecId-c} for example). Applying Lemma \ref{lem:unitary} to $\mH=\fH$, $\mK=\widehat\fH$, and $U=\fF$, we get $B=P$ and $\fF E_Q(\chi)\fF^\ast=E_P(\chi)$ for all $\chi\in\mB(\R)$ if $A=Q$ and, likewise, $B=\widehat H$ and $\fF E_H(\chi)\fF^\ast=E_{\widehat H}(\chi)$ for all $\chi\in\mB(\R)$ if $A=H$. Hence, since $\e_t\in C_b(\R)$ for all $t\in\R$ (where $\e_t(x):=\e^{\ii tx}$ for all $x\in\R$ stems from Theorem \ref{thm:spect} \ref{spect-fourier}) and using that there exists $M_{ac}\in\mM(\R)$ such that $1_{ac}(H)=E_H(1_{M_{ac}})$,
\eqref{PRL-2} leads to
\ba
 \label{PRL-3}
\widehat P_{RL}
=\slim_{t\to\infty} \e^{-\ii t\widehat H} \sign(P)\hspace{0.5mm} \e^{\ii t\widehat H} 1_{ac}(\widehat H).
\ea
Now, recall from the proof of Proposition \ref{prop:V} \ref{prop:V-a} that $\e^{\ii t\widehat H}\dom(P)\subseteq\dom(P)$ for all $t\in\R$. 
Hence, for all $t\in\R^+$, we set $\dom(V^t):=\dom(P)$, we define the operator $V^t:\dom(V^t)\to\widehat\fH$, for all $\Phi\in\dom(V^t)$, by 
\ba
\label{Vt}
V^t\Phi
:=\frac1t\hspace{0.5mm}\e^{-\ii t\widehat H} P \e^{\ii t\widehat H}\Phi,
\ea
and we note that $V^t$ is unbounded for all $t\in\R^+$ (since, for all $t\in\R^+$ and all $n\in\N$, we have $\|V^t\Phi_n\|\ge \sqrt{2}(n+1)/t$, where $\Phi_{n}:=\e^{-\ii t \widehat H}\hat f_n\oplus \hat f_n\in\dom(P)$ and $f_n\in\dom(q)$ is given after \eqref{def:q}).
In order to express \eqref{PRL-3} be means of  \eqref{Vt}, we again use  Lemma \ref{lem:unitary} for $\mH=\mK=\widehat\fH$, $U=\e^{-\ii t\widehat H}$ for all $t\in\R^+$, and $A=P$, and obtain $B=t V^t$ and $\e^{-\ii t\widehat H} E_P(\chi)\e^{\ii t\widehat H}=E_{tV^t}(\chi)$ for all $t\in\R^+$ and all $\chi\in\mB(\R)$.
Moreover, Remark \ref{rem:rA}
yields $E_{tV^t}(\chi)=E_{V^t}(\chi_t)$ for all $t\in\R^+$ and all $\chi\in\mB(\R)$ 
and, hence, we get 
\ba
\label{PRL-4}
\widehat P_{RL}
=\slim_{t\to\infty}\hspace{0.5mm} \sign(V^t) 1_{ac}(\widehat H),
\ea
where we used that $\sign_t=\sign$ for all $t\in\R^+$. Since $\dom(V^t)=\dom(P)$ for all $t\in\R\setminus\{0\}$, since $\dom(P)$ is a core for $\bar V$ because the closure of $V$ is equal to the bounded extension $\bar V$ of $V$,
and since $\lim_{t\to\infty} V^t\Phi=\bar V\Phi$ for all $\Phi\in\dom(P)$ , due to Proposition \ref{prop:V} (and, for example, \cite{Weidmann1980}), we have
\ba
\label{srconv}
\srlim_{t\to\infty} V^t=\bar V.
\ea
Finally, since $\sign=\kappa_0-1_{(-\infty,0]}-1_{(-\infty,0)}$, where $\kappa_0$ is the unity function from \eqref{kp0}, we have $\sign(V^t)=E_{V^t}(\sign)=1-E_{V^t}(1_{(-\infty,0]})-E_{V^t}(1_{(-\infty,0)})$. Hence, since we know that, under Assumption \ref{ass:eig} \ref{ass:eig-1} (which is equivalent to $E_{\bar V}(1_{\{0\}})=0$), \eqref{srconv} implies $\slim_{t\to\infty} E_{V^t}(1_{(-\infty,0]})=E_{\bar V}(1_{(-\infty,0]})$ and $\slim_{t\to\infty} E_{V^t}(1_{(-\infty,0)})=E_{\bar V}(1_{(-\infty,0)})$ (see again \cite{Weidmann1980} for example), we get
\ba
\label{limsignV}
\slim_{t\to\infty}\hspace{0.5mm} \sign(V^t)
=\sign(\bar V).
\ea

\vspace{1mm}
{\it (b)}\, 
Using Proposition \ref{prop:diag} \ref{prop:diag-b} and the Pauli coefficient functions $v_0\in L^\infty(\T)$ and $v\in L^\infty(\T)^3$ of $\bar V$ from  \eqref{Vas2}-\eqref{Vas3}, we can write $\sign(\bar V)=m[w_0]\sigma_0+m[w]\sigma$, where  $w_0\in L^\infty(\T)$ and $w\in L^\infty(\T)^3$ are given by 
\ba
\label{w0-1}
w_0
&:=\frac12(\sign\circ(v_0+|v|)+\sign\circ(v_0-|v|)),\\
\label{w-1}
w
&:=\frac12(\sign\circ(v_0+|v|)-\sign\circ(v_0-|v|))\tilde v,
\ea
and we have $v_0=u_0'$ and 
\ba
\label{|v|}
|v|
=\begin{cases}
|\tilde u u'|, & \mbox{\textup{on} $\mZ_u^c$},\\
\hfill |u'|, & \mbox{\textup{on} $\mZ_u$}.
\end{cases}
\ea
In order to simplify \eqref{w0-1}-\eqref{w-1}, we make the decomposition $\T=(\mZ_v\cap\mZ_u^c)\cup(\mZ_v\cap\mZ_u)\cup(\mZ_v^c\cap\mZ_u^c) \cup(\mZ_v^c\cap\mZ_u)$, where we have $\mZ_v\cap\mZ_u^c=\{k\in\mZ_u^c\,|\, (uu')(k)=0\}$ and $\mZ_v\cap\mZ_u=\{k\in\mZ_u\,|\, |u'|(k)=0\}$.
As for \eqref{w0-1}, we get $w_0=\sign\circ u_0'$ on $\mZ_v\cap\mZ_u^c$ and $\mZ_v\cap\mZ_u$. Moreover, making the further decomposition $\mZ_v^c\cap\mZ_u^c=(\mZ_v^c\cap\mZ_u^c)_+\cup (\mZ_v^c\cap\mZ_u^c)_-$ with $(\mZ_v^c\cap\mZ_u^c)_\pm:=\{k\in\mZ_v^c\cap\mZ_u^c\,|\, \pm (uu')(k)>0\}$, we can write, on $\mZ_v^c\cap\mZ_u^c$,
\ba
w_0
&=\frac12(\sign\circ(u_0'+|\tilde uu'|)+\sign\circ(u_0'-|\tilde uu'|))\nonumber\\
&=\frac12\begin{cases}
\sign\circ(u_0'+\tilde uu')+\sign\circ(u_0'-\tilde uu'), & \mbox{\textup{on} $(\mZ_v^c\cap\mZ_u^c)_+$},\\
\sign\circ(u_0'-\tilde uu')+\sign\circ(u_0'+\tilde uu'), & \mbox{\textup{on} $(\mZ_v^c\cap\mZ_u^c)_-$}
\end{cases}\nonumber\\
&=\frac12(\sign\circ e_+'+\sign\circ e_-'),
\ea
where we recall from Definition \ref{def:eigf} that $e_\pm'=u_0'\pm\tilde uu'$ on $\mZ^c_u$. On the other hand, on $\mZ_v^c\cap\mZ_u$, we can write $w_0=(\sign\circ(u_0'+|u'|)+\sign\circ(u_0'-|u'|))/2$. Hence, since $e_\pm'=u_0'$ on $\mZ_v\cap\mZ_u^c$ and since $f_\pm=u_0'\pm|u'|=u_0'$ on $\mZ_v\cap\mZ_u$, we arrive at \eqref{RL1}.
As for \eqref{w-1}, we first note that 
\ba
\label{tildev}
\tilde v
=\begin{cases}
\sign\circ(\tilde u u')\hspace{0.2mm}\tilde u, & \mbox{\textup{on} $\mZ_v^c\cap\mZ_u^c$},\\
\hfill\displaystyle\frac{u'}{|u'|}, & \mbox{\textup{on} $\mZ_v^c\cap\mZ_u$}, \\
\hfill 0, & \mbox{\textup{on} $\mZ_v$}.
\end{cases}
\ea
Hence,  $w=0$ on $\mZ_v\cap\mZ_u^c$ and on $\mZ_v\cap\mZ_u$ and, on $\mZ_v^c\cap\mZ_u^c$, we have
\ba
w
&=\frac12(\sign\circ(u_0'+|\tilde uu'|)-\sign\circ(u_0'-|\tilde uu'|))\hspace{0.2mm}\sign\circ(\tilde u u')\hspace{0.2mm}\tilde u\nonumber\\
&=\frac12\begin{cases}
\hfill(\sign\circ(u_0'+\tilde uu')-\sign\circ(u_0'-\tilde uu'))\hspace{0.2mm}\tilde u, & \mbox{\textup{on} $(\mZ_v^c\cap\mZ_u^c)_+$},\\
-(\sign\circ(u_0'-\tilde uu')-\sign\circ(u_0'+\tilde uu'))\hspace{0.2mm}\tilde u, & \mbox{\textup{on} $(\mZ_v^c\cap\mZ_u^c)_-$}
\end{cases}\nonumber\\
&=\frac12(\sign\circ e_+'-\sign\circ e_-')\hspace{0.2mm}\tilde u.
\ea
Moreover, on $\mZ_v^c\cap\mZ_u$, we have $w=(\sign\circ f_+-\sign\circ f_-)\widetilde{u'}/2$. Therefore, we arrive at \eqref{RL2} as above.
\eprf

\section{Heat flux}
\label{sec:Ep}
In this section, we determine the expectation value of the macroscopic heat flux observable in general R/L mover states. Moreover, we prove strict positivity of the entropy production in such states and provide examples of physically important models for such systems.

In the following, we make use of the selfdual second quantization $b$ introduced in Definition \ref{def:obs} \ref{def:obs-d}. Moreover, for any separable complex Hilbert space $\mH$, we denote by $\tr:\mL^1(\mH)\to\C$ the usual trace on $\mL^1(\mH)$ (as a special case, the same notation will be used on $\C^{n\times n}$ for all $n\in\N$) and, for all $A\in\mL(\mH)$, we set $\Re(A):=(A+A^\ast)/2$ and $\Im(A):=(A-A^\ast)/(2\i)$. 

\bd[R/L mover heat flux]
\label{def:ep}
Let $H\in\mL(\fH)$ be a Hamiltonian satisfying Assumption \ref{ass:H} \ref{1sc=0} and \ref{L1}, and let $T_0\in\mL(\fH)$ be an initial  2-point operator, $\rho\in\mB(\R)$ a Fermi function, and $\beta_L, \beta_R\in\R$ the inverse reservoir temperatures. Moreover, let $T\in\mL(\fH)$ be the R/L mover 2-point operator for $H$, $T_0$, $\rho$, and $\beta_L, \beta_R$, and let $\omega_T\in\mE_\fA$ be an R/L mover state.
\bn[label=(\alph*),ref={\it (\alph*)}]
\setlength{\itemsep}{0mm}
\item
\label{ep-1}
The 1-particle observable $\Phi\in\mL^1(\fH)$ describing the heat flux from the left reservoir into the sample is defined by 
\ba
\label{def:Phi}
\Phi
:=-\frac12\frac{\rd}{\rd t}\bigg|_{t=0}\e^{\ii t H} H_L\hspace{0.5mm} \e^{-\ii t H}.
\ea

\item
\label{ep-2}
The R/L mover heat flux is defined to be the expectation value of the macroscopic heat flux observable in the R/L mover state $\omega_T$, \ie, 
\ba
\label{def:J}
J
:=\omega_T(b(\Phi)).
\ea
Moreover, we set $J_{pp}:=-\tr(T_{pp}\Phi)$ and $J_{ac}:=-\tr(T_{ac}\Phi)$.

\item
\label{ep-3}
The entropy production rate $\sigma\in\R$ in the R/L mover state is defined by
\ba
\label{ep-sigma}
\sigma
:=(\beta_R-\beta_L) J.
\ea
\en
\ed

\br
\label{rem:deriv}
Since, for all $s\in\R\setminus\{0\}$, we have $\|(\e^{\ii s H}-1)/s-\ii H\|\le \|H\| (\e^{|s|\|H\|}-1)$ and
\ba
\frac{\e^{\ii sH}H_L\e^{-\ii sH}-H_L}{s}-\ii [H,H_L]
=\bigg[ \frac{\e^{\ii sH}-1}{s}-\ii H,H_L\bigg] \e^{-\ii sH}+\ii [H,H_L](\e^{-\ii s H}-1),
\ea
the map $\Psi:\R\to\mL(\fH)$, defined by $\Psi^t:=-\hspace{0.5mm}\e^{\ii t H} H_L \e^{-\ii t H}/2$ for all $t\in\R$, has a well-defined derivative with respect to the operator norm at all points $t\in\R$, \ie, due to the fact that $H\in\mL(\fH)$, the limit defining the derivative in \eqref{def:Phi} exists with respect to the uniform topology on $\mL(\fH)$ (note that, since $\spec(H)$ is compact and since the exponential series converges compactly, the estimate of Proposition \ref{prop:ext} \ref{prop:ext-1}  leads to the same conclusion). 
\er

\br
\label{rem:time}
Applying Remark \ref{rem:cf}, the 1-particle heat flux observable reads
\ba
\label{Phi-2}
\Phi
&=-\frac\ii2 [H,H_L]\\
\label{Phi-3}
&=-\hspace{0.5mm}\Im(H_L (H_{LS}+H_{LR})).
\ea
Hence, since $H_{LR}\in\mL^1(\fH)$ by Assumption \ref{ass:H} \ref{L1}, we get $\Phi\in\mL^1(\fH)$ as stated in Definition \ref{def:ep} \ref{ep-1}. Moreover, \eqref{Phi-2}, and \eqref{H2} for $H$ and \eqref{QGmm}, respectively, imply
\ba
\label{Phi-star}
\Phi^\ast
&=\Phi,\\
\label{GmPhiGm}
\Gamma\Phi\Gamma
&=-\Phi,
\ea 
\ie,  $\Phi$ satisfies \eqref{H1}-\eqref{H2} (and, without being  the generator of the time evolution of the system under consideration, could be called a selfdual observable). Next, we know (see \cite{Araki1971}) that, if $T\in\mL(\fH)$ is a 2-point operator, $\omega_T\in\mE_\fA$ a state with 2-point operator $T$, and $A\in\mL^1(\fH)$ with $\Gamma A\Gamma=-A^\ast$, we have
\ba
\label{expct-sd}
\omega_T(b(A))
=-\hspace{0.5mm}\tr(T A),
\ea
where $b(A)\in\fA$ is the selfdual second quantization of $A$ from Definition \ref{def:obs} \ref{def:obs-d}. 
Due to \eqref{Phi-star}-\eqref{GmPhiGm} and the fact that $ \dot\Psi^0=\Phi$ and $\dot\Psi^t=\e^{\ii t H}\Phi\e^{-\ii t H}$ for all $t\in\R$, where $\Psi$ stems from Remark \ref{rem:deriv} (and the dot stands for the derivative with respect to $t$), we have  $\dot\Psi^t\in\mL^1(\fH)$,  $(\dot\Psi^t)^\ast=\dot\Psi^t$, and  $\Gamma\dot\Psi^t\Gamma=-\dot\Psi^t$ for all $t\in\R$. Hence, \eqref{expct-sd} yields, for all $t\in\R$, 
\ba
\omega_T(b(\dot\Psi^t))
&=-\hspace{0.5mm}\tr(\e^{-\ii t H} T \e^{\ii t H} \Phi)\nonumber\\
&=\omega_T(b(\Phi)),
\ea
where we used the cyclicity of the trace, $[T,H]=0$ from Proposition \ref{prop:R/L2pt} \ref{prop:R/L2pt-b}, and the first part of the proof of Lemma \ref{lem:SpecId} \ref{lem:SpecId-d}. Therefore,  the R/L mover heat flux \eqref{def:J} is independent of the choice $t=0$ in \eqref{def:Phi}. 
\er

\br
Since $\omega_T(A^\ast)=\overline{\omega_T(A)}$ for all $A\in\mE_\fA$, \eqref{Phi-star} and the fact that $b(A)^\ast=b(A^\ast)$ for all $A\in\mL^1(\fH)$ from Remark \ref{rem:prop-b(A)} imply that $J\in\R$.
\er

\br
\label{rem:1stlaw}
Let us denote by $J_R$ the expectation value in the R/L mover state of the macroscopic heat flux observable $b(\Phi_R)$ whose 1-particle observable $\Phi_R$ describes the heat flux from the right reservoir into the sample, i.e., $\Phi_R$ is defined as in \eqref{def:Phi} but with $H_L$ replaced by $H_R$. Setting $K:=H-(H_L+H_R)$, we get $K\in\mL^1(\fH)$ due to \eqref{partsH} and Assumption \ref{ass:H} \ref{L1}. Moreover, since $\Phi+\Phi_R=\ii [H,K]/2$, \eqref{expct-sd} and Proposition \ref{prop:R/L2pt} \ref{prop:R/L2pt-b} yield
\ba
 \label{1stlaw}
J+J_R
&=-\frac\ii 2\hspace{0.2mm}\tr(T[H,K])\nonumber\\
&=-\frac\ii 2\hspace{0.2mm}\tr([T,H]K)\nonumber\\
&=0,
\ea
\ie, we obtain the first law of thermodynamics in the R/L mover state. Moreover, due to \eqref{1stlaw}, the definition of the entropy production rate from \eqref{ep-sigma} boils down to the usual one, \ie, we have $\sigma=-(\beta_L J+\beta_R J_R)$.
\er

In the following, we make use of Assumption \ref{ass:H} \ref{HLR=0} which means that there is no direct coupling between the two reservoirs, \ie, that the range of the Hamiltonian is bounded by the finite number $n_S$ of the sites in the configuration space $\Z_S$ of the confined sample. This assumption is physically meaningful since the coupling interaction of a real physical sample to a thermal reservoir usually acts by short-range forces across the boundaries of the sample (for a lattice spacing of the order of $10^{-10}m$ and a sample dimension of the order of $10^{-3}m$ [see \cite{SGOVR2001} for example], we get $n_S\sim 10^7$). Finally, recall that $\theta\in\mL(\fh)$ is the right translation from Definition \ref{def:iso} \ref{def:iso-a}. 

Under the additional Assumption \ref{ass:H} \ref{HLR=0}, the Hamiltonian can be written as follows. 

\bl[Finite range]
\label{lem:fnt}
Let $H\in\mL(\fH)$ be a Hamiltonian satisfying Assumption \ref{ass:H} \ref{HTheta}, \ref{HLR=0}, and \ref{nontrv} and let $u_0\in L^\infty(\T)$ and $u=[u_1,u_2,u_3]\in L^\infty(\T)^3$ be the Pauli coefficient functions of $\widehat H$. Then, there exists $\nu\in\l1,n_S\r$ such that the Pauli coefficients of $H$ read, for all $\alpha\in\l0,3\r$, 
\ba
\label{h-range}
h_\alpha
=\begin{cases}
\hfill -2\sum_{n\in\l1,\nu\r} \Im(\check u_\alpha(n))\hspace{0.5mm} \Im(\theta^n), & \alpha\in\l0,2\r,\\
\Re(\check u_3(0))1+2\sum_{n\in\l1,\nu\r}\Re(\check u_3(n))\hspace{0.5mm} \Re(\theta^n), & \alpha=3.
\end{cases}
\ea
The smallest number $\nu$ such that \eqref{h-range} holds is called the range of the Hamiltonian $H$. 
\el

\bprf
Recall that the Pauli coefficients $h_0\in\mL(\fh)$ and $h=[h_1,h_2,h_3]\in\mL(\fh)^3$ of the Hamiltonian $H=h_0\sigma_0+h\sigma$ satisfy \eqref{hsa}-\eqref{hac}. Therefore, since $Q_\kp=q_\kp\sigma_0$ for all $\kappa\in\{L,S,R\}$, Assumption \ref{ass:H} \ref{HLR=0} is equivalent to the fact that, for all $\alpha\in\l0,3\r $,
\ba
q_L h_\alpha q_R=0.
\ea
Using Assumption \ref{ass:H} \ref{HTheta}, which is equivalent to the fact that, for all $\alpha\in\l0,3\r $, 
\ba
\label{htheta}
[h_\alpha, \theta]
=0,
\ea
and using that $\delta_x=\theta^x\delta_0$ and $(\theta^x)^\ast=\theta^{-x}$ for all $x\in\Z$ (where we set $\theta^0:=1$ and $\theta^{-x}:=(\theta^{-1})^x$ for all $x\in\N$), we can write, for all $x\in\Z_L$, all $y\in\Z_R$, and all $\alpha\in\l0,3\r $, that $(\delta_{x-y}, h_\alpha\delta_0)
=(\delta_x,h_\alpha\delta_y)
=(q_L\delta_x,h_\alpha q_R\delta_y)
=(\delta_x,q_L h_\alpha q_R\delta_y)
=0$, 
and then, with \eqref{hsa}, that  $(\delta_{y-x}, h_\alpha\delta_0)=0$, too. Moreover, we have $x-y\ge n_S+1\ge 2$ for all  $x\in\Z_R$ and all $y\in\Z_L$, and any $z\in\Z$ with $z\ge n_S+1$ can be written as $z=x-y$ with $x\in\Z_R$ and $y\in\Z_L$ (and analogously if $z\le -(n_S+1)$). Hence, since, for all  $\alpha\in\l0,3\r $, the function $\check u_\alpha\in\fh$ is given, for all $x\in\Z$, by
\ba
\check u_\alpha(x)
&=(\e_x, m[u_\alpha]\e_0)\nonumber\\
&=(\delta_x,h_\alpha\delta_0), 
\ea
we get $\check u_\alpha(x)=0$ for all $\alpha\in\l0,3\r $ and all $x\in\Z$ with $|x|\ge n_S+1$. Therefore, there exists a smallest number $\nu\in\l1,n_S\r$ such that, for all $\alpha\in\l0,3\r $ and all $x\in\Z$ with $|x|\ge\nu+1$, 
\ba
\label{hrange}
\check u_\alpha(x)
=0,
\ea
and $\nu=0$ is excluded due to Assumption \ref{ass:H} \ref{nontrv}. Hence, \eqref{hrange} implies that, for all $\alpha\in\l0,3\r $ and all $y\in\Z$, we have  $h_\alpha\delta_y
=\theta^yh_\alpha\delta_0
=\theta^y\sum_{x\in\l-\nu,\nu\r} \check u_\alpha(x)\delta_x
=\sum_{x\in\l-\nu,\nu\r}\check u_\alpha(x)\theta^x\delta_y$,
\ie, we get, for all $\alpha\in\l0,3\r $, 
\ba
\label{hfr1}
h_\alpha
&=\sum_{x\in\l-\nu,\nu\r}\check u_\alpha(x)\hspace{0.2mm}\theta^x\\
\label{hfr2}
&=\check u_\alpha(0)1+2\sum_{n\in\l1,\nu\r}\Re(\check u_\alpha(n) \theta^n)\\
\label{hfr3}
&=\begin{cases}
\hfill -2\sum_{n\in\l1,\nu\r} \Im(\check u_\alpha(n))\hspace{0.5mm} \Im(\theta^n), & \alpha\in\l0,2\r,\\
\Re(\check u_3(0))1+2\sum_{n\in\l1,\nu\r}\Re(\check u_3(n))\hspace{0.5mm} \Re(\theta^n), & \alpha=3,
\end{cases}
\ea
where we used \eqref{hsa} for \eqref{hfr2} and \eqref{hac} for \eqref{hfr3}. 
\eprf

\br
\label{rem:onehalf}
Let $A\in\mL^1(\fH)$ with $\Gamma A\Gamma=-A^\ast$. Then, for all $B\in\fA$, we define the map $f_B:\R\to\fA$ by $f_B^t:=\e^{\ii t b(A)/2}B\e^{-\ii t b(A)/2}$ for all $t\in\R$, where the exponential is defined through its absolutely convergent series with respect to the $C^\ast$-norm of $\fA$. Hence, the Cauchy product in $\fA$ yields (as in Remark \ref{rem:deriv}) that, for all $s\in\R\setminus\{0\}$,  
\ba
\frac{f_B^{t+s}-f_B^t}{s}-f_{\frac\ii 2[b(A),B]}^t
&=\e^{\ii t\frac 12 b(A)}\bigg[\frac{\e^{\ii s\frac 12 b(A)}-1}{s}-\frac\ii 2b(A),B\bigg]\e^{-\ii (t+s)\frac 12 b(A)}\nonumber\\
&\quad+\e^{\ii t\frac 12 b(A)}  \frac\ii 2[b(A),B]\big(\e^{-\ii s\frac 12 b(A)}-1\big)\e^{-\ii t\frac 12 b(A)}.
\ea
Since, due to \eqref{norm-b(A)}, we again have $\|(\e^{\ii s b(A)/2}-1)/s-\ii b(A)/2\|\le \|A\|_1(\e^{|s|\|A\|_1}-1)$, the map $f_B$ is differentiable everywhere on $\R$ (the dot again stands for the derivative with respect to $t$) and, for all $t\in\R$, we get
\ba
\label{fB-drv}
\dot f_B^t
=f_{\frac\ii 2[b(A),B]}^t.
\ea
Hence, $f_B$ is infinitely differentiable on $\R$,  and since the $n$-th derivative of $f_B$ at the point $s\in\R$ is bounded by $\|B\|\e^{|s|\|A\|_1} \|A\|_1^n$ for all $n\in\N$, Taylor's theorem for $\fA$ implies that $f_B$ is real analytic on $\R$. Moreover, since Definition \ref{def:obs} \ref{def:obs-e} and \ref{def:obs-d} 
imply that $[b(A),B(F)]=2B(AF)$ for all $F\in\fH$ and all $A\in\mL^1(\fH)$ with $\Gamma A\Gamma=-A^\ast$, the Taylor series for $f_{B(F)}$ in $\fA$ and for the map $\R\ni t\mapsto \e^{\ii t A}F\in\fH$ yield, for all $t\in\R$, all $F\in\fH$, and all $A\in\mL^1(\fH)$ with $\Gamma A\Gamma=-A^\ast$, 
\ba
\label{prp-B(F)}
\e^{\ii t\frac 12 b(A)}B(F)\e^{-\ii t \frac 12 b(A)}
=B(\e^{\ii t A}F).
\ea
Similarly, since  $[b(A),b(B)]=2b([A,B])$ for all $A,B\in\mL^1(\fH)$ with $\Gamma A\Gamma=-A^\ast$ and  $\Gamma B\Gamma=-B^\ast$,  the Taylor series for $f_{b(B)}$ in $\fA$ and for the map $\R\ni t\mapsto \e^{\ii t A}B\e^{-\ii t A}\in \mL^1(\fH)$ yield, for all $t\in\R$ and all $A,B\in\mL^1(\fH)$ with $\Gamma A\Gamma=-A^\ast$ and  $\Gamma B\Gamma=-B^\ast$,
\ba
\label{prp-b(A)}
\e^{\ii t\frac 12 b(A)}b(B)\e^{-\ii t \frac 12 b(A)}
=b(\e^{\ii t A}B\e^{-\ii t A}).
\ea
Next, for all $N\in\N$, let us define $q_N:=m[1_{\l-N,N\r}]$ and $Q_N:=q_N\sigma_0\in\mL^0(\fH)$, and set $H_N:=Q_NHQ_N$ for all $N\in\N$. Moreover, for all $\kp,\lm\in\{L,S,R\}$ and all $N\in\N$, we set $H_{\kp,N}:=Q_NH_\kp Q_N$ and  $H_{\kp\lm,N}:=Q_NH_{\kp\lm} Q_N$. Using \eqref{prod}, we note that all the Pauli coefficients of $H_L H_{LS}$ are linear combinations of operators of the form $q_Lh_\alpha q_L h_\beta q_S$, where $\alpha,\beta\in\l0,3\r$ (see also \eqref{s0}-\eqref{s} below). With the help of \eqref{hfr1}, we can write $q_Lh_\alpha q_L h_\beta q_S
=\sum_{x,y\in\l-\nu,\nu\r} \check u_\alpha(x)\check u_\beta(y) q_L (\theta^x q_L) m[1_{\l x_L+x+y,x_R+x+y\r}]\theta^{x+y}$ for all $\alpha,\beta\in\l0,3\r$. Similarly,  we get $q_N(q_Lh_\alpha q_L)q_N (q_L h_\beta q_S) q_N
=\sum_{x,y\in\l -\nu,\nu\r} \check u_\alpha(x)\check u_\beta(y) q_L (\theta^x q_L) m[1_{N,x,y}] m[1_{\l x_L+x+y,x_R+x+y\r}]\theta^{x+y}$ for all $\alpha,\beta\in\l0,3\r$ and all $N\in\N$, where, for all $N\in\N$ and all $x,y\in\l-\nu,\nu\r $, the function $1_{N, x,y}\in\ell^\infty(\Z)$ is defined by $1_{N,x,y}:=1_{\l -N,N\r}1_{\l-N+x,N+x\r}1_{\l-N+x+y,N+x+y\r}$. If we assume that
\ba
\label{N-lbound}
N
\ge |x_L|+|x_R|+2\nu,
\ea
we have $1_{N,x,y}1_{\l x_L+x+y,x_R+x+y\r}=1_{\l x_L+x+y,x_R+x+y\r}$ for all $x,y\in\l-\nu,\nu\r $.  Therefore,  we can write $H_{L,N}H_{LS,N}=H_L H_{LS}$, \ie, using \eqref{Phi-3}, we get, for all $N\in\N$ satisfying \eqref{N-lbound}, 
\ba
\label{HNH}
[H_N,H_{L,N}]
&=-2\ii \hspace{0.2mm}\Im(H_{L,N}H_{LS,N})\nonumber\\
&=-2\ii \hspace{0.2mm}\Im(H_L H_{LS})\nonumber\\
&=[H,H_L],
\ea
where, in the first equality, we used the fact that $[Q_\kp, Q_N]=0$ for all $\kp\in\{L,S,R\}$ and all $N\in\N$ (and Assumption \ref{ass:H} \ref{HLR=0}). Now, due to \eqref{prp-B(F)}, we note that the quasifree dynamics generated by the local Hamiltonian  $H_N\in\mL^0(\fH)$ on the 1-particle Hilbert space $\fH$ is induced, macroscopically, by the selfdual second quantization of $H_N/2$ and the local  macroscopic Hamiltonian of the left reservoir is given by $b(H_{L,N})/2$. Hence, using \eqref{fB-drv}, the commutator identity after \eqref{prp-B(F)}, \eqref{HNH}, and \eqref{Phi-2}, we get, for all $N\in\N$ satisfying \eqref{N-lbound}, 
\ba
\label{flx-1/2}
-\frac{\rd}{\rd t}\bigg|_{t=0}\e^{\ii t\frac 12 b(H_N)}\hspace{0.2mm}\frac12\hspace{0.5mm} b(H_{L,N})\hspace{0.5mm}\e^{-\ii t \frac 12 b(H_N)}
&=b\Big(\hspace{-1mm}-\frac\ii 2[H_N,H_{L,N}]\Big)\nonumber\\
&=b(\Phi), 
\ea
\ie, the fact that macroscopic dynamics is generated by the selfdual second quantization of $H_N/2$ explains the existence of the factor $1/2$ in \eqref{def:Phi}.
\er

\br
Let $\omega_T\in\mE_\fA$ be a gauge-invariant state with 2-point operator $T\in\mL(\fH)$ and let $H\in\mL(\fH)$ be a gauge-invariant Hamiltonian. Then, due to Lemma \ref{lem:gauge} \ref{lem:gauge-b}  (and its proof), there exist $s,h\in\mL(\fh)$ with $0\le s\le 1$ and $h^\ast=h$ such that $T=(1-s)\oplus \zeta  s\zeta$ and $H=h\oplus  (-\zeta h\zeta)$. Hence, the heat flux observable \eqref{def:Phi} has the form
\ba
\Phi
=\frac12\hspace{1mm}(\vi\oplus (-\zeta \vi\zeta)),
\ea 
where $\vi\in\mL(\fh)$ is given by $\vi:=-\ii [h, q_Lhq_L]$. Due to Assumption \ref{ass:H} \ref{HLR=0}, we then have $\vi=-\ii [q_Lhq_S+q_Shq_L,q_Lhq_L]\in\mL^0(\fh)$ and, hence, $\Phi\in\mL^0(\fH)$. Moreover, $\Phi^\ast=\Phi$ and $\Gamma\Phi\Gamma=-\Phi$ which implies, with \eqref{expct-sd}, that 
\ba
\label{flow-gauge}
\omega_T(b(\Phi))
&=\frac12 \hspace{0.5mm}\tr(s\vi)+\frac12 \hspace{0.5mm}\tr(\zeta s\vi\zeta)-\frac12 \hspace{0.5mm}\tr(\vi)\nonumber\\
&=\tr(s\vi), 
\ea
where we used that $\zeta\delta_x=\delta_x$ for all $x\in\Z$,  the commutator form of $\vi$, and the cyclicity of the trace on $\fh$. Note that $\omega_s({\rm d}\Gamma(\vi))=\tr(s\vi)$, where $\omega_s\in\mE_\fA$ is the state defined in Remark \ref{rem:state-gauge} and ${\rm d}\Gamma$ is the usual second quantization.
\er

We next clarify the effect of Assumption \ref{ass:H} \ref{HLR=0} on the assumptions used in the foregoing sections.

\bl[Assumptions]
\label{lem:ass}
Let $H\in\mL(\fH)$ be a Hamiltonian satisfying Assumption \ref{ass:H} \ref{HTheta}, \ref{HLR=0}, and \ref{nontrv}. Moreover, let $u_0\in L^\infty(\T)$ and $u=[u_1,u_2,u_3]\in L^\infty(\T)^3$ be the Pauli coefficient functions of $\widehat H$ and let us define the following mutually exclusive, exhaustive, and non-empty cases:
\ba
\label{ass-cases}
\mbox{Case }
\begin{cases}
\mbox{1}, & \mbox{$u_0=0$, $u\neq 0$, and $uu'=0$}\\
\mbox{2}, & \mbox{$u_0=0$ and $uu'\neq 0$}\\
\mbox{3}, & \mbox{$u_0\neq 0$ and $u=0$}\\
\mbox{4}, & \mbox{$u_0\neq 0$, $u\neq 0$, and $uu'=0$}\\
\mbox{5}, & \mbox{$u_0\neq 0$, $uu'\neq 0$, and $u_0^2\neq u^2$}\\
\mbox{6}, & \mbox{$u_0\neq 0$ and $u_0^2=u^2$}
\end{cases}
\ea
Then:
\bn[label=(\alph*),ref={\it (\alph*)}]
\setlength{\itemsep}{0mm}
\item
\label{lem:ass-b}
In all cases, Assumptions \ref{ass:H} \ref{L1}, \ref{ass:reg} \ref{ass:reg-1} and \ref{ass:reg-3} are satisfied.

\item
\label{lem:ass-a}
We have 
\ba
[1_{pp}(H), 1_{ac}(H), 1_{sc}(H)]
=\begin{cases}
\hfill [1,0,0], & \mbox{Case 1},\\
\hfill [0,1,0], & \mbox{Case 2, 3, 4, and 5},\\
\hfill [1_0(H), 1-1_0(H),0],& \mbox{Case 6},
\end{cases}
\ea
where, in Case 6,  $\dim(\ran(1_0(H)))=\infty$ but $1_0(H)\neq 1$.  In particular, in all cases, Assumption \ref{ass:H} \ref{1sc=0} is satisfied. 

\item
\label{lem:ass-c}
The bounded extension $\bar V\in\mL(\widehat\fH)$ of the asymptotic velocity with respect to $\widehat H$ satisfies
\ba
\label{eigV0}
\eig(\bar V)\cap\{0\}
=\begin{cases}
\hfill\{0\}, & \mbox{Case 1 and 6},\\
\hfill\emptyset, & \mbox{Case 2, 3, 4, and 5}.
\end{cases}
\ea
In particular, in Case 2, 3, 4, and 5, Assumption \ref{ass:eig} \ref{ass:eig-1} holds.
\en
\el

\br
If Assumption \ref{ass:H} \ref{nontrv} does not hold, Remark \ref{rem:nontrv} yields $H=0$, \ie, we have $u_\alpha=0$ for all $\alpha\in\l0,3\r$. Hence, Assumptions \ref{ass:H} \ref{L1}, \ref{ass:reg} \ref{ass:reg-1} and \ref{ass:reg-3} are satisfied. Moreover, since $1_{pp}(H)=1$ and since $\{1_{pp}(H), 1_{ac}(H), 1_{sc}(H)\}$ is a complete orthogonal family of orthogonal projections, Assumption \ref{ass:H} \ref{1sc=0} is also satisfied. Finally, due to \eqref{Vas2}-\eqref{Vas3} (or directly from \eqref{def:assvel-0}), we have $\bar V=0$, \ie, Assumption \ref{ass:eig} \ref{ass:eig-1} does not hold.
\er

\br
Since $u_0u_0'=uu'$ if $u_0^2=u^2$, the first and the third condition of {\it Case 4} imply $u_0^2\neq u^2$ (see  {\it Case 3} in the proof of Lemma \ref{lem:ass} \ref{lem:ass-a}). Hence, the six cases are mutually exclusive and exhaust all the possibilities. 
\er

In the following, we denote by $TP(\T)$ the real trigonometric polynomials on $\T$ (for the structure of this ring, see \cite{Picavet2003} for example). Note that, due to the fundamental theorem of algebra, we have, for all $v\in TP(\T)$,
\ba
\label{ZvTP}
\mbox{$\card(\mZ_v)<\infty$ if and only if $v\neq 0$.}
\ea

\vspace{5mm}

\bprf
Due to Assumption \ref{ass:H} \ref{HTheta}, \ref{HLR=0}, and \ref{nontrv} and Lemma \ref{lem:fnt}, the Pauli coefficients of $\widehat H$ have the form $\hat h_\alpha=m[u_\alpha]$ for all $\alpha\in\l0,3\r$, where the Pauli coefficient functions $u_0\in L^\infty(\T)$ and $u=[u_1,u_2,u_3]\in L^\infty(\T)^3$ are given, for all $\alpha\in\l0,3\r$, by
\ba
\label{Pauli-cf}
u_\alpha
=\begin{cases}
\hfill -2\sum_{n\in\l1,\nu\r }c_{\alpha,n}\sin(n\hspace{0.5mm}\cdot), & \alpha\in\l0,2\r ,\\
c_{3,0}+2\sum_{n\in\l1,\nu\r } c_{3,n}\cos(n\hspace{0.5mm}\cdot), & \alpha=3,
\end{cases}
\ea
\ie, we have $u_\alpha\in TP(\T)$ for all $\alpha\in\l0,3\r$. Here, for all $\alpha\in\l 0,2\r$ and all $x\in\Z$, we set
\ba
\label{c-an}
c_{\alpha,x}
&:=\Im(\check u_\alpha(x)),\\
\label{c-3n}
c_{3,x}
&:=\Re(\check u_3(x)),
\ea
and we note that, due to \eqref{Evu}-\eqref{Oddu}, we have $\check u_\alpha(x)=\ii c_{\alpha,x}$ and $\check u_3(x)=c_{3,x}$ for all $\alpha\in\l0,2\r$ and all $x\in\Z$, respectively.

\vspace{1mm}
\ref{lem:ass-b}\,
In all cases, Assumption \ref{ass:H} \ref{HLR=0} implies Assumption \ref{ass:H} \ref{L1}. Moreover, since $u_\alpha\in TP(\T)$ for all $\alpha\in\l0,3\r$, Assumptions \ref{ass:reg} \ref{ass:reg-1} and \ref{ass:reg-3} are satisfied.

\vspace{1mm}
\ref{lem:ass-a}\,
Using part \ref{lem:ass-b}, \eqref{rem:spec-1} from Remark \ref{rem:spec} yields
\ba
\label{specH}
\spec(\widehat H)
=\ran(e_+)\cup \ran(e_-).
\ea
Moreover, for all  $\lm\in\R$, we define $p_\lm\in TP(\T)$ by $p_\lm :=\det([\widehat H]-\lm 1)$, where $[\widehat H]\in L^\infty(\T,\C^{2\times 2})$ is the identification specified in Remark \ref{rem:spec}. Hence, for all  $\lm\in\R$, we have
\ba
\label{plm}
p_\lm
=(u_0-\lm)^2-u^2,
\ea
and we know that $\lm\in\eig(\widehat H)$ if and only if $|\mZ_{p_\lm}|>0$ (see \cite{HW1996} for example), \ie, with \eqref{ZvTP}, 
\ba
\label{Mlm>0}
\mbox{$\lm\in\eig(\widehat H)$ if and only if $p_\lm=0$}.
\ea
Furthermore, due to Lemma \ref{lem:sqr}, we can write  
\ba
\label{u2-TP}
u^2
=a_0+2\sum_{m\in\l1,2\nu\r } a_m\cos(m\hspace{0.5mm}\cdot),
\ea
where we set $a_n:=\sum_{\alpha\in\l1,3\r}a_{\alpha,n}$ for all $n\in\l0,2\nu\r$ and the coefficients $a_{\alpha,n}\in\R$ for all $\alpha\in\l0,3\r$ and all $n\in\l0,2\nu\r$ are given in Lemma \ref{lem:sqr}.

\vspace{1mm}
{\it Case 1}\quad
Since $uu'=(u^2)'/2$ and since we know that $\{1, 2\sin(n\hspace{0.5mm}\cdot), 2\cos(n\hspace{0.5mm}\cdot)\}_{n\in\N}$  constitutes an orthonormal basis of $\hat\fh$, 
\eqref{u2-TP} implies that $u^2=a_0$, and $a_0>0$ because $u\neq 0$. Hence, we get $e_\pm=\pm\sqrt{a_0}$ and, due to \eqref{specH},
\ba
\spec(\widehat H)
=\{-\sqrt{a_0},\sqrt{a_0}\}.
\ea
Moreover, since the points $\pm\sqrt{a_0}$ are isolated in $\spec(\widehat H)$, we know that $\pm\sqrt{a_0}\in\eig(\widehat H)$
(or by directly using \eqref{Mlm>0}) and, hence, we get $\spec(\widehat H)=\eig(\widehat H)$.
Therefore, since $E_{\widehat H}(1_{\eig(\widehat H)})=1_{pp}(\widehat H)$, since  $E_{\widehat H}(1_{\spec(\widehat H)})=1$, and since $\{1_{pp}(\widehat H), 1_{ac}(\widehat H), 1_{sc}(\widehat H)\}$ is a complete orthogonal family of orthogonal projections, we get $1_{pp}(\widehat H)=1$ and, hence, $1_{ac}(\widehat H)=1_{sc}(\widehat H)=0$. 
Since $1_\mu(\widehat H)=\fF 1_\mu(H)\fF^\ast$ for all $\mu\in\{pp, ac, sc\}$ (which follows from Lemma \ref{lem:unitary}), 
we arrive at $1_{pp}(H)=1$ and $1_{ac}(H)=1_{sc}(H)=0$.

\vspace{0.5mm}
{\it Case 2}\quad
Since $u\neq 0$ due to $uu'\neq0$, \eqref{ZvTP} and \eqref{Zu-2} imply $\card(\mZ_u)<\infty$. Moreover, since $e_\pm'=\pm uu'/|u|$ on $\mZ_u^c$, we get $\mZ_\pm=\{k\in\mZ_u^c\,|\, (uu')(k)=0\}$ and, since $uu'\in TP(\T)$, \eqref{ZvTP} leads to $\card(\mZ_\pm)<\infty$. Therefore, Assumption \ref{ass:nullsets} \ref{ass:reg-4} and \ref{ass:reg-5} are satisfied for $M=\spec(\widehat H)\in\mM(\R)$ and Proposition \ref{prop:ac} and Remark \ref{rem:spec} yield $1_{ac}(\widehat H)=1$. Hence, we get $1_{pp}(\widehat H)=1_{sc}(\widehat H)=0$.

\vspace{0.5mm}
{\it Case 3}\quad
We have $\mZ_u\cap e_\pm^{-1}(M)=u_0^{-1}(M)$ for all $M\in\mM(\R)$. Hence, there exists no $M\in\mM(\R)$ with $\spec(\widehat H)=\ran(u_0)\subseteq M$ such that Assumption \ref{ass:nullsets} \ref{ass:reg-4} holds (compare with Remark \ref{rem:spec}). But note that $\ran(1_{ac}(\widehat H))=\ran(1_{ac}(m[u_0])\oplus 1_{ac}(m[u_0]))$ 
and, hence, $1_{ac}(\widehat H)=1_{ac}(m[u_0])\oplus 1_{ac}(m[u_0])$. 
Since $u_0\in TP(\T)$ has the form \eqref{Pauli-cf}, we also have $u_0'\neq 0$, and \eqref{ZvTP} yields $\card(\mZ_{u_0'})<\infty$. Hence, for the scalar multiplication operator $m[u_0]\in\mL(\hat\fh)$, we know that $1_{ac}(m[u_0])=1$
(we can also readily adapt the proof of Proposition \ref{prop:ac} by replacing \eqref{dec-1} by $e_\pm^{-1}(A')=u_0^{-1}(A')=(u_0^{-1}(A')\cap\mZ_{u_0'})\cup (u_0^{-1}(A')\cap\mZ_{u_0'}^c)$ and by carrying out the further decompositions of $u_0^{-1}(A')\cap\mZ_{u_0'}^c$ analogously). Hence, we get $1_{ac}(\widehat H)=1$ and $1_{pp}(\widehat H)=1_{sc}(\widehat H)=0$.

\vspace{0.5mm}
{\it Case 4}\quad
Since $u^2\in TP(\T)$ and $u^2\neq 0$, \eqref{ZvTP} yields $\card(\mZ_u)<\infty$. Moreover, we have $e_\pm'=u_0'$ on $\mZ_u^c$ and, as in Case 3,  $u_0'\in TP(\T)$ satisfies $u_0'\neq 0$. Hence, \eqref{ZvTP} implies $\card(\mZ_\pm)<\infty$. Therefore, Assumption \ref{ass:nullsets} \ref{ass:reg-4} and \ref{ass:reg-5} are satisfied for $M=\spec(\widehat H)\in\mM(\R)$ and Proposition \ref{prop:ac} and Remark \ref{rem:spec} yield $1_{ac}(\widehat H)=1$. Hence, we get $1_{pp}(\widehat H)=1_{sc}(\widehat H)=0$. 

\vspace{0.5mm}
{\it Case 5}\quad
As in {\it Case 4},  we have  $\card(\mZ_u)<\infty$. We next want to show that $\card(\mZ_\pm)<\infty$. To this end, we make the following three steps which require $u_0\neq 0$ and $u\neq0$ to hold only.
First, since $\card(\mZ_u)<\infty$, there exists $N\in\N$ and $\{a_i, b_i\}_{i\in\l1,N\r}\subseteq\R$ with $a_i<b_i$ for all $i\in\l1,N\r$ such that $(a_i,b_i)\cap (a_j,b_j)=\emptyset$ for all $i,j\in\l1,N\r$ with $i\neq j$ and $\mZ_u^c=\bigcup_{i\in\l1,N\r}(a_i,b_i)$ if $u^2(\pi)=0$ and $\mZ_u^c=(\bigcup_{i\in\l1,N\r}(a_i,b_i))\cup\{-\pi,\pi\}$ if $u^2(\pi)\neq0$. Let $a,b\in\R$ with $a<b$, let $I:=(a,b)\subseteq (a_i,b_i)$ for some $i\in\l1,N\r$, and let $\kp\in\{\pm\}$ be fixed. 
The function $e_\kp$ is differentiable on $I$ and if $e'_\kp=0$ on $I$, there exists $\lm\in\R$ such that 
$p_\lm=0$ on $I$ (see \eqref{plm}). Hence, due to \eqref{ZvTP}, we get $p_\lm=\lm^2-2\lm u_0+u_0^2-u^2=0$ on $\T$. Using \eqref{Evu}-\eqref{Oddu}, we thus find $\lm u_0=0$ on $\T$ which implies $\lm=0$ since  $u_0\neq 0$. Hence, we arrive at $p_0=0$ (on $\T$).
Second, let $p\in TP(\T)$ be defined by 
\ba
\label{def:poly-1}
p
:=u_0'^2u^2-(uu')^2,
\ea
 and let us show that, in general, $p=0$ if and only if  $p_0=0$. If $p_0=0$, we have $p_0'=2(u_0u_0'-uu')=0$ which implies $p=-u_0'^2 p_0=0$. Conversely, if $p=0$, there exists a function $\sigma: \mZ_u^c\to\{-1,1\}$ such that $u_0'=\sigma uu'/|u|$ on $\mZ_u^c$ and, 
in particular,  $u_0'=\sigma uu'/|u|$ on $(a_i,b_i)$ for all $i\in\l1,N\r$. Moreover, for all $i\in\l1,N\r$, there exists $k_i\in (a_i,b_i)$ such that $u_0'(k_i)\neq 0$ (assuming the opposite contradicts $u_0\neq 0$) and, hence, $(uu')(k_i)\neq 0$. Let $i\in\l1,N\r$ be fixed and let $u_0'(k_i)>0$ (the case $u_0'(k_i)<0$ is completely analogous). Since $u_0'$ is continuous on $(a_i,b_i)$, there exists $\veps>0$ such that $(k_i-\veps, k_i+\veps)\subseteq (a_i,b_i)$ and $u_0'(k)>0$ for all $k\in(k_i-\veps, k_i+\veps)$. Since $uu'$ is continuous on $(a_i,b_i)$, too, if $(uu')(k_i)>0$, there exists $\veps'>0$ such that $(k_i-\veps', k_i+\veps')\subseteq (a_i,b_i)$ and $(uu)'(k)>0$ for all $k\in(k_i-\veps', k_i+\veps')$. Setting $\delta:=\min\{\veps,\veps'\}$, we get $\sigma=1$ on $I_i:=(k_i-\delta, k_i+\delta)$ and, hence, $e_-'=0$ on $I_i$. Then, the first step yields $p_0=0$ (if $(uu')(k_i)<0$, we get $e_+'=0$ on $I_i$ and again $p_0=0$ from the first step).
 Third, let $M\subseteq\mZ_u^c$ such that $\card(M)=\infty$ (not finite) and let $\kp\in\{\pm\}$ be fixed. If $e_\kp'=0$ on $M$, we also have $p=0$ on $M$ since $p=u^2e_+'e_-'$ on $\mZ_u^c$. Since $p\in TP(\T)$ and since $\card(M)=\infty$, \eqref{ZvTP} implies that $p=0$. It then  follows from the second step that $p_0=0$.  

Suppose now that $\card(\mZ_\pm)=\infty$. Then, it follows from the third step that $p_0=0$ which contradicts $u_0^2\neq u^2$. Hence, we have  $\card(\mZ_\pm)<\infty$ and, as in {\it Case 4}, Assumption \ref{ass:nullsets} \ref{ass:reg-4} and \ref{ass:reg-5} are satisfied for $M=\spec(\widehat H)\in\mM(\R)$ and Proposition \ref{prop:ac} and Remark \ref{rem:spec} yield $1_{ac}(\widehat H)=1$ and, thus, $1_{pp}(\widehat H)=1_{sc}(\widehat H)=0$. 

\vspace{0.5mm}
{\it Case 6}\quad
Since $p_0=0$, \eqref{Mlm>0} yields $0\in\eig(\widehat H)$. On the other hand, we know from the end of the first step in {\it Case 5} that $p_\lm=0$ implies $\lm=0$ if $u_0\neq 0$, \ie, $\eig(\widehat H)\subseteq\{0\}$ due to \eqref{Mlm>0}. Hence, we get $\eig(\widehat H)=\{0\}$ and $1_{pp}(\widehat H)=E_{\widehat H}(1_{\eig(\widehat H)})=1_0(\widehat H)$. 
Moreover, we know that $\ran(1_0(\widehat H))$ is infinite dimensional (see \cite{HW1996} for example).
But, due to \eqref{specH}, we have $|\spec(\widehat H)|>0$ because \eqref{Evu}-\eqref{Oddu} and $u_0\neq 0$ imply that $e_\pm$ are non constant, 
\ie,  we also get $1_0(\widehat H)\neq 1$.
We next want to apply Proposition \ref{prop:ac} for $M=\spec(\widehat H)\setminus\{0\}\in\mM(\R)$. First, as in {\it Case 4},  we have $\card(\mZ_u)<\infty$ and, hence, Assumption \ref{ass:nullsets} \ref{ass:reg-4} is satisfied. In order to verify Assumption \ref{ass:nullsets} \ref{ass:reg-5}, we write $\mZ_\pm\cap e_\pm^{-1}(M)=\{k\in\mZ_u^c\,|\,\mbox{$e_\pm(k)\neq 0$ and $e_\pm'(k)= 0$}\}$. Since $u_0^2=u^2$ implies $u_0u_0'=uu'$, we get $e_\pm'=\pm u_0'e_\pm/|u|$ on $\mZ_u^c$ and, hence, 
$\mZ_\pm\cap e_\pm^{-1}(M)\subseteq\mZ_u^c\cap\mZ_{u_0'}$.
Since $u_0\neq 0$, we have, as in {\it Case 3}, that 
$\card(\mZ_{u_0'})<\infty$, \ie, Assumption \ref{ass:nullsets} \ref{ass:reg-5} is satisfied. Therefore, Proposition \ref{prop:ac} yields $\ran(1_M(\widehat H))\subseteq\ran(1_{ac}(\widehat H))$, \ie,  we have $1_M(\widehat H)=1_{ac}(\widehat H)1_M(\widehat H)$. On the other hand, we also have $1=E_{\widehat H}(1_{\spec(\widehat H)})=1_{pp}(\widehat H)+1_M(\widehat H)$ since $\spec(\widehat H)=\{0\}\cup M$. Hence, we get $1_M(\widehat H)=1_{ac}(\widehat H)(1-1_{pp}(\widehat H))=1_{ac}(\widehat H)$ since $1_{ac}(\widehat H)1_{pp}(\widehat H)=0$, and we arrive at $1_{sc}(\widehat H)=1-(1_{pp}(\widehat H)+1_{ac}(\widehat H))=0$.

\vspace{1mm}
\ref{lem:ass-c}\,
As above (see \eqref{Mlm>0}), we use that $0\in\eig(\bar V)$ if and only if $|\mZ_{\vi_0}|>0$, 
where $\vi_0\in L^\infty(\T)$ is defined by $\vi_0:=\det([\bar V])=v_0^2-v^2$ and the Pauli coefficient functions are given by $v_0= u_0'$ and $v=(\tilde uu')\tilde u$ on $\mZ_u^c$ and $v=u'$ on $\mZ_u$. Hence, we can write $\mZ_{\vi_0}=A\cup B$, where $A,B\in\mM(\R)$ are defined by $A:=\{k\in\mZ_u^c\,|\, p(k)=0\}$ and $B:=\{k\in\mZ_u\,|\, q(k)=0\}$ and where $p$ stems from \eqref{def:poly-1} and  $q\in TP(\T)$ is given by 
\ba
\label{trigpoly-3}
q
:=u_0'^2-u'^2.
\ea

{\it Case 1}\quad
Since $u^2=a_0>0$ as in {\it Case 1} of part \ref{lem:ass-b}, we have  $\mZ_u=\emptyset$. Hence, we get $v_0=0$, and $v=0$ on $\mZ_u^c=\T$, \ie, $\bar V=0$ and $0\in\eig(\bar V)$.

\vspace{0.5mm}
{\it Case 2}\quad
Since $p\in TP(\T)$ and $p=-(uu')^2\neq 0$, \eqref{ZvTP} yields $\card(\mZ_p)<\infty$ and, hence,  $\card(A)<\infty$. Moreover, since $\card(B)<\infty$ because $u\neq 0$, we get $\card(\mZ_{\vi_0})<\infty$,  \ie, $0\notin\eig(\bar V)$.

\vspace{0.5mm}
{\it Case 3}\quad
Since $u=0$, we have $A=\emptyset$ and $q=u_0'^2$. Hence, we get $\card(B)=\card(\mZ_q)=\card(\mZ_{u_0'})$ and, as in {\it Case 3} of part \ref{lem:ass-b}, we have $\card(\mZ_{u_0'})<\infty$. This implies $0\notin\eig(\bar V)$.

\vspace{0.5mm}
{\it Case 4}\quad
We have $p=u_0'^2u^2\neq 0$ and, hence, $\card(A)\le\card(\mZ_p)<\infty$. Since $\card(B)\le\card(\mZ_u)<\infty$, we get $0\notin\eig(\bar V)$.

\vspace{0.5mm}
{\it Case 5}\quad
Due to the second step in {\it Case 5} of part \ref{lem:ass-b}, we have $p=0$ if and only if $p_0=u_0^2-u^2=0$. Hence, $p\neq 0$ and  $\card(A)\le\card(\mZ_p)<\infty$. Since again $\card(B)\le\card(\mZ_u)<\infty$, we get $0\notin\eig(\bar V)$.

\vspace{0.5mm}
{\it Case 6}\quad
Since $p_0=0$, we have $p=0$ and $|A|=|\mZ_u^c|>0$ since $u\neq 0$, \ie, we get $0\in\eig(\bar V)$.

Hence, for the bounded extension $\bar V\in\mL(\widehat\fH)$ of the asymptotic velocity with respect to $\widehat H$ to satisfy Assumption \ref{ass:eig} \ref{ass:eig-1} is equivalent for the Pauli coefficient functions $u_0\in L^\infty(\T)$ and $u\in L^\infty(\T)^3$  of $\widehat H$ to belong to {\it Case 2, 3, 4 or 5}, \ie, due to part \ref{lem:ass-a}, to the absence of eigenvalues of $H$ (compare also directly with \eqref{def:assvel-0}).
\eprf

For the following, recall from Remark \ref{rem:rA} that, for all $\chi\in\mB(\R)$ and all $r\in\R$, the function $\chi_r\in\mB(\R)$ is given by $\chi_r(x)=\chi(rx)$ for all $r, x\in\R$. In particular, in the following theorem, if $\rho\in\mB(\R)$ is a Fermi function and $\beta_\kappa\in\R$ is the reservoir temperature of reservoir $\kappa\in\{L,R\}$, we have, for all $x\in\R$,
\ba
\rho_{\beta_\kappa}(x)
=\rho({\beta_\kappa}x).
\ea

\vspace{5mm}

We now arrive at our main result. It asserts that the R/L mover heat flux has the following properties. 

\bt[R/L mover heat flux]
\label{thm:hf}
Let $H\in\mL(\fH)$ be a Hamiltonian satisfying Assumption \ref{ass:H} \ref{HTheta}, \ref{HLR=0}, and \ref{nontrv} and let the bounded extension $\bar V\in\mL(\widehat\fH)$ of the asymptotic velocity with respect to $\widehat H$ satisfy Assumption \ref{ass:eig} \ref{ass:eig-1}. Moreover, let $u_0\in L^\infty(\T)$ and $u\in L^\infty(\T)^3$ be the Pauli coefficient functions of $\widehat H$, and let $T\in\mL(\fH)$ be an R/L mover 2-point operator for $H$, for an initial 2-point operator $T_0\in\mL(\fH)$, for a Fermi function $\rho\in\mB(\R)$, and for the inverse reservoir temperatures $\beta_L, \beta_R\in\R$. Then, the heat flux in the R/L mover state $\omega_T\in\mE_\fA$ has the decomposition $J=J_{pp}+J_{ac}$, where:
\bn[label=(\alph*),ref={\it (\alph*)}]
\setlength{\itemsep}{0mm}
\item
\label{thm:hf-1}
The pure point contribution satisfies $J_{pp}=0$.

\item
\label{thm:hf-2}
The absolutely continuous contribution is given by 
\ba
\label{Jacfinal}
J_{ac}
=\frac12\Int \hspace{0.5mm} e(k) |e'(k)|\hspace{0.3mm}\Delta_{\beta_L,\beta_R}(e(k)),
\ea
where $\Delta_{\beta_L,\beta_R}\in\mB(\R)$ 
reads
\ba
\label{Pauli-Delta}
\Delta_{\beta_L,\beta_R}
:=\rho_{\beta_R}-\rho_{\beta_L},
\ea
and $e, e'\in L^\infty(\T)$ are defined by $e:=u_0+|u|$ and $e':=u_0'+\tilde u u'$.
\en
\et

\vspace{5mm}

\br
If $u\neq0$, \ie, in {\it Case 2, 4 and 5} under Assumption \ref{ass:eig} \ref{ass:eig-1}, we have have $e=e_+$ on $\T$ and $e'=e_+'=u_0'+(uu')/|u|$ on $\mZ_u^c$ with $\card(\mZ_u)<\infty$ (as discussed in the first step of {\it Case 5} of the proof of Lemma \ref{lem:ass} \ref{lem:ass-b}). Moreover, $e_+\in C(\T)\cap C^\infty(\T\setminus\mZ_u)$ and $|e_+'|\le |u_0'|+|u'|$ on $\mZ_u^c$. Therefore, due to Remark \ref{rem:recurrent}, 
the integrand of \eqref{Jacfinal} satisfies $e |e'|\Delta_{\beta_L,\beta_R}\circ e\in L^\infty(\T)$. In {\it Case 3}, the only case with $u=0$ under Assumption \ref{ass:eig} \ref{ass:eig-1}, we have $e=e_+=u_0$ and $e'=u_0'$ on $\T=\mZ_u$ (see \eqref{tildeui}), and the same conclusion holds again.
 \er
  
\br
Since, for all $b\in\mL(\fh)$ and all $A=a_0\sigma_0+a\sigma\in\mL(\fH)$ with $a_0\in\mL(\fh)$ and $a=[a_1,a_2,a_3]\in\mL(\fh)^3$, we have $[A,b\sigma_3]=[a_3,b]\sigma_0+[\ii\{a_2,b\},-\ii\{a_1,b\},[a_0,b]]\sigma$, we get
\ba
\mbox{$u_0=0$ if and only if $[H,\xi\sigma_3]=0$,}
\ea
where we used \eqref{Evu}-\eqref{Oddu} and the fact that $[m[u_\alpha],\hat\xi]=2\hspace{0.2mm}m[\Od(u_\alpha)]\hspace{0.1mm}\hat \xi$ and $\{m[u_\alpha],\hat\xi\}=2\hspace{0.2mm}m[\Ev(u_\alpha)]\hspace{0.1mm}\hat \xi$ for all $\alpha\in\l0,3\r $. Hence, in this special case, which occurs frequently in practice, the R/L mover heat flux has the form 
\ba
\label{Pauli-J}
J
=\frac12\Int |(uu')(k)|\hspace{0.5mm}\Delta_{\beta_L,\beta_R}(|u|(k)).
\ea
\er
 
\br
\label{rem:pure-ac}
Since Assumption \ref{ass:eig} \ref{ass:eig-1} is equivalent with {\it Case 2, 3, 4, and 5} of Lemma \ref{lem:ass}, we get, in all these cases, that $1_{ac}(H)=1$.
\er
 
\br
\label{rem:ee'V}
With the help of $K:=(1\sigma_0+m[\tilde u]\sigma)/2\in\mL(\widehat\fH)$ (which is an orthogonal projection if $u\neq 0$), we can write $\hat H\bar VK=(m[ee']\sigma_0+m[ee'\tilde u]\sigma)/2$
and, for all $k\in\mZ_u^c$ in {\it Case 2, 4, and 5} and for all $k\in\T$ in {\it Case 3}, we get
$\tr([\hat H^2K](k))=e(k)^2$ and
\ba
\tr([\hat H\bar VK](k))
=e(k) e'(k),
\ea
where we used the notation of Remark \ref{rem:spec}. This expression highlights the dependence of the absolutely continuous contribution $J_{ac}$ on the asymptotic velocity $\bar V$. 
\er

\br
\label{rem:ass-eig}
Let the Pauli coefficient functions of $\widehat H$ from Theorem \ref{thm:hf} satisfy $u_\alpha=0$ for all $\alpha\in\l0,2\r$ and $u_3=1$, \ie, we have {\it Case 1} for which the proof of Lemma \ref{lem:ass} \ref{lem:ass-c} asserts that $\bar V=0$. Since $\spec(Q)=\spec_{pp}(q)=\Z$ and since $\ran(E_q(1_\lm))=\spa\{\delta_\lm\}$ and $\ran(E_Q(1_\lm))=\spa\{\delta_\lm\oplus 0, 0\oplus\delta_\lm\}$ for all $\lm\in\Z$, we get $\spec(P)=\spec_{pp}(P)=\Z$ and $\ran(E_P(1_\lm))=\spa\{\e_\lm\oplus 0,0\oplus \e_\lm\}$ for all $\lm\in\Z$ as in the proof of Proposition \ref{prop:RLgen} \ref{prop:RLgen-a}. Moreover, \eqref{Vt} and \eqref{qas} yield $V^t\Phi=P\Phi/t$ for all $\Phi\in\dom(P)$ and all $t\in\R^+$ and, with the help of Remark \ref{rem:rA}, we get $E_{V^t}(1_{(-\infty,0)})=E_P(1_{(-\infty,0)})$ and $E_{V^t}(1_{(-\infty,0]})=E_P(1_{(-\infty,0]})$ for all $t\in\R^+$. Hence, since $\sign=\kappa_0-1_{(-\infty,0]}-1_{(-\infty,0)}$, we find, for all $t\in\R^+$,
\ba
E_{V^t}(\sign)
=E_P(\sign).
\ea
Moreover, since, on one hand, $E_P(\sign)\hspace{0.1mm}\e_1\oplus 0=E_P(\sign)E_P(1_1)\hspace{0.2mm}\e_1\oplus 0=\e_1\oplus 0$ because $1_1\sign =1_1$ and, on the other hand, $\sign(\bar V)=0$, we arrive at
\ba
\slim_{t\to\infty}\hspace{0.5mm} \sign(V^t)
\neq\sign(\bar V),
\ea
\ie, in general, Proposition \ref{prop:RLgen} \ref{prop:RLgen-a} does not hold if Assumption \ref{ass:eig} \ref{ass:eig-1} is not satisfied (see \eqref{limsignV} at the end of the proof of Proposition \ref{prop:RLgen} \ref{prop:RLgen-a}).
\er

In the following, we set $E_x:=\delta_x\oplus 0$ for all $x\in\Z$. For all $a\in\mL(\fh)$ and all $b=[b_1,b_2,b_3]\in\mL(\fh)^3$, we also use the notations $\{a,b\}:=[\{a,b_1\},\{a,b_2\},\{a,b_3\}]\in\mL(\fh)^3$ and $b^2:=bb\in\mL(\fh)$, where the latter is defined after \eqref{prod}. Moreover, recall that, for all $\vi\in\hat\fh$, the function $\check\vi\in\fh$ is given by $\check\vi(x)=(\ff^\ast\vi)(x)=(\e_x,\vi)$. If $\vi=[\vi_1,\vi_2,\vi_3]\in\hat\fh^3$, we set $\check\vi(x):=[\check\vi_1(x),\check\vi_2(x),\check\vi_3(x)]\in\C^3$ for all $x\in\Z$.

\vspace{5mm}
 
\bprf 
Due to \eqref{Phi-3} and Assumption \ref{ass:H} \ref{HLR=0}, the 1-particle heat flux observable reads
\ba
\Phi
\label{PhiIm}
=-\hspace{0.5mm}\Im(H_L H_{LS}),
\ea
\ie, we have $\Phi\in\mL^0(\fH)$ since $H_{LS}\in\mL^0(\fH)$. Moreover, Definition \ref{def:ep} \ref{ep-2} and \eqref{Phi-star}-\eqref{expct-sd} imply that $J=J_{pp}+J_{ac}$.

\vspace{1mm}
\ref{thm:hf-1}\,
Since we have $1_{ac}(H)=1$ as discussed in Remark \ref{rem:pure-ac}, we get $T_{pp}=0$ and, hence, from Definition \ref{def:ep} \ref{ep-2}, $J_{pp}=0$.

\vspace{1mm}
\ref{thm:hf-2}\,
Since $\tr(A\hspace{0.5mm}\Im(B))=(\tr(AB)-\overline{\tr(A^\ast B)})/(2\i)$ for all $A\in\mL(\fH)$ and all $B\in\mL^1(\fH)$ and since $T_{ac}^\ast=T_{ac}$ due to Proposition \ref{prop:R/L2pt} \ref{prop:R/L2pt-a}, Definition \ref{def:ep} \ref{ep-2} and \eqref{PhiIm} yield
\ba
J_{ac}
&=- \hspace{0.5mm}\tr(T_{ac}\Phi)\nonumber\\
&=\tr(T_{ac}\hspace{0.5mm} \Im((Q_L H)^2Q_S))\nonumber\\
&=\Im(\tr(T_{ac} (Q_L H)^2Q_S))\nonumber\\
\label{Jac2}
&=\sum_{\substack{x\in\Z_S\\\alpha\in\l0,1\r}}\Im((\Gamma^\alpha E_x, T_{ac} (Q_L H)^2 \Gamma^\alpha E_x)).
\ea
Writing $T_{ac}=r_0\sigma_0+r\sigma$ and  $(Q_L H)^2=s_0\sigma_0+s\sigma$ with $r_0, s_0\in\mL(\fh)$ and $r, s\in\mL(\fh)^3$, we get
\ba
\label{Jac3}
J_{ac}
=2\sum_{x\in\Z_S}\Im(j_{ac}(x)),
\ea
where $j_{ac}:\Z_S\to\C$ is defined, for all $x\in\Z_S$, by 
\ba
\label{jac}
j_{ac}(x)
:=(\delta_x, (r_0s_0+rs)\delta_x),
\ea
and we used that $\sum_{\alpha\in\l 0,1\r}(\Gamma^\alpha E_x,(a_0\sigma_0+a\sigma)(b_0\sigma_0+b\sigma)\Gamma^\alpha E_x)=2(\delta_x,(a_0b_0+ab)\delta_x)$ for all $x\in\Z$, all $a_0,b_0\in\mL(\fh)$, and all $a, b\in\mL(\fh)^3$. 
Next, we want to compute \eqref{jac}. First, since $Q_L H=(q_L h_0)\sigma_0+(q_L h)\sigma$, we can write
 \ba
\label{s0}
s_0
&=(q_L h_0)^2+(q_L h)^2,\\
\label{s}
s
&=\{q_L h_0, q_L h\}+\ii (q_L h)\wedge (q_L h).
\ea
Using \eqref{hfr1}, for all $\alpha, \beta\in\langle 0,3\rangle$ and all $x\in\Z_S$, the main ingredient of \eqref{s0}-\eqref{s} reads 
\ba
\label {qhqh}
q_L h_\alpha q_L h_\beta\delta_x
=\sum_{y,z\in\l-\nu,\nu\r}  1_L(x+y) 1_L(x+y+z)\hspace{0.5mm} \check u_\alpha(z) \hspace{0.3mm}\check u_\beta(y) \hspace{0.3mm}\delta_{x+y+z}.
\ea
As for $r_0$ and $r$, since $1_{ac}(H)=1$ as discussed in Remark \ref{rem:pure-ac}, \eqref{Rac} yields $T_{ac}=\rho(\Delta H)$. Moreover, due to \eqref{RLgen}, the R/L mover generator reads $\widehat\Delta=\beta 1+\delta\hspace{0.2mm}\sign(\bar V)$ and \eqref{signV} leads to $\sign(\bar V)=m[w_0]\sigma_0+m[w]\sigma$, where $w_0\in L^\infty(\T)$ and $w\in L^\infty(\T)^3$ are given by \eqref{RL1} and \eqref{RL2}, respectively. We next discuss {\it Case 2, 4, and 5} and {\it Case 3} separately.

\begin{figure}
\centering
\begin{tikzpicture}
\draw [->, line width=0.25mm] (-4,0) -- (4.0,0);
\draw [->, line width=0.25mm](0,-4.0) -- (0,4.0);
\foreach \a in {1,...,10}
\draw [line width=0.1mm] (\a*-0.3,0) -- (-3.7,3.7-\a*0.3);
\foreach \a in {1,...,7}
\foreach \b in {1,...,7}
{
\filldraw (\a*-0.3,\b*0.3-0.3) circle (0.6mm);
\ifnum \a=\b
\breakforeach
 \fi}
 \draw [line width=0.2mm, rounded corners] (-2.22,-0.12) rectangle (-0.18,1.92);
 \node at (-1.4,3.5) {$\bigcup_{n\in\l1,\nu\r }X_{1,n}$};
 \draw [->,line width=0.2mm] (-1.4,3.2) to[out=270,in=90] (-1.1,1.95);
\foreach \a in {-0.3, -0.6,..., -3.3}
\draw [line width=0.1mm] (\a,0) -- (\a,-3.7);
\foreach \a in {-0.3, -0.6,..., -2.4}
\foreach \b in {-0.3,-0.6,..., -2.4}
\filldraw (\a,\b) circle (0.6mm);
\draw [line width=0.2mm, rounded corners] (-2.22,-2.22) rectangle (-0.18,-0.18);
\node at (-4.5,-1.2) {$\bigcup_{n\in\l1,\nu\r }X_{2,n}$};
\draw [->,line width=0.2mm] (-3.4,-1.1) to[out=0,in=180] (-2.25,-1.35);
\draw [line width=0.1mm] (2.1,2.1) -- (2.1,-2.1) -- (-2.1,-2.1) -- (-2.1,2.1) -- cycle;
\node at (2.35,-0.25) {$\nu$};
\draw [line width=0.5mm] (2.1,-0.1) -- (2.1,0.1);
\draw [line width=0.1mm] (3.0,3.0) -- (3.0,-3.0) -- (-3.0,-3.0) -- (-3.0,3.0) -- cycle;
\node at (3.35,-0.3) {$n_S$};
\draw [line width=0.5mm] (3.0,-0.1) -- (3.0,0.1);
\node at (4.3,0) {$y$};
\node at (0,4.3) {$z$};
\end{tikzpicture}
\caption{The set $X=\bigcup_{n\in\l1,\nu\r }(X_{1,n}\cup X_{2,n})$.}
\label{fig:setB}
\end{figure}
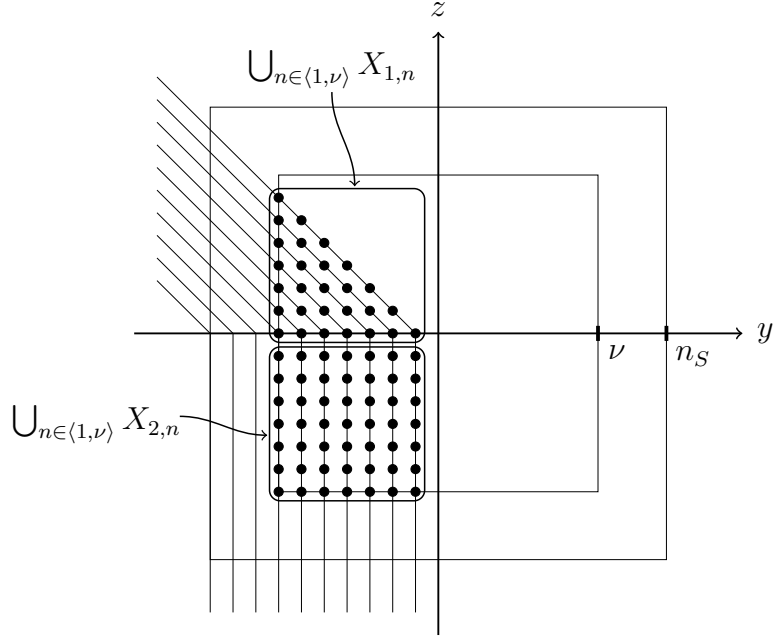

\vspace{0.5mm}
{\it Case 2, 4, and 5}\quad
Since $u\neq 0$ in all these cases, we have $\card(\mZ_u)<\infty$. Hence, \eqref{RL1}-\eqref{RL2} yield, on $\mZ_u^c$, 
\ba
\label{w0Zuc}
w_0
&=\frac12(\sign\circ e_+'+\sign\circ e_-'),\\
\label{wZuc}
w
&=\frac12(\sign\circ  e_+'-\sign\circ e_-') \hspace{0.2mm}\tilde u.
\ea
Moreover, since $\widehat H=m[u_0]\sigma_0+m[u]\sigma$, we get $\widehat\Delta\widehat H=m[b_0]\sigma_0+m[b]\sigma$, where $b_0\in L^\infty(\T)$ and $b\in L^\infty(\T)^3$ have the form $b_0:= \beta u_0+\delta(w_0u_0+wu)$ and $b:=\beta u+\delta(w_0u+u_0w+\ii w\wedge u)$. Using \eqref{w0Zuc}-\eqref{wZuc} and noting that the wedge product vanishes, we get, on $\mZ_u^c$, 
\ba
\label{a0Zuc}
b_0
&=\beta u_0+\frac{\delta}{2}\hspace{0.1mm}((\sign\circ e_+')\hspace{0.1mm}e_+ +(\sign\circ e_-')\hspace{0.1mm}e_-),\\
\label{aZuc}
b&=\beta u+\frac{\delta}{2}\hspace{0.1mm}((\sign\circ e_+')\hspace{0.1mm}e_+ -(\sign\circ e_-')\hspace{0.1mm}e_-)\hspace{0.1mm}\tilde u.
\ea
Since $\widehat T_{ac}
=E_{\widehat\Delta\widehat H}(\rho)$ (see Lemma \ref{lem:unitary}), Proposition \ref{prop:diag} \ref{prop:diag-b} yields 
\ba
\widehat T_{ac}
=m[a_0]\sigma_0+m[a]\sigma, 
\ea
where $a_0\in L^\infty(\T)$ and $a\in L^\infty(\T)^3$ are given by $a_0:=(\rho\circ b_++\rho\circ b_-)/2$ and $a=(\rho\circ b_+-\rho\circ b_-)\hspace{0.2mm}\tilde b/2$, and $b_\pm:=b_0\pm|b|$. Since $b=(\tilde u b)\tilde u$ due to \eqref{aZuc}, we get, on $\mZ_u^c$, 
\ba
\label{fd:a0}
a_0
&=\frac{1}{2}\hspace{0.3mm}(\rho_++\rho_-),\\
\label{fd:a}
a
&=\frac{1}{2}\hspace{0.3mm}(\rho_+-\rho_-)\hspace{0.3mm}\tilde u,
\ea
where $\rho_\pm\in L^\infty( \T)$ are given by
\ba 
\rho_\pm
:=\rho\circ((\beta+\delta\hspace{0.3mm}\sign\circ e_\pm')e_\pm).
\ea
Now, plugging \eqref{qhqh} into \eqref{s0}-\eqref{s}, writing the resulting expression for \eqref{jac} in momentum space, and using that $\hat r_\alpha=m[a_\alpha]$ for all $\alpha\in\l0,3\r $, we get $j_{ac}(x)=\sum_{y,z\in\l-\nu,\nu\r}1_L(x+y) 1_L(x+y+z)  G(y+z,z,y)$ for all $x\in\Z_S$, where $G:\Z^3\to\C$ is defined, for all $x,y,z\in\Z$ , by
\ba
G(x,y,z)
&:=\check a_0(-x)(\check u_0(y)\check u_0(z)+\check u(y)\check u(z))\nonumber\\
&\quad+\check a(-x)(\check u_0(y)\check u(z)+\check u_0(z)\check u(y)+\ii \check u(y)\wedge \check u(z)),
\ea 
and we recall that $\check u(x)\check u(y)$ for all $x,y\in\Z$ stands for the real Euclidean scalar product between $\check u(x)\in\C^3$ and $\check u(y)\in\C^3$ (see after \eqref{prod}). Hence, summing over all $x\in\Z_S$, we get
\ba
\label{sumjac}
\sum_{x\in\Z_S} j_{ac}(x)
=\sum_{(y,z)\in X} \chi(y,z)\hspace{0.5mm}G(y+z,z,y),
\ea
where the staircase type function $\chi:\Z^2\to\l0,n_S\r$ is defined by $\chi(y,z)
:=\sum_{x\in\Z_S} 1_L(x+y) 1_L(x+y+z)$ for all $y,z\in\Z$, and the summation on the right hand side of \eqref{sumjac} is carried out over the set $X
:=\bigcup_{n\in\l1,n_S\r}\chi^{-1}(\{n\})\cap \l-\nu,\nu\r^2$.
Next, let us make the decomposition $X=\bigcup_{n\in\l1,\nu\r }(X_{1,n}\cup X_{2,n})$, where, for all $n\in\l1,\nu\r $, we define $X_{1,n}:=\chi^{-1}(\{n\})\cap (\l-\nu,\nu\r\times \l0,\nu\r)$ and $X_{2,n}:=\chi^{-1}(\{n\})\cap (\l-\nu,\nu\r\times \l-\nu,-1\r)$, see Figure \ref{fig:setB}. 

Since, for all $n\in\l1,\nu\r $, we have $y+z=-n$  for all $(y,z)\in X_{1,n}$ and $y=-n$ for all $(y,z)\in X_{2,n}$, \eqref{Jac3} becomes
\ba
\label{Jac4}
J_{ac}
&=2\hspace{-1mm}\sum_{n\in\l1,\nu\r } n\hspace{-2mm}\sum_{z\in\l0,\nu-n\r}
\Im(G(-n,z,-n-z))\nonumber\\
&\quad+2\hspace{-1mm}\sum_{n\in\l1,\nu\r } n\hspace{-2mm}\sum_{z\in\l-\nu,-1\r}
\Im(G(-n+z,z,-n)).
\ea
In order to determine the imaginary parts on the right hand side of \eqref{Jac4}, we first note that, due to \eqref{Imu=0}, we have $\zeta\check u_\alpha=\xi \check u_\alpha$ for all $\alpha\in\l0,3\r$ (and the same holds for $\check a_\alpha$ for all $\alpha\in\l0,3\r$ due to \eqref{fd:a0}-\eqref{fd:a}). Moreover, \eqref{Evu} and \eqref{Oddu} yield $\xi \check u_\alpha=- \check u_\alpha$ for all  $\alpha\in\l0,2\r$ and $\xi \check u_3=\check u_3$, respectively. Hence, we get $\check u_\alpha=\ii\Im(\check u_\alpha)$ for all $\alpha\in\l0,2\r $ and $\check u_3=\Re(\check u_3)$ which implies, for all $x,y,z\in\Z$,  
\ba
\label{imtau1}
\Im(G(x,y,z))
=\eta_{0,-x}(c_{0,y} c_{0,-z}+c_y c_{-z})+\eta_{-x}(c_{0,y} c_{-z}+c_{0, -z} (Lc_y)+c_y \wedge (Lc_{-z})),
\ea
where $c_{\alpha,x}\in\R$ for all $\alpha\in\l0,3\r$ and all $x\in\Z$ are given in \eqref{c-an}-\eqref{c-3n} and we set $c_x:=[c_{1,x}, c_{2,x}, c_{3,x}]\in\R^3$ for all $x\in\Z$. Moreover, the diagonal matrix $L\in\C^{3\times 3}$ is given by $L:=\diag[1,1,-1]$ and, for all $x\in\Z$, we set  $\eta_{\alpha, x}:=\Im(\check a_\alpha(x))$ for all $\alpha\in\l0,2\r$, $\eta_{3, x}:=\Re(\check a_3(x))$, and $\eta_x:=[\eta_{1,x}, \eta_{2,x}, \eta_{3,x}]\in\R^3$. Hence, \eqref{Jac4} can be written as 
\ba
\label{Jac5}
J_{ac}
&=-
\int_{-\pi}^\pi\frac{\rd k}{2 \pi}\hspace{1mm} \bigg(\mu_1(k)+\mu_2(k)
+\frac{1}{|u(k)|}
\sum_{i\in\l3,8\r}\mu_i(k)\bigg) \rho_+(k)\nonumber\\
&\quad-
\int_{-\pi}^\pi\frac{\rd k}{2 \pi}\hspace{1mm} \bigg(\mu_1(k)+\mu_2(k)
-\frac{1}{|u(k)|}
\sum_{i\in\l3,8\r}\mu_i(k)\bigg) \rho_-(k),
\ea
where, for all $i\in\l1,8\r$, the explicit expressions of the functions $\mu_i\in L^\infty(\T)$ are given in Lemma \ref{lem:exp}. Furthermore, using \eqref{Pauli-cf}, a direct computation yields 
\ba
\label{A1A2}
\mu_1+\mu_2
&=-\frac12(u_0u_0'+uu'),\\
\label{A3A8}
\sum_{i\in\l3,8\r}\mu_i
&=-\frac12(u_0(uu')+u_0'|u|^2),
\ea
where, in \eqref{A1},  \eqref{A3}, \eqref{A5}, and \eqref{A7} of  Lemma \ref{lem:exp}, we used that, for all families $\{f_n\}_{n\in\l1,\nu\r }$ of functions $f_n:\T\to\R$, we have  $\sum_{n\in\l1,\nu\r,\hspace{0.2mm}l\in\l0,\nu-n\r} f_n(k)c_{\alpha_1,l} c_{\alpha_2,n+l}=\sum_{n\in\l1,\nu\r,\hspace{0.2mm}m\in\l0,n-1\r} f_{n-m}(k) 
\allowbreak
c_{\alpha_1,m} c_{\alpha_2,n}$ for all $k\in\T$ and all $\alpha_1,\alpha_2\in\l0,3\r$ 
(that $c_{\alpha,0}=0$ for all $\alpha\in\l0,2\r $, and that all the contributions in \eqref{A3}-\eqref{A8} involving a product of the form $c_{1,n} c_{2,m} c_{3,l}$ add up to zero, where $n,m\in\l1,\nu\r$ and $l\in\l1,\nu\r$ or $l=0$).
Moreover, note that, due to \eqref{A3A8}, the modulus of $\sum_{i\in\l3,8\r}\mu_i/|u|$ in \eqref{Jac5} is bounded from above by $(|u_0||u'|+|u_0'||u|)/2$. Plugging \eqref{A1A2}-\eqref{A3A8} into \eqref{Jac5}, using that $e_\pm e_\pm'=u_0u_0'+uu'\pm (u_0(\tilde uu')+u_0'|u|)$ on $\mZ_u^c$ and that $\card(\mZ_u)<\infty$, \eqref{Jac5} becomes
\ba
\label{Jac-2}
J_{ac}
&=\frac12\Int (e_+(k) e_+'(k)\hspace{0.2mm}\rho_+(k)+e_-(k) e_-'(k)\hspace{0.2mm}\rho_-(k))\\
\label{Jac-3}
&=\Int e_+(k) e_+'(k)\hspace{0.2mm}
\rho((\beta+\delta\sign(e_+'(k)))e_+(k)),
\ea
where, in \eqref{Jac-3}, we also used that $\hat\xi e_-=-e_+$ and $\hat\xi e_-'=e_+'$ on $\mZ_u^c$ due to \eqref{Evu}-\eqref{Oddu}, that \eqref{def:Ff-2} holds, and that $e_+e_+'=(e_+^2)'/2$ on $\mZ_u^c$. 
Finally, plugging $e_+e_+'=(1_{\mZ_{++}}+1_{\mZ_{+-}}+1_{\mZ_{-+}}+1_{\mZ_{--}})e_+e_+'$ into\eqref{Jac-3}, where $\mZ_{\pm\pm}:=\{k\in\mZ_u^c\,|\, \mbox{$\pm e_+(k)>0$ and $\pm e_+'(k)>0$}\}$ and $\mZ_{\pm\mp}:=\{k\in\mZ_u^c\,|\, \mbox{$\pm e_+(k)>0$ and $\mp e_+'(k)>0$}\}$, the integrand in \eqref{Jac-3} can be written, on $\mZ_u^c$, as 
\ba
\label{dec-ee'}
e_+ e_+'\rho((\beta+\delta\sign\circ e_+')e_+)
&=\frac12\hspace{0.5mm}|e_+ e_+'|(\rho\circ (\beta_R|e_+|)-\rho\circ (\beta_L|e_+|))\nonumber\\
&\quad+\frac12\hspace{0.5mm} e_+ e_+'(\rho\circ (\beta_R e_+)+\rho\circ (\beta_Le_+)). 
\ea
Since, due to  Definition \ref{def:Ff}, the restriction of $\rho$ to $[-a,a]$ satisfies $\rho\in L^1([-a,a])$, where $a:=(1+|\beta_L|+|\beta_R|)\sum_{\alpha\in\l0,3\r }\|u_\alpha\|_\infty/2>0$, the first fundamental theorem of calculus
yields that the function $\rho_1: [-a,a]\to \R$, defined by $\rho_1(x):=\int_{-a}^x\rd t\, \rho(t)$ for all $x\in [-a,a]$, is absolutely continuous and an antiderivative of $\rho$ almost everywhere on $[-a,a]$. Similarly, the function $\rho_2: [-a,a]\to \R$, defined by $\rho_2(x):=\int_{-a}^x\rd t\, \rho_1(t)$ for all $x\in [-a,a]$, is an antiderivative of $\rho_1$ on $[-a,a]$. Hence, for all $\kp\in\{L,R\}$ for which $\beta_\kp\neq 0$, using twice partial integration (for absolutely continuous functions), the property $e_+(-\pi)=e_+(\pi)$, and the fact that $e_+e_+'\rho\circ (\beta_\kp e_+)=e_+(\rho_1\circ (\beta_\kp e_+))'/\beta_\kp$ and $e_+'\rho_1\circ (\beta_\kp e_+)=(\rho_2\circ (\beta_\kp e_+))'/\beta_\kp$, the integral of each term in the second expression on the right hand side of \eqref{dec-ee'} vanishes (if $\beta_\kp=0$ for some $\kp\in\{L,R\}$, \eqref{def:Ff-2} yields $e_+e_+'\rho\circ (\beta_\kp e_+)=e_+e_+'/2=(e_+^2)'/4$). Therefore, \eqref{Jac-3} takes the form
\ba
\label{Jac-af}
J_{ac}
=\frac12\Int \hspace{0.5mm}|e_+(k) e_+'(k)|\hspace{0.3mm}\Delta_{\beta_L,\beta_R}(|e_+(k)|),
\ea
where $\Delta_{\beta_L,\beta_R}\in\mB(\R)$ stems from \eqref{Pauli-Delta} and, with the help of \eqref{def:Ff-2}, we arrive at \eqref{Jacfinal}.

\vspace{0.5mm}
{\it Case 3}\quad
If $u=0$ (and $u_0\neq 0$), due to \eqref{RL1}-\eqref{RL2}, \eqref{w0Zuc}-\eqref{wZuc} become $w_0=\sign\circ u_0'$ and $w=0$, respectively. Moreover, we get $b_0=\beta u_0+\delta(\sign\circ u_0')u_0$ and $b=0$, and $a_0=\rho\circ b_0$ and $a=0$ instead of \eqref{a0Zuc}-\eqref{fd:a}. Formula \eqref{Jac5} then reads as $J_{ac}= \int_{-\pi}^\pi\rd k\hspace{1mm}u_0(k)u_0'(k) \rho((\beta+\delta\sign(u_0'(k)))u_0(k))/(2\pi)$ and the same arguments apply to its integrand as the ones used for \eqref{dec-ee'}. Hence, since we defined $e'=u_0+\tilde u u'$ on $\T$, \eqref{Jacfinal}-\eqref{Pauli-Delta} also hold for {\it Case 3}.
\eprf

\vspace{5mm}

We next illustrate the R/L mover heat flux for two well-known examples. The first example is the prominent XY spin chain (see Remark \ref{rem:SDC-XY}).

\bx[XY model]
\label{ex:XY}
Let $c_{2,1}, c_{3,0}\in\R$. 
The XY model with anisotropy $c_{2,1}$ and spatially homogeneous exterior magnetic field $c_{3,0}$ is specified by 
\ba
\label{h0-XY}
h_0
&=0,\\
h_1
&=0,\\
\label{h2-XY}
h_2
&=-2c_{2,1}\Im(\theta),\\
\label{h3-XY}
h_3
&=c_{3,0} 1+\Re(\theta).
\ea
Hence, we have $\nu=1$ and, for all $c_{2,1}, c_{3,0}\in\R$, Assumptions \ref{ass:H} \ref{HTheta}, \ref{HLR=0}, and \ref{nontrv} are satisfied (note that $c_{3,1}=1/2$). Moreover, for all $k\in\T$, we get from \eqref{h0-XY}-\hspace{0.5mm}\eqref{h3-XY} that $u_0=u_1=0$, that $u_2(k)=-2c_{2,1}\sin(k)$, and that $u_3(k)=c_{3,0}+\cos(k)$. Writing $u^2=a_0+2\sum_{m\in\l1,2\nu\r } a_m\cos(m\hspace{0.5mm}\cdot)$ as in \eqref{u2-TP}, Lemma \ref{lem:sqr} \ref{lem:sqr-1} yields
\ba
a_0
&=\frac12+2c_{2,1}^2+c_{3,0}^2,\\
a_1
&=c_{3,0},\\
a_2
&=\frac14-c_{2,1}^2.
\ea
Therefore, the function $uu'=(u^2)'/2\in TP(\T)$ satisfies $uu'=0$ (\ie, {\it Case 1} occurs) if and only if $(c_{2,1},c_{3,0})\in\{(-1/2,0), (1/2,0)\}$  (the Ising (-Lenz) model), \ie, due to \eqref{eigV0}, Assumption \ref{ass:eig} \ref{ass:eig-1} holds if and only if $(c_{2,1},c_{3,0})\in\R^2\setminus\{(-1/2,0), (1/2,0)\}$. In these cases, Theorem \ref{thm:hf} is applicable and, for all Fermi functions $\rho\in\mB(\R)$ and all inverse reservoir temperatures $\beta_L, \beta_R\in\R$, the R/L mover heat flux is given by \eqref{Pauli-J}. Moreover, if the Fermi function $\rho$ is the Fermi-Dirac distribution \eqref{ex:KMS-2}, we get 
\ba
\label{J-AP}
J
=\frac12\int_{-\pi}^\pi\frac{\rd k}{2\pi}\hspace{1mm} |(uu')(k)|\hspace{0.5mm}
\frac{\sinh(\delta |u|(k))}{\cosh(\delta |u|(k))+\cosh(\beta |u|(k))},
\ea
where we used that $1/(1+\e^{-x})-1/(1+\e^{-y})=\sinh((x-y)/2)/(\cosh((x-y)/2)+\cosh((x+y)/2))$ for all $x,y\in\R$. The formula \eqref{J-AP} has been obtained in \cite{AP2003} for $c_{2,1}\in(-1/2,1/2)$ and $c_{3,0}\in\R$.
\ex

\br
\label{rem:local}
For all $x,y\in\Z$, we have $p_{x,y}\sigma_0=(E_x, \cdot)E_y+(\Gamma E_x, \cdot)\Gamma E_y$, $p_{x,y}\sigma_1=(\Gamma E_x, \cdot)E_y+( E_x, \cdot)\Gamma E_y$, $p_{x,y}\sigma_2=-\ii (\Gamma E_x, \cdot)E_y+\ii ( E_x, \cdot)\Gamma E_y$, and $p_{x,y}\sigma_3=(E_x, \cdot)E_y-(\Gamma E_x, \cdot)\Gamma E_y$, where we recall that $p_{x,y}=(\delta_x,\,\cdot\,)\,\delta_y\in\mL^0(\fh)$ for all $x,y\in\Z$ and $E_x=\delta_x\oplus 0$ for all $x\in\Z$. 
Therefore, the selfdual second quantization \eqref{sdsq} yields, for all $x,y\in\Z$, 
\ba
b(p_{x,y}\sigma_0)
&=a_y^\ast a_x+a_y a_x^\ast,\\
b(p_{x,y}\sigma_1)
&=a_y^\ast a_x^\ast+a_y a_x,\\
b(p_{x,y}\sigma_2)
&=-\ii (a_y^\ast a_x^\ast-a_y a_x),\\
b(p_{x,y}\sigma_3)
&=a_y^\ast a_x-a_y a_x^\ast,
\ea
where we used the notation from before Remark \ref{rem:SDC-XY}. Since, for all $N\in\N$ satisfying \eqref{N-lbound}, we have $q_N\theta^n q_N=\sum_{x\in\l-N,N-n\r}p_{x,x+n}$ for all $n\in\l1,\nu\r$, where $q_N$ stems from Remark \ref{rem:onehalf}, we get for the local zeroth Pauli coefficient of $H$ and for all $N\in\N$ satisfying \eqref{N-lbound},
\ba
 b((q_Nh_0q_N)\sigma_0)
 =-2\ii \sum_{n\in\l1,\nu\r } c_{0,n}
\hspace{-1mm}\sum_{x\in\l-N,N-n\r} (a_x^\ast a_{x+n}-a_{x+n}^\ast a_x).
\ea
Moreover, using the generalized Jordan-Wigner transformation (see Remark \ref{rem:SDC-XY}), we can write, for all $x\in\Z$ and all $n\in\N$, 
 \ba
 a_x^\ast a_{x+n}-a_{x+n}^\ast a_x
 =\begin{cases}
\hfill \frac\ii 2 \big(\sigma_1^{(x)}\sigma_2^{(x+1)}- \sigma_2^{(x)}\sigma_1^{(x+1)}\big), & n=1,\\
\frac\ii 2 \big(\sigma_1^{(x)}\big(\prod_{i\in\l1,n-1\r}\sigma_3^{(x+i)}\big) \sigma_2^{(x+n)}- \sigma_2^{(x)}\big(\prod_{i\in\l1,n-1\r}\sigma_3^{(x+i)}\big)\sigma_1^{(x+n)}\big) , & n\ge 2,
\end{cases}
 \ea
 \ie, we get (generalized) Dziyaloshinskii-Moriya type interactions. For the sake of completeness, we display the selfdual second quantization of the local first, second, and third Pauli coefficient of $H$ in the fermionic and the spin picture in Appendix \ref{app:Fermi-Pauli}.
\er

The second example,  whose Hamiltonian has been introduced in \cite{S1971}, is a generalized form of the foregoing XY model. 

\bx[Suzuki model]
Let $\nu\in\l1,n_S\r$, let $\{c_{2,n}\}_{n\in\l1,\nu\r }\subseteq\R$, and let $\{c_{3,n}\}_{n\in\l0,\nu\r}\subseteq\R$. The Suzuki model (also called generalized XY model or $\nu$XY model) is specified by
\ba
\label{h0-S}
h_0
&=0,\\
h_1
&=0,\\
\label{h2-S}
h_2
&=-2\sum_{n\in\l1,\nu\r }c_{2,n} \Im(\theta^n),\\
\label{h3-S}
h_3
&=c_{3,0} 1+2\sum_{n\in\l1,\nu\r } c_{3,n} \Re(\theta^n).
\ea
Hence, Assumptions \ref{ass:H} \ref{HTheta} and \ref{HLR=0} are satisfied. If at least one of the coefficients from $\{c_{2,n}\}_{n\in\l1,\nu\r }$ and  $\{c_{3,n}\}_{n\in\l0,\nu\r}$ is different from zero, Assumptions \ref{ass:H} \ref{nontrv} also holds. Moreover, \eqref{h0-S}-\hspace{0.5mm}\eqref{h3-S} lead to $u_0=u_1=0$, $u_2(k)=-2\sum_{n\in\l1,\nu\r }c_{2,n}\sin(nk)$, $u_3(k)=c_{3,0}+2\sum_{n\in\l1,\nu\r }c_{3,n}\cos(nk)$ for all $k\in\T$, and 
Lemma \ref{lem:sqr} \ref{lem:sqr-2} yields the coefficients in \eqref{u2-TP} for ($n_S$ sufficiently large and) $\nu=2$, 
\ba
a_0
&=2(c_{2,1}^2+c_{2,2}^2+c_{3,1}^2+c_{3,2}^2)+c_{3,0}^2,\\
\label{Sa1}
a_1
&=2(c_{2,1}c_{2,2}+c_{3,1}(c_{3,0}+c_{3,2})),\\
a_2
&=-c_{2,1}^2+c_{3,1}^2+2c_{3,1}(c_{3,0}+c_{3,2}),\\
a_3
&=-2(c_{2,1}c_{2,2}-c_{3,1}c_{3,2}),\\
\label{Sa4}
a_4
&=-c_{2,2}^2+c_{3,2}^2,
\ea
and, for $\nu\ge 3$, the coefficients $a_0,\ldots, a_{2\nu}$ are given in  Lemma \ref{lem:sqr} \ref{lem:sqr-ge3}. Since we have $uu'=-\sum_{m\in\l1,2\nu\r } m a_m\sin(m\hspace{0.5mm}\cdot)$, we get $uu'=0$ (\ie, {\it Case 1} occurs) if and only if $a_m=0$ for all $m\in\l1,2\nu\r$. The solutions of this system of $2\nu$ multivariate homogeneous quadratic equations in $2\nu+1$ variables specify the Suzuki models for which Assumption \ref{ass:eig} \ref{ass:eig-1} is not satisfied (for example, $c_{2,1}=c_{3,1}=0$, $c_{2,2}=c_{3,2}$, and any $c_{3,0}$ for the $\nu=2$ system given by \eqref{Sa1}-\hspace{0.5mm}\eqref{Sa4}).  
In all the other cases, Theorem \ref{thm:hf} is applicable and, for all Fermi functions $\rho\in\mB(\R)$ and all inverse reservoir temperatures $\beta_L, \beta_R\in\R$, the R/L mover heat flux is again given by \eqref{Pauli-J}. 
\ex

Finally, we want to discuss the following example.

\bx[Full range 1 model]
Let $\nu=1$ and $\{c_{0,1}, c_{1,1}, c_{2,1}, c_{3,0}, c_{3,1}\}\subseteq\R$.  The full range 1 model is specified by
\ba
\label{h0-fm}
h_0
&=-2 c_{0,1} \Im(\theta),\\
h_1
&=-2 c_{1,1} \Im(\theta),\\
h_2
&=-2c_{2,1} \Im(\theta),\\
\label{h3-fm}
h_3
&=c_{3,0} 1+2 c_{3,1} \Re(\theta).
\ea
Hence, Assumptions \ref{ass:H} \ref{HTheta} and \ref{HLR=0} are satisfied. If at least one of the coefficients from $\{c_{0,1}, c_{1,1}, c_{2,1}, c_{3,0}, c_{3,1}\}$ is different from zero, Assumptions \ref{ass:H} \ref{nontrv} also holds. Moreover, \eqref{h0-fm}-\hspace{0.5mm}\eqref{h3-fm} lead to $u_\alpha(k)=-2 c_{\alpha,1}\sin(k)$ for all $\alpha\in\l0,2\r$ and all $k\in\T$ and $u_3(k)=c_{3,0}+2c_{3,1}\cos(k)$ for all $k\in\T$. Hence, Lemma \ref{lem:sqr} \ref{lem:sqr-2} yields the coefficients in \eqref{u2-TP}, \ie, 
\ba
a_0
&=2(c_{1,1}^2+c_{2,1}^2+c_{3,1}^2)+c_{3,0}^2,\\
\label{R1a1}
a_1
&=2c_{3,0}c_{3,1},\\
\label{R1a2}
a_2
&=-(c_{1,1}^2+c_{2,1}^2)+c_{3,1}^2.
\ea
Hence, {\it Case 1} occurs if and only if $c_{3,0}=0$ and  $c_{1,1}^2+c_{2,1}^2=c_{3,1}^2\neq 0$ or $c_{3,0}\neq0$ and $c_{1,1}=c_{2,1}=c_{3,1}=0$. Moreover, {\it Case 6} occurs if and only if $c_{3,0}=0$, $c_{3,1}=0$, and $c_{1,1}^2+c_{2,1}^2=c_{0,1}^2\neq 0$.
In all the other cases, Theorem \ref{thm:hf} is applicable and, for all Fermi functions $\rho\in\mB(\R)$ and all inverse reservoir temperatures $\beta_L, \beta_R\in\R$, the R/L mover heat flux is given by \eqref{Pauli-J} (in {\it Case 2}) and \eqref{Jacfinal} (in {\it Case 3, 4, and 5}).
\ex

\vspace{5mm}

For the following, recall the definitions of $e$, $e'$, and $\Delta_{\beta_L,\beta_R}$ from Theorem \ref{thm:hf} \ref{thm:hf-2}.

Using the additional Assumption \ref{ass:mono} \ref{ass:mono-a} and \ref{ass:mono-b} leads to the strict positivity of the R/L mover heat flux.

\bt[Nonvanishing heat flux]
\label{thm:nnv}
Let $H\in\mL(\fH)$ be a Hamiltonian satisfying Assumption \ref{ass:H} \ref{HTheta}, \ref{HLR=0}, and \ref{nontrv} and let the bounded extension $\bar V\in\mL(\widehat\fH)$ of the asymptotic velocity with respect to $\widehat H$ satisfy Assumption \ref{ass:eig} \ref{ass:eig-1}. Moreover, let $u_0\in L^\infty(\T)$ and $u\in L^\infty(\T)^3$ be the Pauli coefficient functions of $\widehat H$, and let $T\in\mL(\fH)$ be an R/L mover 2-point operator for $H$, for an initial 2-point operator $T_0\in\mL(\fH)$, for a Fermi function $\rho\in\mB(\R)$, and for the inverse reservoir temperatures $\beta_L, \beta_R\in\R$.  Moreover, let the Fermi function and the inverse temperatures satisfy Assumption \ref{ass:mono} \ref{ass:mono-a} and \ref{ass:mono-b}, respectively. Then, the heat flux in the R/L mover state $\omega_T\in\mE_\fA$ is nonvanishing and the heat is flowing from the left to the right reservoir, 
\ba
\label{J-2}
J
>0.
\ea
\et

\vspace{5mm}

\bprf
Under Assumption \ref{ass:mono} \ref{ass:mono-a} and \ref{ass:mono-b}, the difference $\Delta_{\beta_L,\beta_R}\circ |e|$ is nonnegative on $\T$ and strictly positive on $\mZ_e^c$. Hence, in order for \eqref{J-2} to hold, it is sufficient to show that $\card(\mZ_e)<\infty$ and $\card(\mZ_{e'})<\infty$ since, from \eqref{Jac-af}, we have 
\ba
\label{J-1}
J
=\frac12\Int \hspace{0.5mm}|e(k) e'(k)|\hspace{0.3mm}\Delta_{\beta_L,\beta_R}(|e(k)|).
\ea
In order to do so, we use the same type of arguments as in the proof of Lemma \ref{lem:ass}.

\vspace{0.5mm}
{\it Case 1 and 6}\quad 
These cases are excluded due to Assumption \ref{ass:eig} \ref{ass:eig-1}. 

\vspace{0.5mm}
{\it Case 2}\quad 
Since $e=|u|$ and $e'=\tilde uu'$, we have $\card(\mZ_e)=\card(\mZ_u)<\infty$ and $\card(\mZ_{e'})\le\card(\mZ_u)+\card(\mZ_{uu'})<\infty$ due to $uu'\neq 0$.

\vspace{0.5mm}
{\it Case 3}\quad 
Since $e=u_0$ and $e'=u_0'$, we have $\card(\mZ_e)=\card(\mZ_{u_0})<\infty$ and $\card(\mZ_{e'})=\card(\mZ_{u_0'})<\infty$ because $u_0=0$ if and only if $u_0'=0$.

\vspace{0.5mm}
{\it Case 4}\quad 
We have $e=u_0+|u|$ and $e'=u_0'$. Suppose that there exists $M\subseteq\mZ_u^c$ with $\card(M)=\infty$ such that $e=0$ on $M$. It then follows that $p_0=0$ on $M$ (see \eqref{plm}) and, hence, $p_0=0$ (on $\T$). Since $u^2=a_0>0$, we get $u_0^2=a_0>0$ which contradicts $u_0\neq 0$ (and $u_0'\neq 0$) due to \eqref{ZvTP}. Hence, $\card(\mZ_e)<\infty$. Moreover, since $e'=u_0'$ and since $u_0'\neq 0$, we also get $\card(\mZ_{e'})<\infty$.

\vspace{0.5mm}
{\it Case 5}\quad 
We have $e=u_0+|u|$ and $e'=u_0'+\tilde uu'$. Suppose that there exists $M\subseteq\mZ_u^c$ with $\card(M)=\infty$ such that $e=0$ on $M$. It then follows, as in {\it Case 4}, that $p_0=0$ which contradicts $u_0^2\neq u^2$ and, hence, $\card(\mZ_e)<\infty$. Moreover, suppose also that there exists $M'\subseteq\mZ_u^c$ with $\card(M')=\infty$ such that $e'=0$ on $M'$. Hence, the third step of {\it Case 5} in the proof of Lemma \ref{lem:ass} \ref{lem:ass-a} implies $p_0=0$ and we again get a contradiction.  Therefore, $\card(\mZ_{e'})<\infty$ holds, too.
\eprf

As an immediate consequence of the Theorem \ref{thm:nnv}, we get the following corollary.

\bc[Strict positivity of the entropy production rate]
Under the conditions of Theorem \ref{thm:nnv}, the entropy production rate in the R/L mover state is strictly positive, 
\ba
\sigma
>0.
\ea
\ec

\bprf
Plug \eqref{J-1} into \eqref{ep-sigma} and use \eqref{J-2}.
\eprf

\br
Since $[\widehat H](k)[K](k)=e(k)[K](k)$ for all $k\in\T$, where $K\in\mL(\widehat\fH)$ stems from Remark \ref{rem:ee'V}, we get $|e(k)|\le |[\widehat H](k)|_0$ for all $k\in\T$, 
where $|\cdot|_0$ stands for the usual operator norm on $\C^{2\times 2}$ induced by the Euclidean vector norm on $\C^2$. Hence, since we know 
that $\|H\|=\|\widehat H\|={\rm ess\,sup}_{k\in\T}|[\widehat H](k)|_0$ (see \cite{BS2006} for example), we have $|e(k)|\le \|H\|$ for all $k\in\T$. Moreover, due Remark \ref{rem:Lbsg} and its consequence for the integral of the derivatives of monotonically increasing functions,
$\rho'$ exists almost everywhere in $[0,\beta_R \|H\|]$ and $\rho(\beta_R|e(k)|)-\rho(\beta_L|e(k)|)\ge \int_{\beta_L|e(k)|}^{\beta_R|e(k)|}\rd x\,\rho'(x)$ for all $k\in\T$. Hence, under the conditions of Theorem \ref{thm:nnv} and the additional Assumption \ref{ass:mono} \ref{ass:mono-aa}, \eqref{ep-sigma} and \eqref{J-1} yield the strictly positive lower bound
\ba
\sigma
&\ge 2c\delta^2\Int e(k)^2 |e'(k)|.
\ea
\er

\vspace{20mm}

{\it Acknowledgments}\quad
We thank the anonymous referee for his careful reading of the manuscript and his constructive comments and suggestions.

\begin{appendix}

\section{Spectral theory}
\label{app:spectral}

In this appendix, we present a brief summary of the approach to spectral theory based on \cite{Hunziker}  (see also \cite{Jost1973}). The spectral properties used in the foregoing sections are direct consequences of the following presentation or can be derived from it in a simple way. Since this approach is somewhat different from the more standard ones, we precisely state the first three claims without giving proofs (Proposition \ref{prop:ext}, Proposition \ref{prop:extC}, and Theorem \ref{thm:spect} below). Subsequently, we rather explicitly carry out the  implications of these claims in view of their applications to the foregoing sections.

In the following, let $\mH$ again stand for any separable complex Hilbert space and $\mL(\mH)$ for the $C^\ast$\hspace{-0.5mm}-algebra of bounded operators on $\mH$. Moreover, equipped with the usual pointwise operations and with the norm given, for all $\chi\in \ell^\infty(\R)$, by
\ba
\label{supnrm}
{|\chi|}_\infty
:=\sup_{x\in\R}|\chi(x)|,
\ea
we denote by $\ell^\infty(\R)$, $C_b(\R)$, and $C_0(\R)$ the $C^\ast$-algebra of bounded complex-valued functions on $\R$, the $C^\ast$-algebra of continuous bounded complex-valued functions on $\R$, and the (non complete) normed $\ast$-algebra of continuous complex-valued functions on $\R$ with compact support, respectively.

\bd[Projection-valued measure]
A $\ast$-algebra homomorphism $E_0:C_0(\R)\to\mL(\mH)$ is called a projection valued measure.
\ed

For the following, let $\kp_0, \kp_{-1}\in C_b(\R)$ and $\kp_1\in C(\R)$ be defined, for all $x\in\R$, by
\ba
\label{kp0}
\kp_0(x)
&:=1,\\
\label{kp1}
\kp_1(x)
&:=x,\\
\label{kp-1}
\kp_{-1}(x)
&:=\frac{1}{1+|x|},
\ea 
where $C(\R)$ stands for the $\ast$-algebra of continuous complex-valued functions on $\R$ (equipped with the same pointwise operations as the foregoing functions spaces). Moreover, for some of the following notions, see also \cite{Kechris}.

\bd[Borel functions]
\label{def:Bf}
\hfill
\bn[label=(\alph*),ref={\it (\alph*)}]
\setlength{\itemsep}{0mm}
\item
\label{Bf:Bc}
Let $\chi:\R\to\C$ and let $(\chi_n)_{n\in\N}$ be a sequence in $\ell^\infty(\R)$. If there exists $C>0$ such that $|\chi_n|_\infty\le C$ for all $n\in\N$ and if $\lim_{n\to\infty}\chi_n(x)=\chi(x)$ for all $x\in\R$, we write 
\ba
\Blim_{n\to\infty}\chi_n
=\chi,
\ea
and we say that $(\chi_n)_{n\in\N}$ is Borel convergent to $\chi\in\ell^\infty(\R)$.

\item
\label{Bf:min}
Let $\mF\subseteq\ell^\infty(\R)$ be a family of functions having the following properties:
\bn[label=(\alph*),ref={\it (\alph*)}]
\setlength{\itemsep}{0mm}
\item[(B1)]
\label{B1}
$C_0(\R)\subseteq\mF$

\item[(B2)]
If $\chi\in\ell^\infty(\R)$ and if $(\chi_n)_{n\in\N}$ in $\mF$ is such that $\Blim_{n\to\infty}\chi_n=\chi$, then $\chi\in\mF$.
\en
The smallest such $\mF$, denoted by $\mB(\R)$, is called the family of bounded Borel functions. It is a normed unital $\ast$-algebra with unity $\kappa_0$, and $C_b(\R)\subseteq\mB(\R)$.
 
 \item
 \label{Bf:Bsets}
 A set $M\subseteq\R$ is called a Borel set if $1_M\in\mB(\R)$. The $\sigma$-algebra of all Borel sets is denoted by $\mM(\R)$.
\en
\ed

\bp[Extension to Borel functions]
\label{prop:ext}
Let $E_0$ be a projection-valued measure. 
\bn[label=(\alph*),ref={\it (\alph*)}]
\setlength{\itemsep}{0mm}
\item
\label{prop:ext-1}
There exists a unique extension of $E_0$ to a $\ast$-algebra homomorphism  $E:\mB(\R)\to\mL(\mH)$ on $\mB(\R)$. Moreover, for all $\chi\in\mB(\R)$, we have
\ba
\label{ext:bnd}
\|E(\chi)\|
\le |\chi|_\infty, 
\ea
and $E(\chi)\ge 0$ for all $\chi\in\mB(\R)$ with $\chi\ge 0$.

\item
\label{prop:ext-2}
Let $(\chi_n)_{n\in\N}$ in $\mB(\R)$ be such that $\Blim_{n\to\infty}\chi_n=\chi$ for some $\chi\in\ell^\infty(\R)$. Then, $\chi\in\mB(\R)$ and the extended projection-valued measure $E$ from {\it (a)} satisfies
\ba
\label{ext:s-conv}
\slim_{n\to\infty} E(\chi_n)
=E(\chi).
\ea
\en
\ep

\bprf
See \cite{Hunziker} (and \cite{Jost1973}).
\eprf

\bd[Resolution of the identity]
\label{def:resolution}
A $\ast$-algebra homomorphism $E:\mB(\R)\to\mL(\mH)$ is called a resolution of the identity if $E(\kappa_0)=1$.
\ed

\bp[Extension to the identity function]
\label{prop:extC}
Let $E$ be a resolution of the identity and let $\mD$ be the dense subspace of $\mH$ defined by 
\ba
\label{domEx}
\mD
:=\ran\hspace{-0.5mm}(E(\kappa_{-1})).
\ea
\bn[label=(\alph*),ref={\it (\alph*)}]
\setlength{\itemsep}{0mm}
\item
\label{domEx-a}
The operator $\bar E(\kappa_1):\mD\to\mH$, defined, for all $\psi\in\mH$, by
\ba
\label{actionEx}
\bar E(\kappa_1) E(\kappa_{-1})\psi
:=E(\kappa_1\kappa_{-1})\psi,
\ea
is selfadjoint.

\item
\label{domEx-b}
For all $\chi\in\mB(\R)$ with $\kappa_1\chi\in\mB(\R)$, we have
\ba
\label{extC:prod}
\bar E(\kappa_1)E(\chi)
=E(\kappa_1\chi). 
\ea
\en
\ep

\bprf
See \cite{Hunziker} (and \cite{Jost1973}).
\eprf

\br
Since $E$ is a resolution of the identity, the set $\mC:=\spa\{ E(\chi)\psi\,|\, \chi\in C_0(\R), \psi\in\mH\}$ is dense in $\mH$. For all $\eta\in C(\R)$, defining the operator $\mE(\eta)$ on $\mC$ by  $\mE(\eta)E(\chi)\psi:=E(\eta\chi)\psi$  for all $\chi\in C_0(\R)$ and all $\psi\in\mH$, the map $C(\R)\ni\eta\mapsto\mE(\eta)$ satisfies $\mE(\eta)\mC\subseteq\mC$ and, on $\mC$, preserves the linear operations and the multiplication. Moreover, we have $\mE(\bar\eta)\subseteq\mE(\eta)^\ast$ for all $\eta\in C(\R)$. Therefore, $\mE(\eta)$ is closable for all $\eta\in C(\R)$ and the closure of $\mE(\eta)$ defines an extension of $E$ to $C(\R)$ which preserves the $\ast$-operation.
\er

In the following,  $\mS(\R)$ stands for the usual Schwartz space of rapidly decreasing functions on $\R$. Note that $\mS(\R)\subseteq C_b(\R)\subseteq\mB(\R)$. Moreover, using the same notation as the one introduced before \eqref{def:fH}, the Fourier transform of $\chi\in\mS(\R)$ is denoted by $\hat\chi\in\mS(\R)$ and we use the convention $\hat\chi(t):=\int_\R\rd x\, \chi(x) \e^{-\ii tx}/\sqrt{2\pi}$ for all $t\in\R$. 

\bt[Spectral theorem]
\label{thm:spect}
Let $A$ be a (not necessarily bounded) selfadjoint linear operator on $\mH$.
\bn[label=(\alph*),ref={\it (\alph*)}]
\setlength{\itemsep}{0mm}
\item
\label{spect-uniq}
There exists a unique resolution of the identity $E$ such that $A=\bar E(\kappa_1)$. 

\item
\label{spect-fourier}
On $\mS(\R)$, the resolution of the identity $E$ from {\it (a)} can be expressed as an inverse Fourier transform, \ie, for all $\chi\in\mS(\R)$ and all $\psi\in\mH$, we have
\ba
\label{ESchwartz}
E(\chi)\psi
=\frac{1}{\sqrt{2\pi}}\int_\R\rd t\, \hat\chi\, (t)\hspace{0.2mm} \e^{\ii t A}\psi,
\ea
where the integral is defined as a Hilbert space-valued improper Riemann integral. Moreover, the strongly continuous unitary 1-parameter group generated by $A$  (through Stone's theorem) satisfies $\e^{\ii t A}=E(\e_t)$ for all $t\in\R$, where $\e_t\in C_b(\R)$ is given by $\e_t(x):=\e^{\ii tx}$ for all $t, x\in\R$.  
\en
\et

\bprf
See \cite{Hunziker} (and \cite{Jost1973}).
\eprf

In the following, if we need to display the dependence on $A$, we sometimes use the notation  $\chi(A):=E(\chi)$ or  $E_A(\chi):=E(\chi)$ for all $\chi\in\mB(\R)$. Moreover, for  all (not necessarily bounded) selfadjoint linear operators $A$ on $\mH$, we denote by $\dom(A)$ the domain of $A$.

\br
\label{rem:det}
Note that $C_0^\infty(\R)\subseteq\mS(\R)$ and that the resolution of the identity is already uniquely determined by its restriction to $C_0^\infty(\R)$ since $C_0^\infty(\R)$ is dense in $C_0(\R)$ with respect to the norm \eqref{supnrm} 
and since \eqref{ext:bnd} holds.
\er

In the following, if $\mH$ and $\mK$ are separable complex Hilbert spaces, $\mD\subseteq\mH$ and $A:\mH\to\mK$ a bounded linear operator, we set $A\mD:=\{A\Psi\,|\,\Psi\in\mD\}\subseteq\mK$.

A first application of this formalism, used in the foregoing sections, is the following standard property.

\bl[Identification]
\label{lem:unitary}
Let $\mH$ and $\mK$ be separable complex Hilbert spaces, let $A$ be a (not necessarily bounded) selfadjoint operator on $\mH$, and let $U:\mH\to\mK$ be unitary. Then:
\bn[label=(\alph*),ref={\it (\alph*)}]
\setlength{\itemsep}{0mm}
\item 
\label{lem:sa}
The operator $B:\dom(B)\subseteq\mK\to\mK$, defined by $\dom(B):=U\dom(A)$ and $B\Phi:=UAU^\ast\Phi$ for all $\Phi\in\dom(B)$, is selfadjoint.

\item 
\label{lem:EUEB}
The map $E^U:\mB(\R)\to\mL(\mK)$, defined by $E^U(\chi):=U E_A(\chi)U^\ast$ for all $\chi\in\mB(\R)$, satisfies $E^U=E_B$.
\en
\el

\bprf
\ref{lem:sa}\,
Since $B$ is densely defined and symmetric, and since, due to the standard criterion for selfadjointness,  $\ran(B\pm\ii 1)=U\ran(A\pm\ii 1)=U\mH=\mK$, 
the operator $B$ is selfadjoint. 

\vspace{1mm}
\ref{lem:EUEB}\,
Note that $E^U$ is a resolution of the identity. Moreover, $\dom(\bar E^U(\kappa_1)) = \ran(E^U(\kappa_{-1})) = U\ran(E_A(\kappa_{-1}))= U\dom(A)= \dom(B)$, where we used that $\dom(A)=\ran((A-\ii 1)^{-1})=\ran(E_A(\kappa_{-1}))$. 
Hence, for all $\Phi\in\dom(B)$, there exists $\Psi\in\mH$ such that $\Phi= E^U(\kappa_{-1})\Psi$ and, from \eqref{actionEx} and Theorem \ref{thm:spect} \ref{spect-uniq}, we get
\ba
\label{EbarUB}
\bar E^U(\kappa_1)\Phi
&= \bar E^U(\kappa_1)E^U(\kappa_{-1})\Psi\nonumber\\
&= E^U(\kappa_1\kappa_{-1})\Psi\nonumber\\
&= UE_A(\kappa_1\kappa_{-1})U^\ast\Psi\nonumber\\
&= U \bar E_A(\kappa_1)U^\ast E^U(\kappa_{-1})\Psi\nonumber\\
&=B\Phi.
\ea
The uniqueness property of the resolution of the identity from Theorem \ref{thm:spect} \ref{spect-uniq} then yields the conclusion.
\eprf

In the following, we also denote by $\zeta$ the operation of complex conjugation from Definition \ref{def:obs} \ref{def:obs-b} when applied to $\chi\in\ell^\infty(\R)$. Moreover, let $\ell^\infty(\R,\R)$, $C_0(\R,\R)$, and $\mB(\R,\R)$ stand for the bounded real-valued functions on $\R$, the continuous real-valued functions on $\R$ with compact support, and the smallest family of functions satisfying the conditions {\it (B1')} and {\it (B2')}, respectively, where {\it (B1')} and {\it (B2')} stand for the modified conditions {\it (B1)} and {\it (B2)} from Definition \ref{def:Bf} \ref{Bf:min} in which $\ell^\infty(\R)$, $C_0(\R)$, and $\mB(\R)$ have been replaced by $\ell^\infty(\R,\R)$, $C_0(\R,\R)$, and $\mB(\R,\R)$, respectively. 

\bl[Spectral identities]
\label{lem:SpecId}
Let $A\in\mL(\fH)$ be selfadjoint and $\chi\in\mB(\R)$. Then:
\bn[label=(\alph*),ref={\it (\alph*)}]
\setlength{\itemsep}{0mm}
\item
\label{lem:SpecId-a}
$\Gamma\chi(A)\Gamma=(\zeta\chi)(\Gamma A\Gamma)$

\item
\label{lem:SpecId-b}
$\chi(\psi(A))=(\chi\circ\psi)(A)$ for all $\psi\in\mB(\R,\R)$

\item
\label{lem:SpecId-c}
For all orthogonal families of orthogonal projections $\{P, Q\}\subseteq\mL(\fH)$  satisfying $[P,A]=[Q,A]=0$ and all $r,s\in\R$, we have
\ba
\label{lem:SpecId-cc}
\chi(r AP+s AQ)(P+Q)
=\chi(r A)P+\chi(s A)Q.
\ea

\item 
\label{lem:SpecId-d}
If $B\in\mL(\fH)$ is selfadjoint, $[A,B]=0$, and $\psi\in\mB(\R)$, we have
\ba
\label{chiApsiB}
[\chi(A),\psi(B)]
=0.
\ea
\en
\el

\vspace{-5mm}

\br
\label{rem:recurrent}
On the right hand side of the equations in Lemma \ref{lem:SpecId} \ref{lem:SpecId-a} and \ref{lem:SpecId-b}, we used that $\zeta\chi\in\mB(\R)$ for all $\chi\in\mB(\R)$ and $\chi\circ\psi\in\mB(\R)$ for all $\chi\in\mB(\R)$ and all $\psi\in\mB(\R,\R)$, respectively. In order to verify these properties, we make use of the following recurrent argument which we detail for the slightly more involved case \ref{lem:SpecId-b} only.
So, let $\chi\in C_0(\R)$ be fixed and set 
\ba
\mF_\chi
:=\{\psi\in\ell^\infty(\R,\R)\,|\, \chi\circ\psi\in\mB(\R)\}.
\ea
Since $\chi\circ\psi\in C_0(\R)$ for all $\psi\in C_0(\R,\R)$, we have  $C_0(\R,\R)\subseteq\mF_\chi$, \ie, $\mF_\chi$ satisfies {\it (B1')}. Moreover, let $(\psi_n)_{n\in\N}$ be a sequence in $\mF_\chi$ with $\Blim_{n\to\infty}\psi_n=\psi$ for some $\psi\in\ell^\infty(\R,\R)$. Since $\chi\circ\psi_n\in\mB(\R)$ for all $n\in\N$ and since $\Blim_{n\to\infty}\chi\circ\psi_n=\chi\circ\psi$ because $\chi\in C_0(\R)$, we get $\chi\circ\psi\in\mB(\R)$ (since $\mB(\R)$ satisfies {\it (B2)}),  \ie, $\mF_\chi$ satisfies {\it (B2')}. Hence, $\mB(\R,\R)\subseteq\mF_\chi$. 
Next, let $\psi\in\mB(\R,\R)$ be fixed and set
\ba
\mG_\psi
:=\{\chi\in\ell^\infty(\R)\,|\, \chi\circ\psi\in\mB(\R)\}.
\ea
Since $\mB(\R,\R)\subseteq\mF_\chi$ for all $\chi\in C_0(\R)$, the family $\mG_\psi$ satisfies {\it (B1)}. Moreover, let $(\chi_n)_{n\in\N}$ be a sequence in $\mG_\psi$ with $\Blim_{n\to\infty}\chi_n=\chi$ for some $\chi\in\ell^\infty(\R)$. Hence, we get $\Blim_{n\to\infty}\chi_n\circ\psi=\chi\circ\psi$ and $\chi\circ\psi\in\mB(\R)$, \ie, $\mG_\psi$ satisfies {\it (B2)}, too. Therefore, we arrive at $\mB(\R)\subseteq\mG_\psi$ for all $\psi\in\mB(\R,\R)$.
\er

\br
\label{rem:rA}
Due to the fact that  $E_A(1_{\spec(A)})=1$ for all selfadjoint operators $A\in\mL(\mH)$,
where $\spec(A)\in\mM(\R)$ 
is a (non-empty) compact subset of $\R$, \eqref{extC:prod} yields $A=E_A(\kappa_11_{\spec(A)})$ because $\kappa_11_{\spec(A)}\in\mB(\R)$ since  $1_{\spec(A)}=1_{[-a,a]}1_{\spec(A)}$ with $a:=\|A\|$ and since $\kappa_11_{[-a,a]}\in\mB(\R)$ (being the limit of a Borel convergent sequence of functions in $C_0(\R)$). Since $rA=E_A(\psi)$ with $\psi:=r\kappa_1 1_{\spec(A)}\in\mB(\R)$  for all $r\in\R$, Lemma \ref{lem:SpecId} \ref{lem:SpecId-b} yields
\ba
\label{rA}
\chi(rA)
=\chi_r(A),
\ea
where we define $\chi_r(x):=\chi(rx)$ for all $\chi\in\ell^\infty(\R)$ and all $r, x\in\R$, and $\chi_r\in\mB(\R)$ for all $\chi\in\mB(\R)$ and all $r\in\R$ (by arguing as in Remark \ref{rem:recurrent}).
Note that, using \eqref{ESchwartz}, a change of variables, Remark \ref{rem:det}, and a minimality type argument as in Remark \ref{rem:recurrent} (see also the proof of Lemma \ref{lem:SpecId} \ref{lem:SpecId-c} and \ref{lem:SpecId-d} below), \eqref{rA} also holds for unbounded selfadjoint operators.
\er

We next turn to the proof of Lemma \ref{lem:SpecId}. For illustration, we use the uniqueness type argument (as in the proof of Lemma \ref{lem:unitary} \ref{lem:EUEB}) for  \ref{lem:SpecId-a} and  \ref{lem:SpecId-b} and the minimality type argument (as in Remark \ref{rem:recurrent}) for  \ref{lem:SpecId-c} and  \ref{lem:SpecId-d}.

\vspace{5mm}

\bprf 
\ref{lem:SpecId-a}\,
Let us define the map $E^\Gamma:\mB(\R)\to\mL(\fH)$ by $E^\Gamma(\chi):=\Gamma E_A(\zeta\chi)\Gamma$ for all $\chi\in\mB(\R)$. In order to show that $E^\Gamma=E_{\Gamma A\Gamma}$, we first note that $E^\Gamma$ is a resolution of the identity because $\Gamma\in\bar\mL(\fH)$ is an antiunitary involution. Moreover, we have $\dom(\bar E^\Gamma(\kappa_1))= \ran(E^\Gamma(\kappa_{-1}))= \Gamma\ran(E_A(\kappa_{-1}))= \Gamma\dom(A)=\fH$.
Hence, for all $F\in\fH$, there exists $G\in\fH$ such that $F= E^\Gamma(\kappa_{-1})G$ and we compute that $\bar E^\Gamma(\kappa_1)F=\Gamma A\Gamma F$ as in \eqref{EbarUB}. Theorem \ref{thm:spect} \ref{spect-uniq} then implies that $\Gamma E_A(\zeta\chi)\Gamma=E_{\Gamma A\Gamma}(\chi)$ for all $\chi\in\mB(\R)$. 

\vspace{1mm}
\ref{lem:SpecId-b}\,
Let $\psi\in\mB(\R,\R)$ and define the map $E^\psi:\mB(\R)\to\mL(\fH)$ by $E^\psi(\chi):=E_A(\chi\circ\psi)$ for all $\chi\in\mB(\R)$. It then follows that $E^\psi$ is a resolution of the identity
and that $\dom(\bar E^\psi(\kappa_1))=\ran(E^\psi(\kappa_{-1}))=\ran(E_A(\kappa_{-1}\circ\psi))$. Moreover, since $\fH=\dom(A)=\ran((A-\ii 1)^{-1})$ and since $\ran((A-\ii 1)^{-1})\subseteq\ran(E_A(\kappa_{-1}\circ\psi))$ due to the fact that $(A-\ii 1)^{-1}=E_A((\kappa_1-\i)^{-1})=E_A((\kappa_{-1}\circ\psi)(\kappa_1-\i)^{-1}(\kappa_{-1}\circ\psi)^{-1})$ and $(\kappa_1-\i)^{-1}(\kappa_{-1}\circ\psi)^{-1}\in\mB(\R)$,
we get $\dom(\bar E^\psi(\kappa_1))=\fH$. 
Hence, for all $F\in\fH$, there exists $G\in\fH$ such that $F=E^\psi(\kappa_{-1})G$ and we compute that $\bar E^\psi(\kappa_1)F=\psi(A) F$.

\vspace{1mm} 
\ref{lem:SpecId-c}\,
Let $r,s\in\R$ be fixed and note that $rAP+sAQ\in\mL(\fH)$ is selfadjoint. Moreover, since $[AP, AQ]=0$, we have $\e^{\ii t(rAP+sAQ)}=\e^{\ii trAP}\e^{\ii tsAQ}=\e^{\ii trA}P+\e^{\ii tsA}Q+1-(P+Q)$ for all $t\in\R$, where we used the (in $\mL(\fH)$ converging) exponential series for the propagators. Hence, for all $\chi\in\mS(\R)$, \eqref{ESchwartz} leads to
\ba
\label{SpecId-1}
E_{rAP+sAQ}(\chi)
=E_{rA}(\chi)P+E_{sA}(\chi)Q+\chi(0)(1-(P+Q)).
\ea
Next, let $\chi\in C_0(\R)$ and let $(\chi_n)_{n\in\N}$ be a  sequence in $C_0^\infty(\R)\subseteq\mS(\R)$ which converges to $\chi$ with respect to the norm \eqref{supnrm} (such a sequence exists due to Remark \ref{rem:det}). Since 
 $E_{rAP+sAQ}(\chi)-(E_{rA}(\chi)P+E_{sA}(\chi)Q+\chi(0)(1-(P+Q)))
=E_{rAP+sAQ}(\chi-\chi_n)+E_{rA}(\chi_n-\chi)P+E_{sA}(\chi_n-\chi)Q+(\chi_n(0)-\chi(0))(1-(P+Q))$, \eqref{ext:bnd} yields the estimate $\|E_{rAP+sAQ}(\chi)-(E_{rA}(\chi)P+E_{sA}(\chi)P+\chi(0)(1-(P+Q)))\|\le 2(1+\|P\|+\|Q\|) |\chi-\chi_n|_\infty$. Hence, \eqref{SpecId-1} also holds for all $\chi\in C_0(\R)$. In order to show that \eqref{SpecId-1} holds for all $\chi\in\mB(\R)$, too, we set
\ba
\mF
:=\{\chi\in\mB(\R)\,|\, E_{rAP+sAQ}(\chi)=E_{rA}(\chi)P+E_{sA}(\chi)Q+\chi(0)(1-(P+Q))\},
\ea
and $\mF$ satisfies {\it (B1)} from Definition \ref{def:Bf} \ref{Bf:min}. Next, let $(\chi_n)_{n\in\N}$ be a sequence in $\mF$ with $\mB-\lim_{n\to\infty}\chi_n=\chi$ for some $\chi\in\ell^\infty(\R)$. Since $\chi_n\in\mB(\R)$ for all $n\in\N$ and since $\mB(\R)$ satisfies {\it (B2)}, we have $\chi\in\mB(\R)$. Moreover, \eqref{ext:s-conv} yields $\slim_{n\to\infty} E_{rAP+sAQ}(\chi_n)=E_{rAP+sAQ}(\chi)$, the analogous properties hold for $E_{rA}(\chi_n)P$ and $E_{sA}(\chi_n)Q$, and  $\lim_{n\to\infty}\chi_n(0)(1-(P+Q))F=\chi(0)(1-(P+Q))F$ for all $F\in\fH$ since $(\chi_n)_{n\in\N}$ is pointwise convergent. Hence, $\mF$ also satisfies {\it (B2)}. Therefore, we get $\mB(\R)\subseteq\mF$, \ie, \eqref{SpecId-1} holds for all $\chi\in\mB(\R)$. Multiplying \eqref{SpecId-1} from the right by $P+Q$ yields \eqref{lem:SpecId-cc}.

\vspace{1mm}
\ref{lem:SpecId-d}\,
For all $\chi\in\mS(\R)$, \eqref{ESchwartz} yields 
\ba
\label{EAB}
[E_A(\chi),B]
=0. 
\ea
Now, let $\chi\in C_0(\R)$ and let $(\chi_n)_{n\in\N}$ be a sequence in $C_0^\infty(\R)$ which converges to $\chi$ with respect to the norm \eqref{supnrm}. Hence, since $[E_A(\chi),B]=[E_A(\chi-\chi_n),B]$ for all $n\in\N$, \eqref{ext:bnd} yields $\|[E_A(\chi),B]\|\le 2\|B\| |\chi-\chi_n|_\infty$, \ie, \eqref{EAB} also holds for all $\chi\in C_0(\R)$. In order to show that \eqref{EAB} holds for all $\chi\in\mB(\R)$, too, we set
\ba
\mF
:=\{\chi\in\mB(\R)\,|\, [E_A(\chi),B]=0\},
\ea
and $\mF$ satisfies {\it (B1)}. In order to verify {\it (B2)}, let $(\chi_n)_{n\in\N}$ be a sequence in $\mF$ with $\Blim_{n\to\infty}\chi_n=\chi$ for some $\chi\in\ell^\infty(\R)$. Hence, $\chi\in\mB(\R)$ and, writing again $[E_A(\chi),B]=[E_A(\chi-\chi_n),B]$ for all $n\in\N$, we get $\|[E_A(\chi),B]F\|\le  \|E_A(\chi-\chi_n)B F\|+\|B\| \|E_A(\chi-\chi_n)F\|$ for all $F\in\fH$ and all $n\in\N$. Therefore, due to \eqref{ext:s-conv}, we get $\chi\in\mF$, \ie, $\mF$ also satisfies {\it (B2)} which implies $\mB(\R)\subseteq\mF$ (note that we did not use the selfadjointness of $B$ yet). In order to show \eqref{chiApsiB}, we get, as for \eqref{EAB}, for fixed $\chi\in\mB(\R)$ and all $\psi\in\mS(\R)$,
\ba
\label{EAEB}
[E_A(\chi),E_B(\psi)]
=0.
\ea
Then, proceeding as before, \eqref{EAEB} holds for all $\psi\in C_0(\R)$. Finally, setting $\mG:=\{\psi\in\mB(\R)\,|\, [E_A(\chi),E_B(\psi)]=0\}$, we verify that $\mG$ satisfies {\it (B1)} and {\it (B2)} and get $\mB(\R)\subseteq\mG$.
\eprf

\section{Matrix multiplication operators}
\label{app:mmo}

In this appendix, we derive the properties used in the foregoing sections of matrix multiplication operators in momentum space.

In the following, we make use of the notation introduced after Assumption \ref{ass:reg} and of the spectral theory from Appendix \ref{app:spectral}. Moreover, $\sinc\in C^\infty(\R)$ (the  infinitely differentiable complex valued functions on $\R$) stands for the usual cardinal sine function defined by $\sinc(x):=\sin(x)/x$ for all $x\in\R\setminus\{0\}$ and $\sinc(0):=1$. 

\bp[Functional calculus]
\label{prop:diag}
Let $u_0\in L^\infty(\T)$ and $u:=[u_1,u_2,u_3]\in  L^\infty(\T)^3$ satisfy Assumption \ref{ass:reg} \ref{ass:reg-1} and define $U\in\mL(\widehat\fH)$ by $U:=m[u_0]\sigma_0+m[u]\sigma$. Then:
\bn[label=(\alph*),ref={\it (\alph*)}]
\setlength{\itemsep}{0mm}
\item
\label{prop:diag-a}
For all $t\in\R$, we have 
\ba
\label{prop:prop}
\e^{\ii t U}
=m[p_0^t]\sigma_0+m[p^t]\sigma, 
\ea
where,  for all $t\in\R$, we define $p_0^t\in L^\infty(\T)$ and $p^t\in L^\infty(\T)^3$ by
\ba
\label{diag-p0}
p_0^t
&:=\exp\circ(\ii t u_0)\cos\circ(t|u|),\\
\label{diag-p}
p^t
&:=\ii t\exp\circ(\ii t u_0) \sinc\circ(t|u|) u.
\ea

\item
\label{prop:diag-b}
For all $\chi\in\mB(\R)$, we have 
\ba
\label{chiU}
\chi(U)=m[v_0]\sigma_0+m[v]\sigma,
\ea 
where $v_0\in L^\infty(\T)$ and $v\in L^\infty(\T)^3$ are given by
\ba
\label{fc1}
v_0
&:=\frac12\left(\chi\circ e_++\chi\circ e_-\right),\\
\label{fc2}
v
&:=\frac12\left(\chi\circ e_+-\chi\circ e_-\right) \tilde u,
\ea
and we recall that $e_\pm=u_0\pm |u|\in L^\infty(\T)$.
\en
\ep

\bprf
\ref{prop:diag-a}\,
We first note that, due to Assumption \ref{ass:reg} \ref{ass:reg-1} and Remark \ref{rem:sa}, we have $U^\ast=U$. For the following, let $t\in\R$ be fixed. Since $m[u_0]\sigma_0, m[u]\sigma\in\mL(\widehat\fH)$ and since $[m[u_0]\sigma_0, m[u_\alpha]\sigma_\alpha]=0$ for all $\alpha\in\l1,3\r$, we have
$\e^{\ii t U}=\e^{\ii t m[u_0]\sigma_0}\e^{\ii t m[u]\sigma}$ and $\e^{\ii t m[u]\sigma}=\lim_{N\to\infty} P_N$, where, for all $N\in\N$, we set $P_N:=\sum_{n\in\l0,N\r}(\ii t)^n/(n!) (m[u]\sigma)^n\in\mL(\widehat\fH)$ and the limit exists with respect to the uniform topology on $\mL(\widehat\fH)$. Moreover, since, due to  \eqref{prod}, we have $(m[u]\sigma)^{2n}=m[|u|^{2n}]\sigma_0$ and $(m[u]\sigma)^{2n+1}=m[|u|^{2n}u]\sigma$ for all $n\in\N$, we can write $P_N=C_N+S_N$, where $C_N, S_N\in\mL(\widehat\fH)$ are defined by $C_N:=\sum_{n\in\l0,N\r}(-1)^n t^{2n}/((2n)!) (m[|u|^{2n}]\sigma_0)$ and $S_N:=\ii \sum_{n\in\l0,N\r}(-1)^n t^{2n+1}/((2n+1)!) (m[|u|^{2n}u]\sigma)$ for all $N\in\N$. Since, for all $\Phi=\vi_1\oplus\vi_2\in\widehat\fH$ and all $N\in\N$, we have $\|m[\cos\circ(t|u|)]\sigma_0\Phi-C_N\Phi\|^2=\sum_{i\in\l1,2\r }\int_{-\pi}^\pi\rd k/(2\pi) |f_i^N(k)|^2$, where, for all $i\in\l1,2\r $ and all $N\in\N$, the function $f_i ^N\in\hat\fh$ is defined by
\ba
f_i^N
:=\cos\circ(t|u|)\vi_i-\sum_{n\in\l0,N\r}\frac{(-1)^n t^{2n}}{(2n)!} |u|^{2n} \vi_i,
\ea
and since, for all $i\in\l1,2\r $ and almost all $k\in\T$, we have that  $\lim_{N\to\infty}f_i^N(k)=0$ and $|f_i^N|^2\le \cosh^2(|t|\||u|\|_\infty)|\vi_i|^2\in L^1(\T)$, Lebesgue's dominated convergence theorem implies $\slim_{N\to\infty} C_N = m[\cos\circ(t|u|)]\sigma_0$. Moreover, since $(C_N)_{N\in\N}$ is a Cauchy sequence with respect to the uniform topology on $\mL(\widehat\fH)$, we get $\lim_{N\to\infty} C_N=m[\cos\circ(t|u|)]\sigma_0$ in $\mL(\widehat\fH)$. Finally, for all $\Phi=\vi_1\oplus\vi_2 \in\widehat\fH$ and all $N\in\N$, we have $\|m[\ii t \sinc\circ(t|u|) u]\sigma\Phi-S_N\Phi\|^2\le 3\sum_{i\in\l1,2\r}\sum_{\alpha\in\l1,3\r}\int_{-\pi}^\pi\rd k/(2\pi) |g_{i,\alpha}^N(k)|^2$, where, for all $i\in\l1,2\r$, all $\alpha\in\l1,3\r$, and all $N\in\N$, the function $g_{i,\alpha}^N\in\hat\fh$ is defined by $g_{i,\alpha}^N
:=\ii t\sinc\circ(t|u|)u_\alpha\vi_i-\ii \sum_{n\in\l0,N\r}(-1)^n t^{2n+1}/((2n+1)!) |u|^{2n}u_\alpha \vi_i$. Hence, the contribution $S_N$ can be treated analogously to $C_N$. 

\ref{prop:diag-b}\,
Let $\chi\in\mS(\R)$ and plug \eqref{prop:prop}-\eqref{diag-p} into \eqref{ESchwartz}. Then, for all $\Phi=\vi_1\oplus\vi_2 \in\widehat\fH$, we get
\ba
\label{EU}
E_U(\chi)\Phi
&=\frac{1}{\sqrt{2\pi}}\int_\R \rd t \hspace{0.5mm}
f_1^t\oplus f_2^t
+\frac{1}{\sqrt{2\pi}}\int_\R \rd t \hspace{0.5mm}
(g_{1,3}^t+g_{2,1}^t -\ii g_{2,2}^t)\oplus (g_{1,1}^t+\ii g_{1,2}^t -g_{2,3}^t),
\ea
where, for all $i\in\l1,2\r$ and all $\alpha\in\l1,3\r$, the maps $f_i:\R\to\hat\fh$ and $g_{i,\alpha}:\R\to\hat\fh$ are defined, for all $t\in\R$, by
\ba
f_i^t
&:=\hat\chi(t) \exp\circ(\ii t u_0)\cos\circ(t|u|)\vi_i,\\
g_{i,\alpha}^t
&:=\ii t \hat\chi(t) \exp\circ(\ii t u_0)\sinc\circ(t|u|)u_\alpha\vi_i,
\ea
and we note that $f_i, g_{i,\alpha}\in C(\R,\hat\fh)$ for all $i\in\l1,2\r$ and all $\alpha\in\l1,3\r$. 
Using the analogous arguments as the ones which lead to \eqref{intb1} in the proof of Proposition \ref{prop:V} \ref{prop:V-a}, 
we get
\ba
\label{intft}
\frac{1}{\sqrt{2\pi}}\int_\R \rd t\hspace{0.5mm} f_i^t
&=\frac12\hspace{0.5mm}(\chi\circ(u_0+|u|)+\chi\circ(u_0-|u|))\vi_i,\\
\label{intgt}
\frac{1}{\sqrt{2\pi}}\int_\R \rd t\hspace{0.5mm} g_{i,\alpha}^t
&=\begin{cases}
\displaystyle\frac12\hspace{0.5mm}(\chi\circ(u_0+|u|)-\chi\circ(u_0-|u|)) \frac{u_\alpha}{|u|} \vi_i, & \mbox{\textup{on} $\mZ_u^c$},\nonumber\\
\hfill 0, & \mbox{\textup{on} $\mZ_u$},
\end{cases}\\
&=\frac12\hspace{0.5mm}(\chi\circ(u_0+|u|)-\chi\circ(u_0-|u|)) \tilde u_\alpha\vi_i,
\ea 
where we used Euler's formula, the fact that Fourier transform is a bijection on $\mS(\R)$, and the notation \eqref{tildeui}. 
Hence, \eqref{chiU}-\eqref{fc2} holds for all $\chi\in\mS(\R)$. We next proceed by using the minimality type argument (as in Remark \ref{rem:recurrent}). To this end, let $\chi\in C_0(\R)$ and let $(\chi_n)_{n\in\N}$ be a sequence in $C_0^\infty(\R)$ which converges to $\chi$ with respect to the norm \eqref{supnrm}. 
Hence, we can write, for all $\Phi\in\widehat\fH$ and all $n\in\N$,
\ba 
\label{EC0}
E_U(\chi)\Phi-(m[v_0]\sigma_0+m[v]\sigma)\Phi 
&=E_U(\chi-\chi_n)\Phi+(m[v_0(\chi_n-\chi)]\sigma_0+m[v(\chi_n-\chi)]\sigma)\Phi,
\ea
where, for all $\chi\in\mB(\R)$, we denote \eqref{fc1} and \eqref{fc2} by $v_0(\chi)$ and $v(\chi)$, respectively. Due to \eqref{ext:bnd}, the first term on the right hand side of \eqref{EC0} is bounded by $\|E_U(\chi-\chi_n)\Phi\|\le |\chi-\chi_n|_\infty \|\Phi\|$ for all $\Phi\in\widehat\fH$ and all $n\in\N$. In order to estimate the second and third term in \eqref{EC0}, we note that $\|v_0(\chi_n-\chi)\|_\infty\le (\|(\chi_n-\chi)\circ(u_0+|u|)\|_{\infty}+\|(\chi_n-\chi)\circ(u_0-|u|)\|_{\infty})/2\le |\chi-\chi_n|_\infty$ for all $n\in\N$.
Similarly, for all $\alpha\in\l1,3\r$ and all $n\in\N$, we have $\|v_\alpha(\chi_n-\chi)\|_\infty \le  |\chi-\chi_n|_\infty$ since $\|\tilde u_\alpha\|_\infty\le 1$ for all $\alpha\in\l1,3\r$. Applying \eqref{Ubnd}, we get, for all $\Phi\in\widehat\fH$ and all $n\in\N$, 
\ba
\label{EC0-2est}
\|(m[v_0(\chi_n-\chi)]\sigma_0+m[v(\chi_n-\chi)]\sigma)\Phi\|
&\le\sum_{\alpha\in\l0,3\r }\|v_\alpha(\chi_n-\chi)\|_\infty \|\Phi\|\nonumber\\
&\le 4 |\chi-\chi_n|_\infty \|\Phi\|.
\ea
Hence, 
\eqref{chiU}-\eqref{fc2} also holds for all $\chi\in C_0(\R)$ (see Remark \ref{rem:recurrent} for the composition of Borel functions).
In order to show that \eqref{chiU}-\eqref{fc2} also holds for all $\chi\in\mB(\R)$, we again use the minimality type argument (as in Remark \ref{rem:recurrent}). To this end, we set
\ba
\mF:=\{\chi\in\mB(\R)\,|\, E_U(\chi)=m[v_0]\sigma_0+m[v]\sigma \mbox{ with \eqref{fc1} and \eqref{fc2}}\},
\ea
and $\mF$ satisfies {\it (B1)}.  Next, let $(\chi_n)_{n\in\N}$ be a sequence in $\mF$ with $\Blim_{n\to\infty}\chi_n=\chi$ for some $\chi\in\ell^\infty(\R)$ and, hence, $\chi\in\mB(\R)$. Making again the decomposition \eqref{EC0}, \eqref{ext:s-conv} applied to the first term on the right hand side of \eqref{EC0} yields $\slim_{n\to\infty} E_U(\chi-\chi_n)=0$. As for the second term in \eqref{EC0}, we have, for all $\Phi=\vi_1\oplus\vi_2 \in\widehat\fH$ and all $n\in\N$, that $\|m[v_0(\chi_n-\chi)]\sigma_0\Phi\|^2
\le\sum_{i\in\l1,2\r }\sum_{\sigma\in\{\pm 1\}}\int_{-\pi}^\pi\rd k/(2\pi)|f_{i,\sigma}^n(k)|^2/2$, where, for all $i\in\l1,2\r $, all $\sigma\in\{\pm1\}$, and all $n\in\N$, the function $f_{i,\sigma}^n\in\hat\fh$ is given by 
\ba
f_{i,\sigma}^n
:=((\chi_n-\chi)\circ(u_0+\sigma|u|))\vi_i.
\ea
Since  the sequence $(\chi_n)_{n\in\N}$ is Borel convergent to $\chi$, we get, for all $i\in\l1,2\r $, all $\sigma\in\{\pm1\}$, all $n\in\N$, and almost all $k\in\T$, that $\lim_{n\to\infty}f_{i,\sigma}^n(k)=0$ and $|f_{i,\sigma}^n|^2\le 4C^2 |\vi_i|^2\in L^1(\T)$, where the constant $C>0$ stems from Definition \ref{def:Bf} \ref{Bf:Bc}. 
Hence, Lebesgue's dominated convergence theorem implies that  $\slim_{n\to\infty} m[v_0(\chi_n-\chi)]\sigma_0=0$. Similarly, as for the third term on the right hand side of \eqref{EC0}, we have, for all $\Phi=\vi_1\oplus\vi_2 \in\widehat\fH$ and all $n\in\N$, that $\|m[v(\chi_n-\chi)]\sigma\Phi\|^2
\le(3/2)\sum_{i\in\l1,2\r }\sum_{\alpha\in\l1,3\r}\sum_{\sigma\in\{\pm 1\}}\int_{-\pi}^\pi\rd k/(2\pi)|g_{i,\alpha, \sigma}^n(k)|^2$, where, for all $i\in\l1,2\r $, all $\alpha\in\l1,3\r$, all $\sigma\in\{\pm1\}$, and all $n\in\N$, the function $g_{i,\alpha,\sigma}^n\in\hat\fh$ is given by $g_{i,\alpha,\sigma}^n
:=[(\chi_n-\chi)\circ(u_0+\sigma|u|)]\tilde u_\alpha\vi_i$. We again get, for all $i\in\l1,2\r $, all $\alpha\in\l1,3\r$, all $\sigma\in\{\pm1\}$, all $n\in\N$, and almost all $k\in\T$, that $\lim_{n\to\infty}g_{i,\alpha,\sigma}^n(k)=0$ and $|g_{i,\alpha,\sigma}^n|^2\le 4C^2 |\vi_i|^2\in L^1(\T)$, where we used that $\|\tilde u_\alpha\|_\infty\le 1$ for all $\alpha\in\l1,3\r$. Hence, Lebesgue's dominated convergence theorem also implies that  $\slim_{n\to\infty} m[v(\chi_n-\chi)]\sigma=0$. Therefore, \eqref{chiU}-\eqref{fc2} holds for $\chi$, \ie, $\mF$ also satisfies {\it (B2)}, and we get $\mB(\R)\subseteq\mF$.
\eprf

\br  
\label{rem:diag}
Instead of applying the explicit form of the propagator \eqref{prop:prop}-\eqref{diag-p}, we can also diagonalize $U$ with the help of its eigenvalue functions $e_\pm\in L^\infty(\T)$ and its (not yet normalized) eigenvector functions $\Phi_\pm\in\widehat\fH$ given by 
\ba 
\Phi_\pm
&:=(u_3\pm |u|)\oplus (u_1+\ii u_2),\\
e_\pm
&=u_0\pm |u|,
\ea
\ie, we have $U\Phi_\pm=m[e_\pm]\sigma_0\Phi_\pm$.
\er

In the following, we denote by $L^2(\T, \C^2)$ the space of vector-valued functions $\T\to\C^2$ whose entry functions belong to $L^2(\T)$. Analogously, $L^\infty(\T, \C^{2\times 2})$ stands for the space of matrix-valued functions $\T\to\C^{2\times 2}$ whose entry functions belong to $L^\infty(\T)$. Moreover, if $f:D\to\C$ is a function defined on $D\subseteq\R$, we use the notation $f(M):=\{f(x)\,|\, x\in M\}$ for all $M\subseteq D$ and, for all $Y\subseteq\C$, the preimage of $Y$ under $f$ is denoted by $f^{-1}(Y):=\{x\in D\,|\, f(x)\in Y\}$. 

The following proposition provides us with a useful sufficient condition for a matrix multiplication operator to be absolutely continuous on some spectral domain.

\bp[Absolute continuity]
\label{prop:ac}
Let $u_0\in L^\infty(\T)$ and $u=[u_1,u_2,u_3]\in  L^\infty(\T)^3$ satisfy Assumption \ref{ass:reg} \ref{ass:reg-1} and \ref{ass:reg-3} and define $U\in\mL(\widehat\fH)$ by $U:=m[u_0]\sigma_0+m[u]\sigma$. Moreover, let $M\in\mM(\R)$ and let Assumption \ref{ass:nullsets} \ref{ass:reg-4} and \ref{ass:reg-5} hold. Then, 
\ba
\ran(1_M(U))
\subseteq\ran(1_{ac}(U)).
\ea
\ep

\br
\label{rem:spec}
Since we know (see \cite{HW1996} for example) that, under the assumptions of Proposition \ref{prop:ac}, the property
$\spec(U)=\bigcup_{k\in\T}\spec([U](k))$ holds, where $[U]\in L^\infty(\T, \C^{2\times 2})$ is given by $\Sigma U\Sigma^\ast=m[[U]]$, where the natural unitary identification operator $\Sigma\in\mL(\widehat\fH, L^2(\T,\C^2))$ reads as $(\Sigma\Phi)(k):=[\vi_1(k),\vi_2(k)]\in\C^2$ for all $\Phi=\vi_1\oplus \vi_2\in\widehat\fH$ and almost all $k\in\T$, and where, for all $A\in L^\infty(\T, \C^{2\times 2})$, the multiplication operator $m[A]\in\mL( L^2(\T,\C^2))$ is given by $(m[A][\vi_1,\vi_2])(k):=A(k)[\vi_1(k),\vi_2(k)]$ for all $[\vi_1,\vi_2]\in L^2(\T,\C^2)$ and almost all $k\in\T$, we get 
\ba
\label{rem:spec-1}
\spec(U)
&=\bigcup_{k\in\T}\{e_+(k), e_-(k)\}\nonumber\\
&=\ran(e_+)\cup\ran(e_-).
\ea
Hence, if the set $M\in\mM(\R)$ from Proposition \ref{prop:ac} has the property $\spec(U)\subseteq M$, we get $e_\kp^{-1}(M)=\T$ for all $\kp\in\{\pm\}$. Moreover, $\ran(1_M(U))=\widehat\fH$ since $E_U(1_{\spec(U)})=1$ (see Remark \ref{rem:rA}). Therefore, if Assumption \ref{ass:nullsets} \ref{ass:reg-4} and \ref{ass:reg-5} hold, Proposition \ref{prop:ac} yields that $U$ is absolutely continuous,  \ie, that $1_{ac}(U)=1$.
\er

For the following, recall the definitions \eqref{Zu}-\eqref{def:Zpm}.

\vspace{5mm}

\bprf 
Since $U^\ast=U\in\mL(\widehat\fH)$ and since we know that
\ba
\ran(1_{ac}(U))
=\{\Phi\in\widehat\fH\,|\, \mbox{$E_U(1_A)\Phi=0$ for all $A\in\mM(\R)$ with $|A|=0$}\},
\ea
we want to show that $E_U(1_A)E_U(1_M)\Psi=E_U(1_{A\cap M})\Psi=0$ for all $A\in\mM(\R)$ with $|A|=0$ and all $\Psi\in\widehat\fH$. To this end, using Proposition \ref{prop:diag} \ref{prop:diag-b}, we write that, for all $\Psi=\psi_1\oplus  \psi_2\in\widehat\fH$ and all $A\in\mM(\R)$, 
\ba
\label{intbnd}
\|E_U(1_{A\cap M})\Psi\|^2
&=(\Psi, E_U(1_{A\cap M})\Psi)\nonumber\\
&\le\frac12\sum_{\kappa\in\{\pm\}}\sum_{i\in\l1,2\r }\int_{e_\kappa^{-1}({A\cap M})}\frac{\rd k}{2\pi}\hspace{1.5mm} |\psi_i(k)|^2\nonumber\\
&\quad+\frac12\sum_{\kappa\in\{\pm\}}\sum_{\alpha\in\l1,2\r } \sum_{\substack{i,j\in\l1,2\r \\ i\neq j}}\int_{e_\kappa^{-1}({A\cap M})}\frac{\rd k}{2\pi}\hspace{1.5mm} |\tilde u_\alpha(k)| |\psi_i(k)| |\psi_j(k)| \nonumber\\
&\quad+\frac12\sum_{\kappa\in\{\pm\}}\sum_{i\in\l1,2\r }\int_{e_\kappa^{-1}({A\cap M})}\frac{\rd k}{2\pi}\hspace{1.5mm} |\tilde u_3(k)| |\psi_i(k)|^2,
\ea
where, due to Assumption \ref{ass:reg} \ref{ass:reg-3}, we have $e_\pm\in C(\T)$ and, hence, $e_\pm^{-1}(A\cap M)\in\mM(\T)$ for all $A\in\mM(\R)$.
Moreover, all the integrals on the right hand side of \eqref{intbnd} exist since $\vi_i\in\hat\fh$ for all $i\in\l1,2\r$ and $|\tilde u_\alpha|_\infty\le 1$ for all $\alpha\in\l1,3\r$.
In order to make the left hand side of \eqref{intbnd} vanish for all $A\in\mM(\R)$ with $|A|=0$ and all $\Psi\in\widehat\fH$, it is sufficient to show that $|e_\pm^{-1}(A\cap M)|=0$ for all $A\in\mM(\R)$ with $|A|=0$. In order to do so, let us stick to the case of $e_+$ in the following (the case of $e_-$ being strictly analogous). Let us start off by making the decompositions $\T=\mZ_u\cup\mZ_u^c$ and $\mZ_u^c=\mZ_+\cup (\mZ_u^c\setminus\mZ_+)$ (recall that $\mZ_+\subseteq\mZ_u^c$). Hence, writing  $A':=A\cap M$ and $e_+^{-1}(A')=e_+^{-1}(A')\cap\T$, we get 
\ba
\label{dec-1}
e_+^{-1}(A')
=(e_+^{-1}(A')\cap\mZ_u)
\cup (e_+^{-1}(A')\cap\mZ_+)
\cup(e_+^{-1}(A')\cap (\mZ_u^c\setminus\mZ_+)).
\ea
Moreover, since $\mZ_u^c=(\mZ_u^c\cap\{-\pi\})\cup (\mZ_u^c\cap\{\pi\})\cup(\mZ_u^c\cap\mathring\T)$, where $\mathring\T:=(-\pi,\pi)$, the last term on the right hand side of \eqref{dec-1} has the form
\ba
\label{dec-2}
e_+^{-1}(A')\cap (\mZ_u^c\setminus\mZ_+)
&=(e_+^{-1}(A')\cap ((\mZ_u^c\cap\{-\pi\})\setminus\mZ_+))
\cup(e_+^{-1}(A')\cap ((\mZ_u^c\cap\{\pi\})\setminus\mZ_+))\nonumber\\
&\quad\cup (e_+^{-1}(A')\cap B),
\ea
where we set $B:=(\mZ_u^c\cap\mathring\T)\setminus\mZ_+$.
Denoting the restriction of  $e_+$ to $\mZ_u^c\cap\mathring\T$ by $f$, we have 
\ba
\label{intrsc}
B
=\{k\in \mZ_u^c\cap\mathring\T\,|\,f'(k)\neq 0\},
\ea
and $B$ is open (in $\R$) since $B$ is open relative to $\mZ_u^c\cap\mathring\T$ and since $\mZ_u^c\cap\mathring\T$ is open.
Since we know that there exists a countable family of compact intervals $\{I_n\}_{n\in\N}$ in $\R$ satisfying $\mathring I_n\cap \mathring I_{n'}=\emptyset$ for all $n,n'\in\N$ with $n\neq n'$ and $B=\bigcup_{n\in\N}I_n$,  
the last term in \eqref{dec-2} reads
\ba
e_+^{-1}(A')\cap B
=\bigcup_{n\in\N} (e_+^{-1}(A') \cap I_n).
\ea
Denoting by $g\in C^1(B)$ the restriction of $e_+$ to $B$, we have $g'(k)\neq 0$ for all $k\in B$. Hence, for all $k\in B$, the inverse function theorem guarantees the existence of an open set $U_k\subseteq B$ with $k\in U_k$ and of an open set $V_{g(k)}\subseteq\R$ with $g(k)\in V_{g(k)}$ such that the restriction of $g$ to $U_k$, denoted by $g_k$,  is a bijection between $U_k$ and $V_{g(k)}$. Moreover, since, for all $n\in\N$, we have $I_n\subseteq\bigcup_{k\in I_n}U_k$ and since $I_n$ is compact, there exists $N_n\in\N$ and $\{k_{n,1},\ldots, k_{n,N_n}\}\subseteq I_n$ such that $I_n\subseteq\bigcup_{m\in\l1,N_n\r} U_{k_{n,m}}$ which implies
\ba
\label{dec-3}
e_+^{-1}(A')\cap B
\subseteq\bigcup_{n\in\N}\hspace{0.5mm}\bigcup_{m\in\l1,N_n\r} (e_+^{-1}(A') \cap U_{k_{n,m}}).
\ea
Since, for all $n\in\N$ and all $m\in \l1,N_n\r$, it holds that 
\ba
\label{dec-4}
e_+^{-1}(A') \cap U_{k_{n,m}}
&=\{k\in U_{k_{n,m}}\,|\, g_{k_{n,m}}(k)\in A'\cap V_{g(k_{n,m})}\}\nonumber\\
&=g_{k_{n,m}}^{-1}(A'\cap V_{g(k_{n,m})}), 
\ea
the properties \eqref{dec-1}-\eqref{dec-2} and \eqref{dec-3}-\eqref{dec-4} yield $e_+^{-1}(A') \subseteq \bigcup_{i\in\l1,5\r} K_i$, where, for all $i\in\l1,5\r$, the sets $K_i\in\mM(\T)$ are defined by
\ba
\label{ss1}
K_1
&:=e_+^{-1}(A')\cap\mZ_u,\\
\label{ss2}
K_2
&:= e_+^{-1}(A')\cap\mZ_+,\\
\label{ss3}
K_3
&:=e_+^{-1}(A')\cap ((\mZ_u^c\cap\{-\pi\})\setminus\mZ_+),\\
\label{ss4}
K_4
&:=e_+^{-1}(A')\cap ((\mZ_u^c\cap\{\pi\})\setminus\mZ_+),\\
\label{ss5}
K_5
&:=\bigcup_{n\in\N}\hspace{0.5mm}\bigcup_{m\in\l1,N_n\r}g_{k_{n,m}}^{-1}(A'\cap V_{g(k_{n,m})}).
\ea
Using Assumption \ref{ass:nullsets} \ref{ass:reg-4} and \ref{ass:reg-5}, we get $|K_1|=|e_+^{-1}(A)\cap(\mZ_u\cap e_+^{-1}(M))|\le |\mZ_u\cap e_+^{-1}(M)|=0$ and $|K_2|\le |\mZ_+\cap e_+^{-1}(M)|=0$, respectively.
Moreover, we have $\card(K_i)\in\{0,1\}$ for all $i\in\l3,4\r$. Hence, it remains to estimate \eqref{ss5}. We first note that, since, for all $n\in\N$ and all $m\in\l1,N_n\r$, the set $V_{g(k_{n,m})}$ is open, there exists again a countable family of compact intervals $\{J_{n,m,p}\}_{p\in\N}$ in $\R$ satisfying $\mathring J_{n,m,p}\cap \mathring J_{n,m,p'}=\emptyset$ for all $p,p'\in\N$ with $p\neq p'$ and $V_{g(k_{n,m})}=\bigcup_{p\in\N}J_{n,m,p}$.
Furthermore, since $A'\in\mM(\R)$ has the property $|A'|\le |A|=0$, we know that, for all $\veps>0$,  there exists a countable family of compact intervals $\{L_q\}_{q\in\N}$ in $\R$ such that $A'\subseteq\bigcup_{q\in\N}L_q$ and $\sum_{q\in\N}|L_q|<\veps$.
Hence, for all $n\in\N$ and all $m\in\l1,N_n\r$, we get
\ba
\label{incl-K}
g_{k_{n,m}}^{-1}(A'\cap V_{g(k_{n,m})})
\subseteq \bigcup_{p, q\in\N} g_{k_{n,m}}^{-1}(J_{n,m,p}\cap L_q).
\ea
Furthermore, the inverse function theorem also guarantees that, for all $n\in\N$ and all $m\in\l1,N_n\r$,  the inverse of $g_{k_{n,m}}$, denoted  by $h_{k_{n,m}}$,  satisfies $h_{k_{n,m}}\in C^1(V_{g(k_{n,m})})$ and, hence, $h_{k_{n,m}}$ is Lipschitz continuous on the compact interval $J_{n,m,p}$ with a Lipschitz constant $C_{n,m,p}>0$. Therefore, for all $n\in\N$, all $m\in\l1,N_n\r$, and all $p, q\in\N$, we get
\ba
\sup_{x,y\in J_{n,m,p}\cap L_q} |h_{k_{n,m}}(x)-h_{k_{n,m}}(y)|
\le C_{n,m,p} |L_q|, 
\ea
and the set $g_{k_{n,m}}^{-1}(J_{n,m,p}\cap L_q)=h_{k_{n,m}}(J_{n,m,p}\cap L_q)\in\mM(\T)$ is contained in a compact interval of length $C_{n,m,p}|L_q|$ (and $C_{n,m,p}$ is independent of $\veps$). Since $\sum_{q\in\N}|L_q|<\veps$, we get $|g_{k_{n,m}}^{-1}(J_{n,m,p}\cap L_q)|=0$ for all $n\in\N$, all $m\in\l1,N_n\r$, and all $p, q\in\N$.
Hence, using \eqref{ss5}-\eqref{incl-K}, we arrive at $|K_5|=0$.
\eprf

\section{Real trigonometric polynomials}
\label{app:TP}

In this appendix, we carry out the computations of the squares in $TP(\T)$ used in the foregoing sections (for more information on the structure of this ring, see \cite{Picavet2003} for example). To this end, let $\nu\in\N$ and,  for all $\alpha\in\l0,3\r$, let $u_\alpha\in L^\infty(\T)$ be given by 
\ba
\label{app:TP-1}
u_\alpha
=\begin{cases}
\hfill -2\sum_{n\in\l1,\nu\r }c_{\alpha,n}\sin(n\hspace{0.5mm}\cdot), & \alpha\in\l0,2\r ,\\
c_{3,0}+2\sum_{n\in\l1,\nu\r } c_{3,n}\cos(n\hspace{0.5mm}\cdot), & \alpha=3,
\end{cases}
\ea
where $c_{\alpha, n}\in\R$ for all $\alpha\in\l0,2\r$ and all $n\in\l1,\nu\r$ and $c_{3, n}\in\R$ for all $n\in\l0,\nu\r$ (recall that in \eqref{Pauli-cf}, we have $\nu\in\l1,n_S\r$). 

\bl[Squares]
\label{lem:sqr}
For all $\alpha\in\l0,3\r$, the squares of $u_\alpha\in L^\infty(\T)$ from  \eqref{app:TP-1} read as
\ba
\label{app:TP-2}
u_\alpha^2
=a_{\alpha,0}+2\hspace{-1mm}\sum_{m\in\l1,2\nu\r } a_{\alpha,m}\cos(m\hspace{0.5mm}\cdot),
\ea
where $a_{\alpha,m}\in\R$ for all $\alpha\in\l0,3\r$ and all $m\in\l0,2\nu\r$. Moreover, setting $b_{\alpha,m}:=[a_{\alpha,m}, a_{3,m}]\in\R^2$ for all $\alpha\in\l0,2\r$ and all $m\in\l0,2\nu\r$, we have:
\bn[label=(\alph*),ref={\it (\alph*)}]
\setlength{\itemsep}{0mm}
\item
\label{lem:sqr-1}
For $\nu=1$, for all $\alpha\in\l0,2\r$ and all $m\in\l0,2\r$, 
\ba
\label{sqr-1}
b_{\alpha,m}
=\begin{cases}
\hfill[2 c_{\alpha,1}^2, c_{3,0}^2+2c_{3,1}^2], & m=0,\\
\hfill[0, 2c_{3,0}c_{3,1}], & m=1,\\
\hfill[-c_{\alpha,1}^2, c_{3,1}^2], & m=2.
\end{cases}
\ea

\item
\label{lem:sqr-2}
For $\nu=2$, for all $\alpha\in\l0,2\r$ and all $m\in\l0,4\r$, 
\ba
b_{\alpha,m}
=\begin{cases}
\hfill[2 (c_{\alpha,1}^2+c_{\alpha,2}^2), c_{3,0}^2+2(c_{3,1}^2+c_{3,2}^2)], & m=0,\\
\hfill[2c_{\alpha,1}c_{\alpha,2}, 2(c_{3,0}c_{3,1}+c_{3,1}c_{3,2})], & m=1,\\
\hfill[-c_{\alpha,1}^2, 2c_{3,0}c_{3,2}+c_{3,1}^2], & m=2,\\
\hfill[-2c_{\alpha,1}c_{\alpha,2}, 2c_{3,1}c_{3,2}], & m=3,\\
\hfill[-c_{\alpha,2}^2, c_{3,2}^2], & m=4.
\end{cases}
\ea

\item
\label{lem:sqr-ge3}
For all $\nu\ge 3$, all $\alpha\in\l0,2\r$ and all $m\in\l0,2\nu\r$, 
\ba
\label{sqr-3}
b_{\alpha,m}
=\begin{cases}
\hfill[2 \sum_{n\in\l1,\nu\r}c_{\alpha,n}^2, c_{3,0}^2+2\sum_{n\in\l1,\nu\r}c_{3,n}^2], & m=0,\\
\hfill[2\sum_{n\in\l1,\nu-1\r}c_{\alpha,n}c_{\alpha,n+1}, 2(c_{3,0}c_{3,1}+\sum_{n\in\l1,\nu-1\r}c_{3,n}c_{3,n+1})], & m=1,\\
\hfill[2\sum_{n\in\l1,\nu-m\r}c_{\alpha,n}c_{\alpha,m+n}-\sum_{n\in\l1,m-1\r}c_{\alpha,n}c_{\alpha,m-n},\\
\hfill 2(c_{3,0}c_{3,m}+\sum_{n\in\l1,\nu-m\r}c_{3,n}c_{3,m+n})+\sum_{n\in\l1,m-1\r}c_{3,n}c_{3,m-n}], & m\in\{2,\nu-1\},\\
\hfill[-\sum_{n\in\l1,\nu-1\r}c_{\alpha,n}c_{\alpha,\nu-n}, 2 c_{3,0}c_{3,\nu}+\sum_{n\in\l1,\nu-1\r}c_{3,n}c_{3,\nu-n}],& m=\nu,\\
\hfill[-\sum_{n\in\l m-\nu,\nu\r}c_{\alpha,n}c_{\alpha,m-n}, \sum_{n\in\l m-\nu,\nu\r}c_{3,n}c_{3,m-n}], & m\in\{\nu+1,2\nu\}.
\end{cases}
\ea
\en
\el

\bprf
Note that $2\sin(x)\sin(y)=\cos(x-y)-\cos(x+y)$ and $2\cos(x)\cos(y)=\cos(x-y)+\cos(x+y)$ for all $x,y\in\R$ 
and that, for all $\nu\ge 2$ and all $\{b_{i,j}\}_{i,j\in\l 1,\nu\r}\subseteq\R$, we have 
\ba
\label{trigo-diff}
\sum_{i,j\in \l 1,\nu\r}b_{i,j}
&=\sum_{\substack{m\in\l 0,\nu-1\r\\i\in\l m+1,\nu\r}}b_{i,i-m}
+\sum_{\substack{m\in\l-\nu+1,-1\r\\i\in\l 1,\nu+m\r}}b_{i,i-m},\\
\label{trigo-sum}
&=\sum_{\substack{m\in\l2,\nu+1\r\\i\in\l1,m-1\r}}b_{i,m-i}
+\sum_{\substack{m\in\l\nu+2,2\nu\r\\i\in\l m-\nu,\nu\r}}b_{i,m-i}.
\ea
Squaring \eqref{app:TP-1} and applying \eqref{trigo-diff}-\eqref{trigo-sum} to the terms in the foregoing trigonometric expressions whose arguments are differences ($i-(i-m)=m$) and sums ($i+(m-i)=m$), respectively, we arrive at \eqref{sqr-1}-\eqref{sqr-3}.
\eprf

\section{Heat flux contributions}
\label{app:cntrJac}

In this appendix, we collect the explicit expressions for the contributions to $J_{ac}$ appearing in the proof of Theorem \ref{thm:hf} \ref{thm:hf-2}. 

\bl[Expansion]
\label{lem:exp}
Let the assumptions of Theorem  \ref{thm:hf} hold. Then, for all $i\in\l1,8\r$, the functions $\mu_i\in L^\infty(\T)$ appearing in \eqref{Jac5} are given by
\ba
\label{A1}
\mu_1
&:=\sum_{\substack{n\in\l1,\nu\r \\l\in\l0,\nu-n\r}} n \sin(n\hspace{0.5mm}\cdot)\hspace{0.5mm}(c_{0,l}c_{0,n+l}+c_{1,l} c_{1,n+l}+c_{2,l}c_{2,n+l}+c_{3,l}c_{3,n+l}),\\
\label{A2}
\mu_2
&:=-\sum_{n,m\in\l1,\nu\r } n \sin((n+m)\hspace{0.5mm}\cdot) \hspace{0.5mm}(c_{0,n}c_{0,m}+c_{1,n}c_{1, m} +c_{2,n}c_{2,m}-c_{3,n}c_{3,m}),\\
\label{A3}
\mu_3
&:=-2\sum_{\substack{n,m\in\l1,\nu\r \\l\in\l0,\nu-n\r}} n \sin(n\hspace{0.5mm}\cdot) \sin(m\hspace{0.5mm}\cdot)\hspace{0.5mm}
c_{1,m}(c_{0,l}c_{1,n+l}+c_{0,n+l}c_{1,l}-c_{2,l}c_{3,n+l}-c_{2,n+l}c_{3,l}),\\
\label{A4}
\mu_4
&:=2\sum_{n,m,l\in\l1,\nu\r } n \sin((n+l)\hspace{0.5mm}\cdot) \sin(m\hspace{0.5mm}\cdot)\hspace{0.5mm}c_{1,m}(c_{0,n}c_{1,l}+c_{0,l}c_{1,n}+c_{2,n}c_{3,l}-c_{2,l}c_{3,n}),
\ea

\ba
\label{A5}
\mu_5
&:=-2\sum_{\substack{n,m\in\l1,\nu\r\\l\in\l0,\nu-n\r}} n \sin(n\hspace{0.5mm}\cdot) \sin(m\hspace{0.5mm}\cdot)\hspace{0.5mm}c_{2,m}(c_{0,l}c_{2,n+l}+c_{0,n+l}c_{2,l}+c_{1,l}c_{3,n+l}+c_{1,n+l}c_{3,l}),\\
\label{A6}
\mu_6
&:=2\sum_{n,m,l\in\l1,\nu\r } n \sin((n+l)\hspace{0.5mm}\cdot) \sin(m\hspace{0.5mm}\cdot)\hspace{0.5mm}c_{2,m}(c_{0,n}c_{2,l}+c_{0,l}c_{2,n} -c_{1,n}c_{3,l}+c_{1,l}c_{3,n}),\\
\label{A7}
\mu_7
&:=-\sum_{\substack{n\in\l1,\nu\r\\l\in\l0,\nu-n\r}} n \cos(n\hspace{0.5mm}\cdot)\hspace{0.5mm}c_{3,0}(c_{0,l}c_{3,n+l}-c_{0,n+l}c_{3,l} +c_{1,l}c_{2,n+l}-c_{1,n+l}c_{2,l})\nonumber\\
&\hspace{5.5mm}-2\sum_{\substack{n,m\in\l1,\nu\r\\l\in\l0,\nu-n\r}} n \cos(n\hspace{0.5mm}\cdot)  \cos(m\hspace{0.5mm}\cdot) \hspace{0.5mm}c_{3,m}(c_{0,l}c_{3,n+l}-c_{0,n+l}c_{3,l} +c_{1,l}c_{2,n+l}-c_{1,n+l}c_{2,l}),\\
\label{A8}
\mu_8
&:=\sum_{n, m\in\l1,\nu\r } n \cos((n+m)\hspace{0.5mm}\cdot)\hspace{0.5mm}c_{3,0}(c_{0,n}c_{3,m}+c_{0,m}c_{3,n} -c_{1,n}c_{2,m}+c_{1,m}c_{2,n})\nonumber\\
&\hspace{5.5mm}+2\sum_{n,m\hspace{-0.5mm},l\in\l1,\nu\r }n \cos((n+l)\hspace{0.5mm}\cdot) \cos(m\hspace{0.5mm}\cdot)\hspace{0.5mm}c_{3,m}(c_{0,n}c_{3,l}+c_{0,l}c_{3,n} -c_{1,n}c_{2,l}+c_{1,l}c_{2,n}).
\ea
\el

\bprf
Noting that, for all $x\in\Z$ and all $\alpha\in\l1,2\r$, 
\ba
\eta_{0,x}
&=-\frac12\Int\sin(kx)(\rho_+(k)+\rho_-(k)),\\
\eta_{\alpha,x}
&=-\frac12\Int\sin(kx)\tilde u_\alpha(k)(\rho_+(k)-\rho_-(k)),\\
\eta_{3,x}
&=\frac12\Int\cos(kx)\tilde u_3(k)(\rho_+(k)-\rho_-(k)),
\ea
plugging \eqref{imtau1} for $G(-n,z,-n-z)$ and $G(-n+z,z,-n)$ into \eqref{Jac4}, using \eqref{Pauli-cf}, and separating the terms with respect to $\rho_+$ and $\rho_-$, we arrive at \eqref{A1}-\eqref{A8}.
\eprf

\section{Hamiltonian densities}
\label{app:Fermi-Pauli}
In this appendix, we display the selfdual second quantization of the local first, second and third Pauli coefficient of $H$ in the fermionic and the spin picture (the selfdual second quantization of the zeroth Pauli coefficient of $H$ is given in Remark \ref{rem:local}).

For the following, recall that $q_N=m[1_{\l-N,N\r}]\in\mL^0(\fh)$ for all $N\in\N$ as defined after \eqref{prp-b(A)}.
 
\bl[Hamiltonian densities]
Let $H\in\mL(\fH)$ be a Hamiltonian satisfying Assumption \ref{ass:H} \ref{HTheta}, \ref{HLR=0}, and \ref{nontrv}. Then:

\bn[label=(\alph*),ref={\it (\alph*)}]
\setlength{\itemsep}{0mm}
\item
The  selfdual second quantizations of the local first, second and third Pauli coefficient of $H$ in the fermionic  picture are given, for all $N\in\N$ satisfying \eqref{N-lbound}, by
 \ba
 b((q_N h_1q_N)\sigma_1)
&=-2\ii \sum_{n\in\l1,\nu\r } c_{1,n}
\hspace{-1mm}\sum_{x\in\l-N,N-n\r } (a_x^\ast a_{x+n}^\ast-a_{x+n} a_x),\\
b((q_N h_2q_N)\sigma_2)
&=-2\sum_{n\in\l1,\nu\r } c_{2,n}
\hspace{-1mm}\sum_{x\in\l-N,N-n\r } (a_x^\ast a_{x+n}^\ast+a_{x+n} a_x),\\
b((q_N h_3q_N)\sigma_3)
&=c_{3,0}
\hspace{-1mm}\sum_{x\in\l-N,N\r} (2a_x^\ast a_x-1)
+2\sum_{n\in\l1,\nu\r } c_{3,n}
\hspace{-1mm}\sum_{x\in\l-N,N-n\r } (a_x^\ast a_{x+n}+a_{x+n}^\ast a_x). 
 \ea
 
\item
In the spin picture we have, for all $x\in\Z$ and all $n\in\N$, 
 \ba
 \lefteqn{a_x^\ast a_{x+n}^\ast-a_{x+n} a_x}\nonumber\\
 &=\begin{cases}
\hfill -\frac\ii 2 \big(\sigma_1^{(x)}\sigma_2^{(x+1)}+ \sigma_2^{(x)}\sigma_1^{(x+1)}\big), & n=1,\\
-\frac\ii 2 \big(\sigma_1^{(x)}\big(\prod_{i\in\l1,n-1\r}\sigma_3^{(x+i)}\big) \sigma_2^{(x+n)}+ \sigma_2^{(x)}\big(\prod_{i\in\l1,n-1\r}\sigma_3^{(x+i)}\big)\sigma_1^{(x+n)}\big) , & n\ge 2,
\end{cases}\\
\label{nXY-1}
\lefteqn{a_x^\ast a_{x+n}^\ast+a_{x+n} a_x}\nonumber\\
&=\begin{cases}
\hfill -\frac12 \big(\sigma_1^{(x)}\sigma_1^{(x+1)}- \sigma_2^{(x)}\sigma_2^{(x+1)}\big), & n=1,\\
-\frac12 \big(\sigma_1^{(x)}\big(\prod_{i\in\l1,n-1\r}\sigma_3^{(x+i)}\big) \sigma_1^{(x+n)}- \sigma_2^{(x)}\big(\prod_{i\in\l1,n-1\r }\sigma_3^{(x+i)}\big)\sigma_2^{(x+n)}\big) , & n\ge 2,
\end{cases}
\ea

\ba
\label{nXY-2}
\lefteqn{a_x^\ast a_{x+n}+a_{x+n}^\ast a_x}\nonumber\\
&=\begin{cases}
\hfill -\frac12 \big(\sigma_1^{(x)}\sigma_1^{(x+1)}+ \sigma_2^{(x)}\sigma_2^{(x+1)}\big), & n=1,\\
-\frac12 \big(\sigma_1^{(x)}\big(\prod_{i\in\l1,n-1\r }\sigma_3^{(x+i)}\big) \sigma_1^{(x+n)}+ \sigma_2^{(x)}\big(\prod_{i\in\l1,n-1\r }\sigma_3^{(x+i)}\big)\sigma_2^{(x+n)}\big) , & n\ge 2.
\end{cases}
\ea
 Moreover, we have $2a_x^\ast a_x-1=\sigma_3^{(x)}$ for all $x\in\Z$.
 \en
 \el
 
 \bprf
 See Remark \ref{rem:local} and, for example, \cite{AP2003}.
\eprf
 
 \end{appendix}



\begin{thebibliography}{10} 

\bibitem{AJR2020} Andr\'eys S, Joye A, and Raqu\'epas R 2020
{\it Fermionic walkers driven out of equilibrium}
arXiv:2009.00604 [math-ph] (preprint)

\bibitem{Araki1968} Araki H 1968
{\it On the diagonalization of a bilinear Hamiltonian by a  Bogoliubov transformation}
Publ. RIMS Kyoto Univ. Ser. A 4 387-412

\bibitem{Araki1971} Araki H 1971
{\it On quasifree states of CAR and Bogoliubov automorphisms}
Publ. RIMS Kyoto Univ. 6 385-442

\bibitem{Araki1984} Araki H 1984 
{\it On the XY-model on two-sided infinite chain}
Publ. RIMS Kyoto Univ. 20 277-296

\bibitem{Araki1987} Araki H 1987 
{\it Bogoliubov automorphisms and Fock representations of canonical 
anticommutation relations}
Contemp. Math. 62 23-141

\bibitem{AH2000} Araki H and Ho T G 2000 
{\it Asymptotic time evolution of a partitioned infinite two-sided isotropic XY-chain}
Proc. Steklov Inst. Math. 228 191-204 

\bibitem{AM1985} Araki H and Matsui T 1985
{\it Ground states of the XY model}
Commun. Math. Phys. 101 213-245

\bibitem{A2011} Aschbacher W H 2011
{\it Broken translation invariance in quasifree fermionic correlations
out of equilibrium}
J. Funct. Anal. 260 3429-3456

\bibitem{A2016} Aschbacher W H 2016
{\it On a quantum phase transition in a steady state out of equilibrium}
J. Phys. A: Math. Theor. 49 415201 1-21

\bibitem{A2019} Aschbacher W H 2019
{\it A rigorous scattering approach to quasifree fermionic systems out of equilibrium}
J. Non-Equilib. Thermodyn. 44 (3) 261-275

\bibitem{AJPP2007} Aschbacher W H, Jak\v si\'c V, Pautrat Y, and Pillet C-A 2007
{\it Transport properties of quasi-free fermions}
J. Math. Phys. 48 032101 1-28

\bibitem{AP2003} Aschbacher W H and Pillet C-A 2003
{\it Non-equilibrium steady states of the XY chain} 
J. Stat. Phys. 112 1153-1175

\bibitem{AJP2006} Attal S, Joye A, and Pillet C-A (Eds.) 2006
{\it Open quantum systems I, II, III}
Lect. Notes Math. 1880, 1881, 1882
(Springer)

\bibitem{BW} Baumg\"artel H and Wollenberg M 1983
{\it Mathematical scattering theory} 
(Birkh\"auser)

\bibitem{BS2006} B\"ottcher A and Silbermann B 2006
{\it Analysis of Toeplitz operators}
(Springer)

\bibitem{BR} Bratteli O and Robinson D W 1987, 1997
{\it Operator algebras and quantum statistical mechanics 1, 2}
(Springer)

\bibitem{CSP1969} Culvahouse J W, Schinke D P, and Pfortmiller L G 1969
{\it Spin-spin interaction constants from the hyperfine structure of coupled ions}
Phys. Rev. 177 454-464

\bibitem{Emch2009} Emch G G 1972
{\it Algebraic methods in statistical mechanics and quantum field theory}
(Reprint, Dover, 2009)

\bibitem{HW1996} Hardt V and Wagenf\"uhrer E 1996
{\it Spectral properties of a multiplication operator}
Math. Nachr. 178 135-156

\bibitem{Hunziker} Hunziker W (n.d.) 
Lecture notes, ETH Z\"urich 

\bibitem{Jost1973} Jost R 1973
{\it Quantenmechanik II}
(Verlag der Fachvereine an der ETH Z\"urich)

\bibitem{Katznelson2004} Katznelson Y 2004
{\it An introduction to harmonic analysis}
(Cambridge University Press)

\bibitem{Kechris} Kechris A S 1995
{\it Classical descriptive set theory}
(Springer)

\bibitem{LSM1961} Lieb E, Schultz T, and Mattis D 1961 
{\it Two soluble models of an antiferromagnetic chain}
Ann. Physics 16 407-466

\bibitem{MO2003} Matsui T and Ogata Y 2003
{\it Variational principle for non-equilibrium steady states of the XX
model}
Rev. Math. Phys. 15 905-923

\bibitem{MK2004} Mikeska H-J and Kolezhuk A K 2004
{\it One-dimensional magnetism} in Schollw\"ock U, Richter J, Farnell D J J, and 
Bishop R F (Ed.) 
{\it Quantum Magnetism} Lect. Notes Phys. 645 1-83
(Springer)

\bibitem{Nambu1950} Nambu Y 1950
{\it A note on the eigenvalue problem in crystal statistics}
Progr. Theor. Phys. 5 1 1-13

\bibitem{Picavet2003} Picavet G and Picavet-L'Hermite M 2003
{\it Trigonometric polynomial rings}
Lecture notes Pure Appl. Math. (Marcel Dekker) 231 419-433.

\bibitem{Primas} Primas H 1983
{\it Chemistry, quantum mechanics, and reductionism}
(Springer)

\bibitem{Ruelle2001} Ruelle D 2001
{\it Entropy production in quantum spin systems}
Commun. Math. Phys. 224 3-16

\bibitem{Sewell2014} Sewell G 1986
{\it Quantum theory of collective phenomena}
(Reprint, Dover, 2014)

\bibitem{Simon1977} Simon B 1977
{\it Notes on infinite determinants of Hilbert space operators}
Adv. Math. 24 244-273

\bibitem{SGOVR2001} Sologubenko A V, Giann\`o K, Ott H R, Vietkine A, and Revcolevschi A 2001
{\it Heat transport by lattice and spin excitations in the spin-chain compounds ${\rm SrCuO_2}$ and ${\rm Sr_2CuO_3}$}
Phys. Rev. B 64 054412 1-11

\bibitem{S1971} Suzuki M 1971 
{\it Relationship among exactly soluble models of critical phenomena, I}
Prog. Theor. Phys. 46 (5) 1337-1359

\bibitem{Weidmann1980} Weidmann J 1980
{\it Linear operators in Hilbert spaces}
(Springer)

\end{thebibliography}
\end{document}